\documentclass[useAMS,usenatbib]{mn2e}
\usepackage{latexsym}
\usepackage{amsmath}
\usepackage{amssymb}
\usepackage{fnpos}
\usepackage{scrextend}
\usepackage{scalerel}
\usepackage{hyperref}
\usepackage{tabularx}
\usepackage{xspace}
\usepackage{enumerate}
\usepackage{graphicx}
\usepackage{caption}
\usepackage{gensymb}
\usepackage{epstopdf}
\usepackage{longtable}
\usepackage{lscape}
\usepackage{subcaption}
\usepackage{pdfpages}

\newcommand{\eal}[2]{\ifmmode{\mathrm{#1\,#2}}\else{#1\textsc{$\,$\lowercase{#2}}}\fi\xspace}
\newcommand{\feal}[2]{\ifmmode{\mathrm{#1\,#2}}\else{[#1\textsc{$\,$\lowercase{#2}}]}\fi\xspace}
\newcommand{\hfeal}[2]{\ifmmode{\mathrm{#1\,#2}}\else{#1\textsc{$\,$\lowercase{#2}}]}\fi\xspace}

\title[nova V407 Lup]{Multiwavelength observations of V407 Lupi (ASASSN-16kt) --- a very fast nova erupting in an intermediate polar}

\author[Aydi et al.]{E. Aydi$^{1,2}$\thanks{E-mail: eaydi@saao.ac.za}, M. Orio$^{3,4}$, A. P. Beardmore$^{5}$, J.-U. Ness$^{6}$, K. L. Page$^{5}$, N. P. M. Kuin$^{7}$,  \newauthor F. M. Walter$^{8}$, D. A. H. Buckley$^{1}$, S. Mohamed$^{1,2}$, P. Whitelock$^{1,2}$, J. P. Osborne$^{5}$, 
\newauthor J. Strader$^{9}$, L. Chomiuk$^{9}$, M. J. Darnley$^{10}$, A. Dobrotka$^{11}$, A. Kniazev$^{1,12,13}$, 
\newauthor B. Miszalski$^{1,12}$, G. Myers$^{14}$, N. Ospina$^{15}$, M. Henze$^{16}$, S. Starrfield$^{17}$, \newauthor and C. E. Woodward$^{18}$\\\\
$^{1}$South African Astronomical Observatory, P.O. Box 9, 7935 Observatory, South Africa\\
$^{2}$Department of Astronomy, University of Cape Town, Private Bag X3, Rondebosch 7701, South Africa\\
$^{3}$INAF--Osservatorio di Padova, vicolo dell’ Osservatorio 5, I-35122 Padova, Italy\\
$^{4}$Department of Astronomy, University of Wisconsin, 475 N. Charter Str., Madison, WI 53704, USA\\
$^{5}$X-ray and Observational Astronomy Group, Department of Physics \& Astronomy, University of Leicester, LE1 7RH, UK\\
$^{6}$\textit{XMM-Newton} Observatory SOC, European Space Astronomy Centre, Camino Bajo del Castillo s/n, Urb. Villafranca del Castillo,\\
E-28692 Villanueva de la Ca\~nada, Madrid, Spain\\
$^{7}$University College London, Mullard Space Science Laboratory, Holmbury St. Mary, Dorking RH5 6NT, UK\\
$^{8}$Department of Physics and Astronomy, Stony Brook University, Stony Brook NY 11794-3800, USA\\
$^{9}$Center for Data Intensive and Time Domain Astronomy, Department of Physics and Astronomy, Michigan State University, East Lansing,\\
MI 48824\\
$^{10}$Astrophysics Research Institute, Liverpool John Moores University, IC2 Liverpool Science Park, Liverpool, L3 5RF, UK\\
$^{11}$Advanced Technologies Research Institute, Slovak University of Technology in Bratislava, Paulinska 16, 91724 Trnava, Slovak Republic\\
$^{12}$Southern African Large Telescope Foundation, PO Box 9, Observatory 7935, South Africa\\
$^{13}$Special Astrophysical Observatory of RAS, Nizhnij Arkhyz, Karachai-Circassia 369167, Russia\\
$^{14}$American Association of Variable Star Observers, 5 Inverness Way, Hillsborough, CA 94010\\
$^{15}$Via Pietro Pomponazzi, 33, 35124 Padova, Italy\\
$^{16}$Department of Astronomy, San Diego State University, San Diego, CA 92182, USA\\
$^{17}$School of Earth and Space Exploration, Arizona State University Tempe, Arizona 85287-1404, USA\\
$^{18}$Minnesota Institute for Astrophysics, University of Minnesota, Minneapolis, MN 55455, USA\\
}

\begin{document}
\date{Accepted 2018 July 1. Received 2018 June 8; in original form 2018 April 17}
\pagerange{\pageref{firstpage}--\pageref{lastpage}} \pubyear{2018}
\maketitle

\label{firstpage}
\begin{abstract}
We present a detailed study of the 2016 eruption of nova V407 Lupi (ASASSN-16kt), including optical, near-infrared, X-ray, and ultraviolet data from SALT, SMARTS, SOAR, \textit{Chandra}, \textit{Swift}, and \textit{XMM-Newton}. Timing analysis of the multiwavelength light-curves shows that, from 168 days post-eruption and for the duration of the X-ray supersoft source phase, two periods at 565\,s and 3.57\,h are detected. We suggest that these are the rotational period of the white dwarf and the orbital period of the binary, respectively, and that the system is likely to be an intermediate polar. The optical light-curve decline was very fast ($t_2 \leq$ 2.9\,d), suggesting that the white dwarf is likely massive ($\gtrsim 1.25$\,M$_{\odot}$). The optical spectra obtained during the X-ray supersoft source phase exhibit narrow, complex, and moving emission lines of \eal{He}{II}, also characteristics of magnetic cataclysmic variables. The optical and X-ray data show evidence for accretion resumption while the X-ray supersoft source is still on, possibly extending its duration.
\end{abstract}


\section{Introduction}
\label{Intro}
Classical novae (CNe) are stellar eruptions that take place within the surface layers of accreting white dwarfs (WDs) in cataclysmic variable (CV) systems. These are interacting binaries consisting of a WD primary accreting from a secondary that typically fills its Roche lobe (see, e.g., \citealt{Warner_1995}). The hydrogen-rich material accreted by the WD accumulates on its surface causing an increase in pressure and density to a level where a thermonuclear runaway (TNR) is triggered \citep{Starrfield_1989,Starrfield_etal_2008,Jose_Shore_2008,Starrfield_etal_2016}. Due to this, a part of the accreted envelope is ejected. The fast expanding envelope moves together with the optical photosphere \citep{Hachisu_etal_2006,Hachisu_Kato_2014}, in what is known as the ``fireball stage''. During this the brightness of the star increases by 8 up to 15 mag \citep{Payne-Gaposchkin_1964} and reaches its maximum visual brightness when the optical photosphere attains its maximum radius \citep{Warner_1995, Hachisu_Kato_2014}. 
Following the TNR, the remaining hydrogen-rich, accreted envelope continuously burns on the WD surface and may become visible when the expanding ejecta become optically thin. 
 
After maximum the optical depth of the expanding ejecta decreases progressively and therefore the optical photosphere starts shrinking. This shifts  the emission to higher energies, and would peak in the soft X-ray band and a SuperSoft Source (SSS) emerges, unless the ejecta become optically thin before that happens, in which case the SSS emergence is due to the visibility of the WD photosphere, which is at that time extended due to the ongoing nuclear burning \citep{Krautter_2008,Osborne_2015}. As the hydrogen in the surface layers is consumed, the SSS ``turns off'' and the system eventually evolves back to quiescence (for a series of reviews of CNe see \citealt{Bode_etal_2008}). 

Observing CNe across many wavelengths is essential to provide a full picture of the eruption and its characteristics, such as: the different physical parameters (e.g., temperatures and velocities), the gas ejection mechanisms, the radiative processes, the nuclear reactions, the shocks in the ejecta/wind, the dust formation, and the properties of the progenitor. 
However, despite their importance systematic multiwavelength studies are still lacking and only a few CNe have been followed panchromatically and in detail (see, e.g., \citealt{Shore_etal_2013,Schwarz_etal_2015,Bode_etal_2016,Shore_etal_2016,Darnley_etal_2016,Li_etal_2017_nature,Mason_etal_2018,Aydi_etal_2018}). Therefore, observing these events in multiple frequency bands is essential for a better understanding of these individual events and ultimately adding to the wealth of knowledge in the field.

While most nova eruptions occur in non-magnetic CVs, where the WD is usually weakly magnetized (magnetic field of the WD is $\lesssim$10$^{6}$\,G), a few novae have been seen to occur in magnetic CVs (mCVs). These systems form a sub-group of CVs in which the WD is highly magnetized, and they include two main sub-types: polars which have the strongest magnetic fields ($\gtrsim$ 10$^7$\,G) and intermediate polars (IPs) with weaker fields (10$^6$\,--\,10$^7$\,G). The strong magnetic field of polars causes the spin (rotational) period of the WD and the orbital period of the binary to synchronize. This is not the case for IPs where the optical, X-ray, and ultraviolet (UV) light-curves might show multiple periodicities, modulated on the orbital period of the binary (with typical values of 3 to 6\,h), the spin period of the WD (with values ranging between $\sim$ 0.5 up to $\sim$ 70\,min), and the sideband periods (typically dominated by the beat period). Although in such systems the mass-accretion mechanism is dominated at some point by the magnetic field of the WD, the accretion onto the surface of WD can still result in a nova eruption. For a comprehensive review of mCVs and their mass-transfer/ -accretion mechanisms see \citet{Warner_1995} and \citet{Hellier_2001}. 

The only nova known to have erupted in a polar system is V1500 Cyg \citep{Kaluzny_Semeniuk_1987,Stockman_etal_1988}. The eruption, which occurred in 1975, was one of the most luminous on record, along with CP Pup \citep{Warner_1985} and nova SMCN 2016-10a \citep{Aydi_etal_2018}. On the other hand, a few novae have been either confirmed (GK Per, DQ Her, V4743 Sgr, and Nova Scorpii 1437 AD) or suggested (V533 Her, DD Cir, V1425 Aql, M31N 2007-12b, and V2491 Cyg) to occur in IPs (see, e.g., \citealt{Osborne_etal_2001,King_etal_2002,Warner_Woudt_2002,Bianchini_etal_2003,Woudt_Warner_2003,Woudt_Warner_2004,Leibowitz_etal_2006,Dobrotka_Ness_2010,Pietsch_etal_2011,Zemko_etal_2016,Potter_Buckley_2018} and references therein). The effect of the magnetic field on the nova eruption and progress is not well understood.

In this paper we present a multiwavelength study of nova V407 Lupi which was discovered by the All-Sky Automated Survey for Supernovae (ASAS-SN)\footnote{\url{http://www.astronomy.ohio-state.edu/asassn/index.shtml}} on HJD 2457655.5 (2016 September 24.0 UT; \citealt{ATel_9538}) at $V$ = 9.1 and is located at equatorial coordinates of $(\alpha, \delta)_{\rm J2000.0}$ = (15$^{\mathrm{h}}$29$^{\mathrm{m}}$01.82$^{\mathrm{s}}$, --44$^{\circ}$49$'$40$\arcsec$.89) and Galactic coordinates of ($l, b$) = (330$^{\circ}$.09, 9$^{\circ}$.573). The last ASAS-SN pre-discovery observation reported the source at $V >$ 15.5 on HJD 2457651.5. Therefore, in the following we assume HJD 2457655.5 (2016 September 24.0 UT) is $t_0$ (eruption start). Our study consists of a set of comprehensive, multiwavelength observations obtained with the aim of studying in detail the post-eruption behaviour of the nova at optical, near-infrared (NIR), ultraviolet (UV), and X-ray wavelengths. Note that this object has not been seen in eruption before. This paper is structured as follows: in Section~\ref{obs_sec} we present the observations and data reduction. The analysis of the photometric and spectroscopic results are given in Sections~\ref{photo_sec} and~\ref{spec_sec}, respectively. We present the discussion in Section~\ref{Disc_sec}, while Section~\ref{Concl_sec} contains a summary and the conclusions.

\section{Observations and data reduction}
\label{obs_sec}
\subsection{SMARTS observations and data reduction}
Since 2016 September 26 (day 2) the eruption has been monitored using the Small and Moderate Aperture Research Telescope System (SMARTS) to obtain optical \textit{BVRI} and NIR \textit{JHK} photometric observations. The integration times at \textit{JHK} were 15\,s (three 5\,s dithered images) before 2016 October 6 (day 12) and 30\,s thereafter. Optical observations are single images of 30\,s, 25\,s, 20\,s, and 20\,s integrations, respectively, in \textit{BVRI} prior to 2016 October 6 (day 12), and uniformly 50\,s thereafter. The procedure followed for reducing the SMARTS photometry is detailed in \citet{Walter_etal_2012}. Fig.~\ref{Fig:BVRI_JHK_LC} represents the SMARTS \textit{BVRI} and \textit{JHK} observations. See Table~\ref{table:SMARTS} for a log of the observations.

\subsection{SALT high-resolution Echelle spectroscopy}  
We used the SALT High Resolution Spectrograph (HRS; \citealt{Barnes_etal_2008}; \citealt{Bramall_etal_2010}; \citealt{Bramall_etal_2012}; \citealt{Crause_etal_2014}) to obtain observations on the nights of 2017 March 6; April 20; May 31; June 13, 29; July 05, 24, 29; and August 06 (respectively, days 164, 209, 250, 263, 279, 285, 304, 309, and 317). HRS, a dual beam, fibre-fed Echelle spectrograph housed in a temperature stabilized vacuum tank, was used in the low-resolution (LR) mode to obtain all the observations. This mode provides two spectral ranges: blue (3800\,--\,5550\,$\mathrm{\AA}$) and red (5450\,--\,9000\,$\mathrm{\AA}$) at a resolution of $R \sim 15000$. A weekly set of HRS calibrations, including four ThAr + Ar arc spectra and four spectral flats, is obtained in all the modes (low, medium, and high-resolution). All the HRS observations are 1800\,s exposures. See Table~\ref{table:SALT} for a log of the observations.

The primary reduction was conducted using the SALT science pipeline \citep{Crawford_etal_2010} which includes over-scan correction, bias subtraction, and gain correction. The rest of the reduction and relative flux calibration was done using the standard MIDAS FEROS \citep{Stahl_etal_1999} and $echelle$ \citep{Ballester_1992} packages. The reduction procedure is described in detail in \citet{kniazev_etal_2016}. Note that absolute flux calibration is not feasible with SALT data. As part of the SALT design, the effective area of the telescope constantly changes with the moving pupil during the track and exposures. 

\subsection{SOAR medium-resolution spectroscopy}
We performed optical spectroscopy of the nova on 2017 June 20 (day 270) using the Goodman spectrograph \citep{Clemens_etal_2004} on the 4.1\,m Southern Astrophysical Research (SOAR) telescope. We obtained a single exposure of 600\,s, using  a 600\,l\,mm$^{-1}$ grating and a 1.07{\arcsec} slit to provide a resolution of $\sim$ 1.3\,$\mathrm{\AA}$ over the range of 3200\,--\,6800\,$\mathrm{\AA}$. The spectrum was reduced using the standard routine and a relative flux calibration has been applied.   

We also obtained two spectra using the Goodman spectrograph on the nights of 2018 January 20 and 21 (days 484 and 485), each of 20\,min exposure. For these two observations we used a setup with a 2100\,l\,mm$^{-1}$ grating and a 0.95\arcsec slit, yielding a resolution of about 0.88 $\mathrm{\AA}$ full width at half maximum (FWHM) (56\,km\,s$^{-1}$) over a wavelength range of 4280\,--\,4950 \AA$\mathrm{\AA}$. The spectra were reduced and optimally extracted in the usual manner.

\subsection{\textit{Swift} X-ray and UV observations}

\subsubsection{\textit{Swift} XRT observations}

The Neil Gehrels \textit{Swift} Observatory (hereafter, \textit{Swift}; \citealt{Gehrels_etal_2004}) first observed V407 Lup on 2016 September
26, two days after the discovery. The high optical
brightness of the nova at this time meant that the UV/Optical
Telescope (UVOT; \citealt{Roming_etal_2005}) could not be used, and the X-ray
Telescope (XRT; \citealt{Burrows_etal_2005}) needed to be operated in Windowed
Timing (WT) mode for the initial observation, in order to minimize the effects of optical loading,
whereby a large number of optical photons can pile-up to appear as a
false X-ray signature\footnote{\url{http://www.swift.ac.uk/analysis/xrt/optical\_loading.php}}. No X-ray source was detected at this time ($<$ 0.02 count\,s$^{-1}$; grade
0 -- single pixel events), or in the individual subsequent daily observations taken in Photon
Counting (PC) mode between 2016 October 2 and 9 (days 8\,--\,15). Coadding
these PC data, there is a source detected at the 99~per~cent confidence
level, with a count rate of (1.4$^{+0.5}_{-0.4}$) $\times 10^{-3}$ count\,s$^{-1}$
(grade 0). There was also a suggestion of a detection in the October 10
(day 16) data alone. These detections could not be confirmed at a higher
significance before V407 Lup became too close to the Sun for \textit{Swift} to observe on 2016 October 13 (day 19).

Observations recommenced on 2017 February 21 (day 150), finding a bright, supersoft X-ray source \citep{ATel_10632}, with a count rate of 56.1\,$\pm$\,0.3\,count\,s$^{-1}$. By this
time the UVOT could also be safely operated, and a UV source with \textit{uvw2}
= 13.49\,$\pm$\,0.02\,mag was measured. Daily observations were performed
between 2017 February 22 and 28 (days 151\,--\,157), followed by observations
approximately every two to three days until 2017 April 24 (day 212).
Observations were continued, though with a slightly lower cadence of
every four days, until 2017 September 10 (day 351), with the cadence then
decreasing to every eight days until 2017 October 13 (day 384), when the
Sun constraint began again. A final dataset was obtained once the source again emerged from the Sun constraint, on 2018 January 23 (day 486). All observations were typically 0.5-1\,ks
in duration, with the exception of the final observation which was 3\,ks. In addition to this regular monitoring a high cadence
campaign was performed between 2017 July 4\,--\,8 (days 283\,--\,287), aimed at pinning down the
periodicity seen in the UVOT light-curve (see Section~\ref{timing_study}). In this case,
$\sim$~500\,s snapshots were taken approximately every 6\,h. Because of
complications caused by a nearby bright star, no UVOT data were
actually collected, and so the campaign was repeated from 2017 July 28\,--\,August 1 (days 307\,--\,311) using an offset pointing to avoid this problem. The \textit{Swift} XRT observations log is given in Table~\ref{table:XRT_log}.

The {\it Swift} data were processed with the standard HEASoft tools (version
6.20)\footnote{\url{https://heasarc.gsfc.nasa.gov/FTP/software/ftools/release/archive/Release_Notes_6.20}}, and analysed using the most up-to-date calibration files. All observations between 2017 February 21 and August 11 (days 150 and 321 post-eruption), inclusive, were obtained using WT mode, because of the high brightness of the X-ray source. These data were
extracted using a circular region of a radius of 20 pixels (1 pixel = 2.36\,arcsec)
for the source, and a background annulus as described at
http://www.swift.ac.uk/analysis/xrt/backscal.php. Observations from 2017 August 15 (day 325) onwards were taken with the XRT in PC mode; the
first three of these suffered from pile-up, so an annulus (outer
radius 30 pixels; inner exclusion radius decreasing from seven to
three pixels) was used when extracting the source counts. The later
data were analysed using circular regions, decreasing in radius from
20 to 10 pixels as the source further faded. Background counts were
estimated from near-by, source-free circular regions of 60 pixels
radius.

The X-ray spectra were binned to provide a minimum of 1 count\,bin$^{-1}$ to facilitate Cash statistic \citep{Cash_1979} fitting within {\sc xspec} \citep{Arnaud_1996}.

\subsubsection{\textit{Swift} UVOT observations}

Observations with the $Swift$ UVOT instrument started once the nova could be observed 
again after its passage behind the Sun. On 2017 February 26 (day 156), the nova was observed in the 
\textit{uvw2} ($\lambda_{\mathrm{central}}$ = 1928\,$\mathrm{\AA}$) filter; the next day in the \textit{uvm2} ($\lambda_{\mathrm{central}}$ = 2246\,$\mathrm{\AA}$) filter. Photometric observations continued until 
2017 October 13 (day 385).  A log of the observations is given in Table~\ref{table:XRT_log}.

Weekly observations with the UV grism started on 2017 March 7, until April 23 (day 165 until day 212) when they were discontinued as the brightness was too low. The UV grism provides a spectral range of $\sim$ 1700\,--\,5100\,$\mathrm{\AA}$. The log of the observations are given in Table~\ref{uvgrism_obs_table}. The eight grism observations were reduced using the UVOTPY software  \citep{Kuin_2014} using the calibration described in \citet{Kuin_etal_2015} with a recent update to the sensitivity affecting the response of the  UV grism mainly below 2000\,$\mathrm{\AA}$ \footnote{see:\url{http:/mssl.ucl.ac.uk/npmk/Grism/}}. The net continuum counts in the UV part of the individual spectra are very low, so the first four and last four spectra were summed, which also improved the signal to noise (S/N) of the weaker lines significantly. No reddening correction has been applied to the spectra.

\subsection{\textit{XMM-Newton} X-ray and UV observations}

V407 Lup was observed by \textit{XMM-Newton} from 2017 March 11, 11:45 to 17:08 UT, 168.5 days post-eruption with an exposure duration of 23\,ks \citep{ATel_10756}. The \textit{XMM-Newton} observatory consists of five different X-ray instruments behind three mirrors plus an optical monitor (OM;\citealt{Mason_etal_2001,Talavera_2009}), which all observe simultaneously. For a full description of the X-ray instruments onboard of \textit{XMM-Newton} see \citet{Jansen_etal_2001}, \citet{den_Herder_etal_2001}, \citet{Struder_etal_2001}, \citet{Turner_etal_2001}, and \citet{Aschenbach_2002}.

In this work, we only used the spectra and light-curves from the Reflection Grating Spectrometer (RGS;\citealt{den_Herder_etal_2001}), the light-curves from the European Photon Imaging Camera (EPIC)/pn \citep{Struder_etal_2001}, and the OM light-curves/grism spectra (see Table~\ref{table:XMM-Newton} for a log of the observations). 

The OM took five exposures, one of which with the visible grism provided a spectrum between 3000 and 6000\,$\mathrm{\AA}$ (\textit{XMM-Newton} could not take a $U$-grism spectrum because of a contaminating nearby star, so we only obtained a $V$-grism spectrum); and four with Science User Defined imaging plus fast mode, which provided a UV light-curve with the \textit{uvw1} ($\lambda \sim 2910\pm 830$\,$\mathrm{\AA}$) filter. 

The RGS provides a spectral range of 6\,--\,38\,$\mathrm{\AA}$, fully covering the Wien tail of SSS spectra (30\,--\,80\,eV) while the Rayleigh Jeans tail is not visible (owing to interstellar absorption). It also provides X-ray light-curves in the same energy range. RGS consists of two instruments RGS1 and RGS2. One of the nine CCDs in RGS2 is suffering from some technical problems. We Combine the RGS1 and RGS2 spectra in order to overcome the problem of the broken CCD in the RGS2 instrument. The EPIC/pn instrument was operated in Timing Mode with medium filter providing another X-ray light-curve. 

\subsection{\textit{Chandra} X-ray observations}

V407 Lup was observed with the \textit{Chandra X-ray Observatory} (hereafter, \textit{Chandra}; \citealt{Weisskopf_etal_2000}), using the High Resolution Camera (HRC; \citealt{Murray_etal_1997}) and the Low Energy Transmission Grating (LETG; \citealt{Brinkman_etal_2000_Feb,Brinkman_etal_2000}) on 2018 August 30 (day 340.6), with an exposure time of 34\,ks. The LETG instrument provides a spectrum over a range of $\sim$ 1.2\,--\,175\,$\mathrm{\AA}$ (0.07\,--\,10.33\,keV), however most of the flux from nova V407 Lup is in the range 15\,--\,50\,$\mathrm{\AA}$ (0.24\,--\,0.82\,keV). The light-curve provided by the HRC instrument has a range of 0.06\,--\,10\,keV.


\begin{figure*}
\begin{center}
  \includegraphics[width=\textwidth]{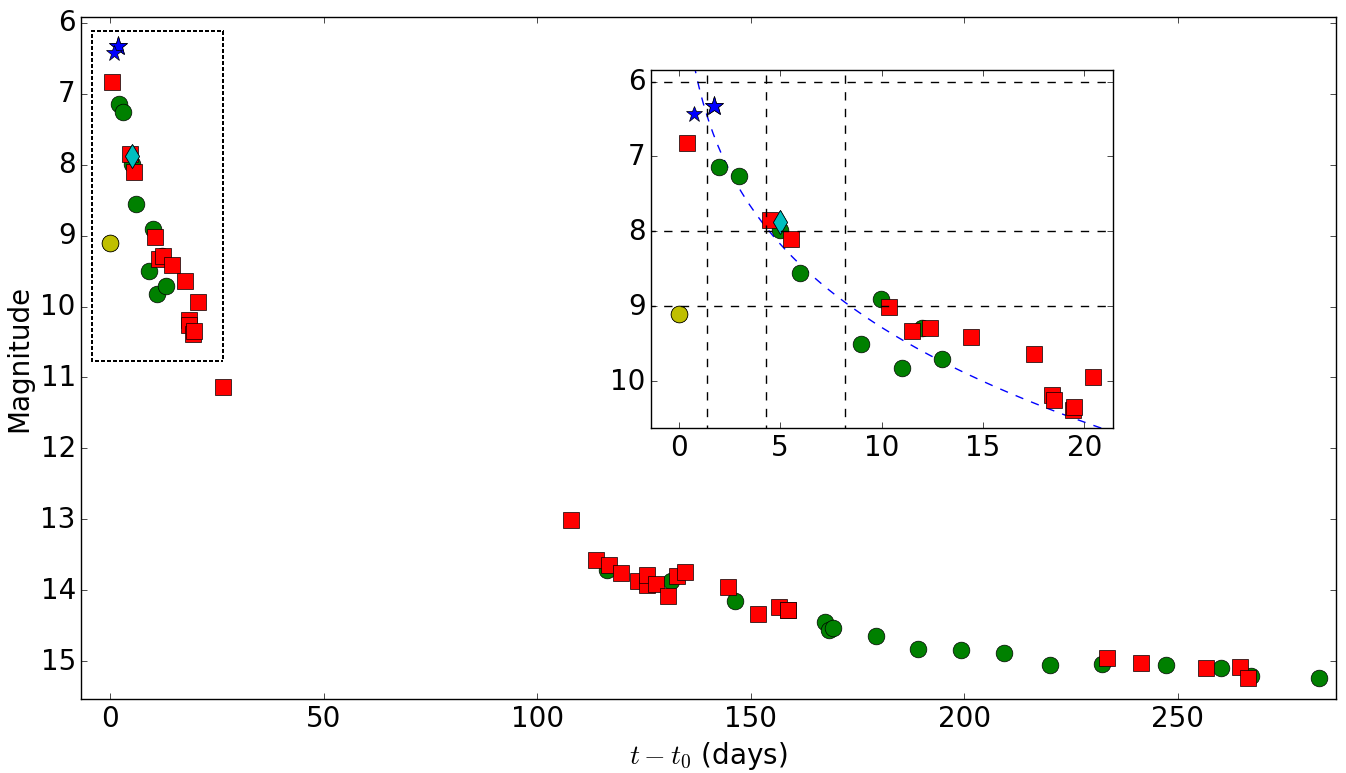}
\caption{$V$-band light-curve using data from SMARTS (green circles), AAVSO (red squares), and ATels from \citet{ATel_9538} (yellow circle), \citet{ATel_9550} (blue stars), and \citet{ATel_9564} (cyan diamond). A zoom-in plot around the first $\sim$ 25 days is added for clarity. Within this plot: the blue dashed line represents a power-law fit to the data (before day 25), excluding the discovery measurement; the black horizontal dashed lines represent from top to bottom $V_{\mathrm{max}}$, $V_{\mathrm{max}}$ + 2, and $V_{\mathrm{max}}$ + 3, respectively; the black vertical dashed lines represent from left to right, $t_{\mathrm{max}}$, $t_2$, and $t_3$, respectively.}
\label{Fig:LC_V}
\end{center}
\end{figure*}

\section{Photometric results and analysis}
\label{photo_sec}
\subsection{Optical light-curve parameters}
\label{Optical_photo_sec}

Several parameters characterize nova light-curves including the rise rate, the rise time to maximum light, the maximum light, the decline rate, and the decline behaviour (see e.g. \citealt{Hounsell_etal_2010,Cao_etal_2012}). Nova V407 Lup was not extensively observed during its rise to maximum. 

Based on $V$-band and visual (\textit{Vis}) measurements from SMARTS and the American Association of Variable Star Observers (AAVSO)\footnote{\url{https://www.aavso.org/}}, we see that the nova reached $V$ = 6.8, 0.4\,d  after $t_0$. Then it was reported at $V = 6.4$ on HJD 2457656.24 \citep{ATel_9564} and \textit{Vis} =  6.3 on HJD 2457656.57, reaching \textit{Vis} = 5.6 on  HJD 2457656.90 (see Table~\ref{table:photo_max}). Half a day later, the nova was at $V= 6.33$ and \textit{Vis} =  6.5, then dropped to $V = 7.85$ in around two days, indicating that the decline had started. Hence, we assume HJD 2457656.90 as $t_{\mathrm{max}}$ (day 1.4).

Caution is required in interpreting magnitudes of novae. This is particularly so for \textit{Vis} where $\rm H \alpha$ may contribute to the flux detected by eye, but  not to CCD $V$-magnitudes. Furthermore, normal transformation relations for CCD magnitudes cannot be used for objects with strong emission lines. Unfortunately, no spectra were taken around maximum light, so the contribution and development of line emission is unknown until later times (day 5; \citealt{Izzo_etal_2018}). In the following, we assume that maximum light was $V_{\mathrm{max}} \leq 6.0$ on HJD 2457656.90. Since novae do not usually have strong line emission at maximum light  \citep{Van_den_Bergh_Younger_1987} and
given that simultaneous $V$ and \textit{Vis} measurements show very similar magnitudes, before and after maximum light, it may be that $V\sim$ \textit{Vis} =5.6, on HJD 2457656.90, is a reasonable estimate. However, we use $V_{\mathrm{max}} \leq 6.0$ for the rest of the analysis to avoid overestimating the maximum brightness. 
 
We construct a $V$-band light-curve by combining the published photometry from different telescopes and instruments and the SMARTS data (see Fig.~\ref{Fig:LC_V}), noting the above caveat. This shows a rapid rise and a sharp peak followed by a fast decline. Such behaviour is characteristic of the S-class nova light-curves \citep{Strope_etal_2010}, which is considered stereotypical for a nova. However, the broadband light-curves show day-to day variability after $t_3$ similar to that of an O-class light-curves (see Fig.~\ref{Fig:BVRI_JHK_LC}). It is worth noting that there is no presence for a plateau in the tail of the light-curve (after $\sim$ 3\,--\,6\,mag from maximum) similar to that seen in recurrent novae (see \citealt{Schaefer_2010}).

\begin{figure}
\begin{center}
  \includegraphics[width=\columnwidth]{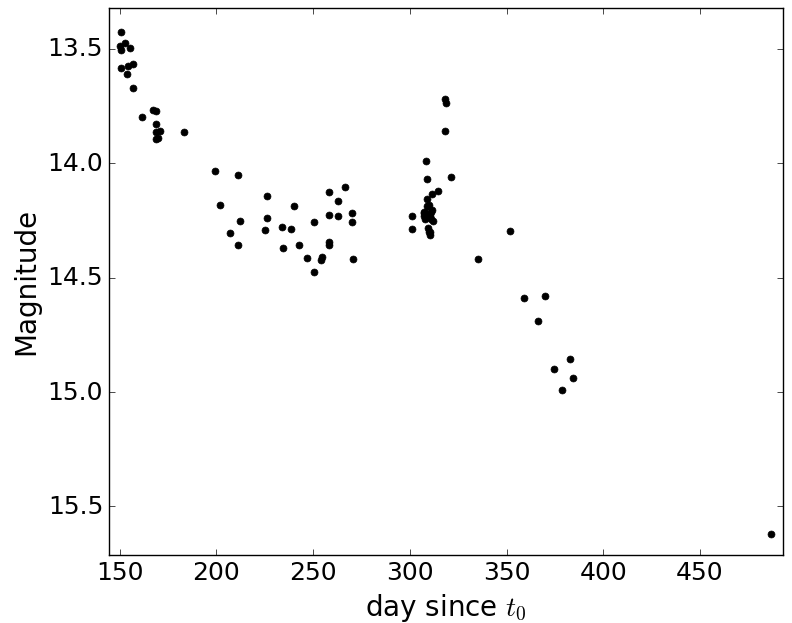}
\caption{The UVOT \textit{uvw2} ($\lambda_{\mathrm{central}}$ = 1928\,$\mathrm{\AA}$) light-curve plotted against day since $t_0$.}
\label{Fig:UVOT_LC}
\end{center}
\end{figure}

Although there is a gap in the light-curve between day 26 and day 107 (due to solar constraints), we fit a power-law to the light-curve (only early decline - before day 25) and derive $t_2 \leq$ 2.9\,$\pm$\,0.5\,d, $t_3 \leq 6.8$ \,$\pm$\,1.0\,d and a power index of $\sim$ 0.16\footnote{\citet{Izzo_etal_2018} have derived longer values of $t_2$ and $t_3$. This was due to misinterpretation of their light-curve (private communication with L. Izzo).}. We took into account the use of measurements from different instruments by increasing ($\times$ 2) the uncertainty on $t_2$ and $t_3$. 

With $t_2\leq$ 2.9\,d and a decline rate of $\sim$ 0.69\,mag\,d$^{-1}$ over $t_2$, nova V407 Lup is a ``very-fast" nova in the classification of \citet{Payne-Gaposchkin_1964} and is one of the fastest known examples. Only a few other novae have shown a decline time $t_2 \lesssim$ 3.0 days, including M31N 2008-12a, U Sco, V1500 Cyg, V838 Her, V394 CrA, and V4160 Sgr (see, e.g.,\citealt{Young_etal_1976,Schaefer_2010,Munari_etal_2011}; and table~5 in \citealt{Darnley_etal_2016}). 

\begin{figure*}
\begin{center}
  \includegraphics[width=140mm]{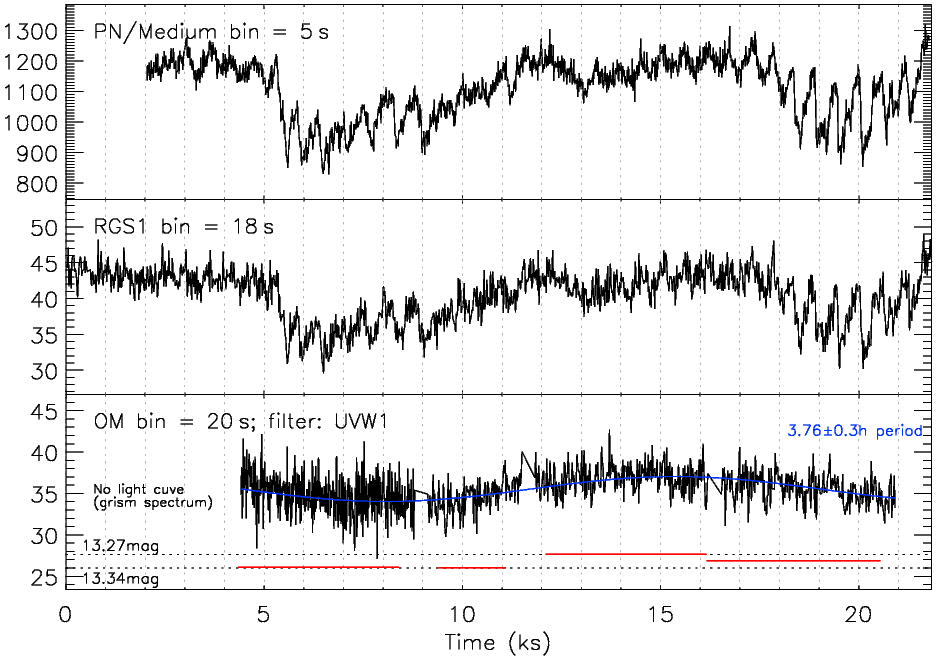}
\caption{\textit{First panel:} the \textit{XMM-Newton} EPIC/pn X-ray light-curve. \textit{Second panel:} the RGS1 X-ray light-curve. The X-ray light-curves show two broad ($\sim$\,2\,--\,3\,ks wide) dips separated by 12.6\,ks, consistent with the 3.57\,h modulation seen in the \textit{Swift}-UVOT data (see Section~\ref{timing_study}). \textit{Third panel:} the \textit{uvw1} OM fast mode (black) and imaging mode (red) light-curves. The first OM exposure is a grism spectrum, hence the light-curves only start at $\sim$ 5\,ks. The blue line represents a best fit sine curve to the OM data with a period of 3.76$\pm$0.3\,h. This period is consistent, within uncertainty, with the period seen in the \textit{Swift}-UVOT data (see Section~\ref{timing_study} for further discussion on the timing analysis). The y-axis in all the panels are counts per second.}
\label{Fig:XMM_LC}
\end{center}
\end{figure*}

\subsection{The \textit{Swift} UVOT light-curve}

The UVOT \textit{uvw2} light-curve (Fig.~\ref{Fig:UVOT_LC}) declined slowly after day 150 before flattening off at around a \textit{uvw2} magnitude of 14.2 by around day 200, with signs of a 0.1\,mag variability superimposed on the overall decline. We find periodicities in the light-curve, which we discuss in Section~\ref{timing_study}. The UV data rebrightened after day 300, to reach a \textit{uvw2} magnitude of 13.7 around day 320. Then the brightness started to decline again, reaching a \textit{uvw2} magnitude of 14.9 at day $\sim$ 385. Fig.~\ref{Fig:multi_LC} shows a direct comparison between the \textit{Swift} X-ray, \textit{Swift} UV, and SMARTS optical light-curves.

\subsection{The \textit{XMM-Newton} and \textit{Chandra} UV and X-ray light-curves}
\label{XMM_LC_sec}
\begin{figure}
\centering
\includegraphics[width = \columnwidth]{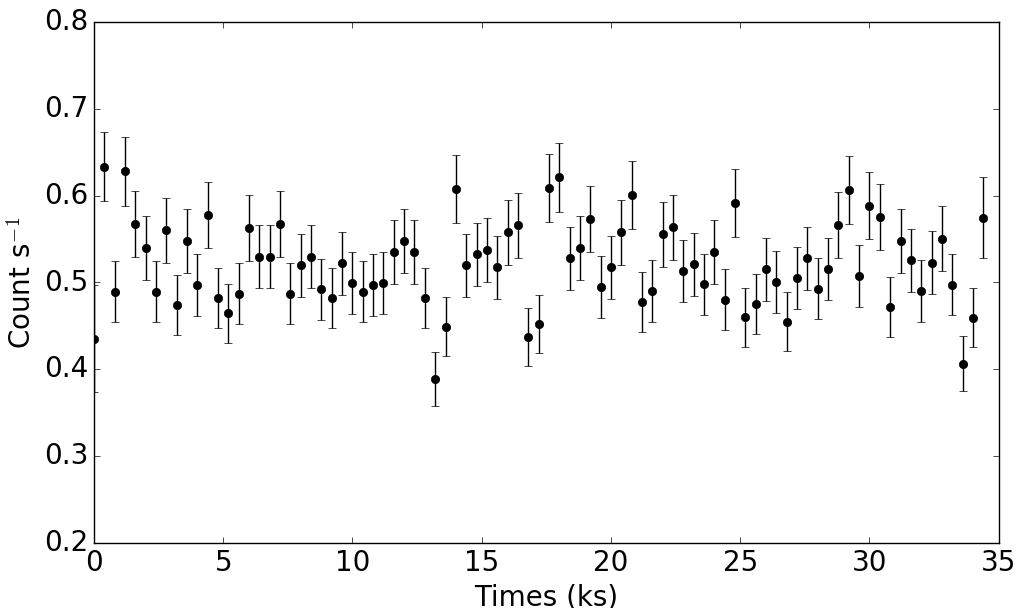}
\caption{The \textit{Chandra} High Resolution Camera (HRC) X-ray light-curve. The HRC energy range is 0.06\,--\,10\,keV. There is a clear short-term variability in the light-curve and it is discussed in Section~\ref{timing_study}.} 
\label{Fig:chandra_HRC}
\end{figure}

The \textit{XMM-Newton} RGS1 X-ray light-curve, the EPIC/pn X-ray light-curve, and the OM UV fast and imaging modes light-curves are all presented in Fig.~\ref{Fig:XMM_LC}. The \textit{XMM-Newton} RGS1 light-curve shows evidence of variability on different timescales. This includes two broad ($\sim$\,2\,--\,3\,ks wide) dips separated by $\sim$\,12.6\,ks $\sim$ 3.5\,h. The OM fast mode UV light-curve shows a variation characterized by a 3.76$\pm$0.3\,h period, which is also consistent (within uncertainty) with the duration between the dips in the RGS1 light-curve. However, no other periodicity is found in the OM light-curve. The \textit{Chandra} HRC light-curve (Fig.~\ref{Fig:chandra_HRC}) also shows evidence for short-term variability, which we discuss in Section~\ref{timing_study}.

\subsection{Timing analysis}
\label{timing_study}

\begin{figure}
\begin{center}
 \includegraphics[width=\columnwidth]{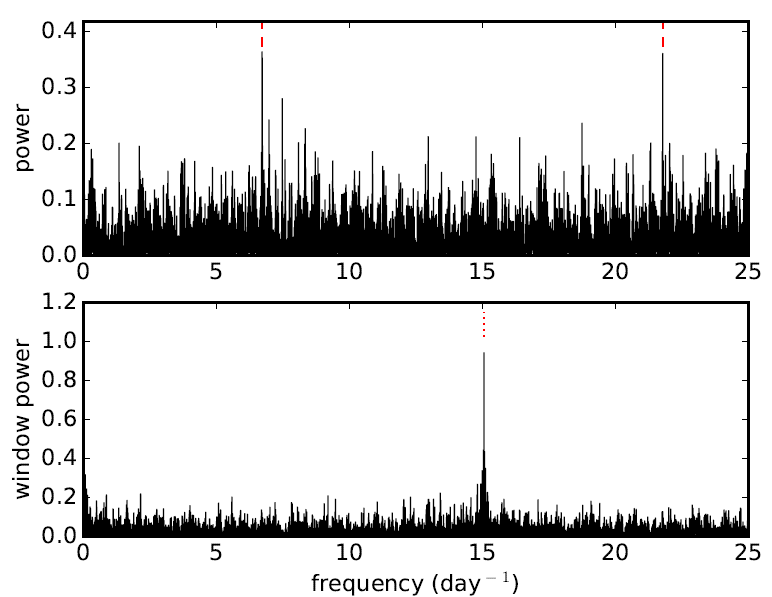}
\caption{Upper panel: the Lomb Scargle periodogram of the detrended \textit{uvw2} light-curve, revealing significant modulation at frequencies of 6.7168\,d$^{-1}$ (7.77$\times 10^{-5}$\,s$^{-1}$; $P = 3.573$\,h) and 21.7696\,d$^{-1}$ ($2.52\times10^{-4}$\,s$^{-1}$; $P = 1.102$\,h; with estimated uncertainties of 0.0003\,d$^{-1}$), marked with red dashed lines. Lower panel: periodogram of the observing window, showing \emph{Swift}'s orbital frequency (dotted line).}
\label{Fig:LSP_UV}
\end{center}
\end{figure}

Since the \emph{Swift}/UVOT \textit{uvw2} light-curve (Fig.~\ref{Fig:UVOT_LC}) shows evidence for intrinsic variability superimposed on the long term decline, we searched for periodicities in the data. Fig.~\ref{Fig:LSP_UV} shows a Lomb-Scargle periodogram (LSP) of the barycentric corrected \textit{uvw2} light-curve after subtracting a third order polynomial to remove the long term trend. The most significant peaks occur at periods of $3.5731\pm 0.0014$\,h and $1.1024\pm 0.0002$\,h, which are aliases of each other caused by \emph{Swift}'s 1.6\,h orbit. One of these periods might represent the orbital period of the binary ($P_{\mathrm{orb}}$). The low cadence of the UVOT data means we cannot break the degeneracy between the periods (the high-cadence observation interval between days 307 and 311 failed to resolve this issue). The amplitude of the modulation is approximately 0.1\,mag.

\begin{figure}
\begin{center}
  \includegraphics[width=\columnwidth]{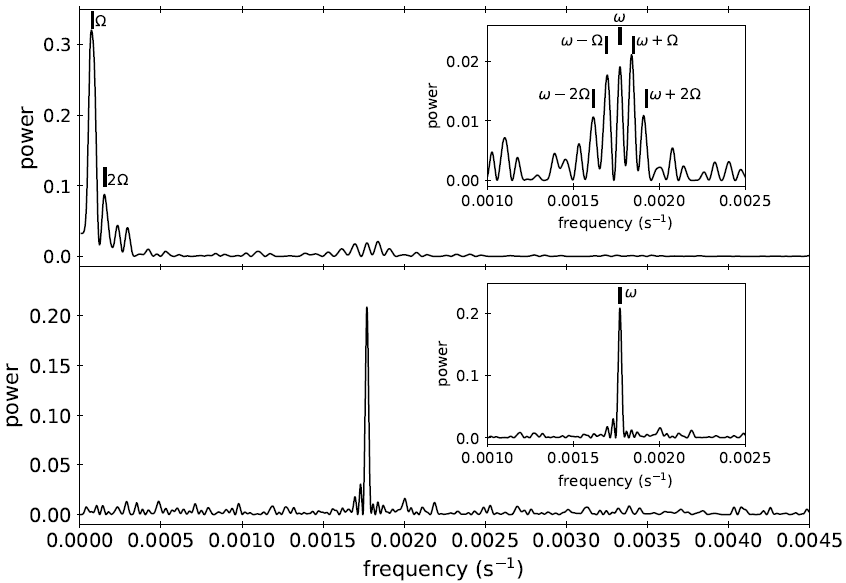}
\caption{Lomb Scargle periodograms of the \textit{XMM-Newton} RGS1 (top) and the \textit{Chandra} HRC (bottom) light-curves. Various frequencies are indicated where $\Omega = 1/ P_{\mathrm{orb}}$ and $\omega = 1/ P_{\mathrm{spin}}$ (see text for more details). We present zoom-in plots around the 1.77$\times10^{-3}$\,s$^{-1}$ in each periodogram.}
\label{Fig:LSP_X-ray}
\end{center}
\end{figure}

We performed a similar Lomb-Scargle analysis of the \emph{XMM-Newton} RGS1 and \emph{Chandra} HRC data (see Fig.~\ref{Fig:LSP_X-ray}). The light-curve of the former shows considerable flux variations, while the flux in the latter is approximately constant over the duration of the observation. The \emph{XMM-Newton} periodogram is dominated by a low frequency of $\Omega$ = 7.77$\times 10^{-5}$\,s$^{-1}$ (3.57\,h). This signal only covers $\sim$ two cycles of the 3.57\,h period and cannot be believed in isolation. However, as this signal is consistent with the long period seen in the UVOT data, we identify it as the same 3.57\,h modulation.

\begin{figure}
\begin{center}
  \includegraphics[width=\columnwidth]{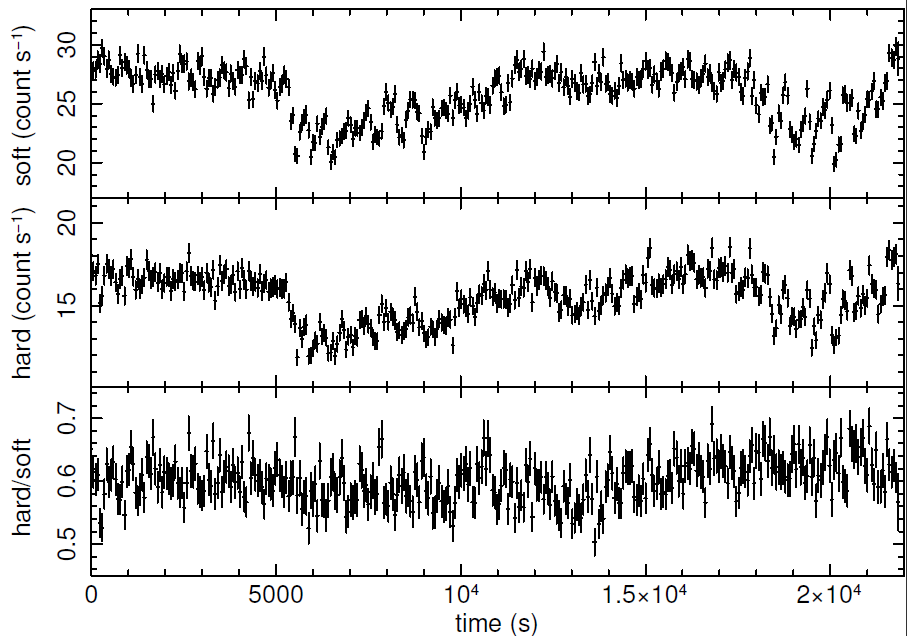}
  \includegraphics[width=\columnwidth]{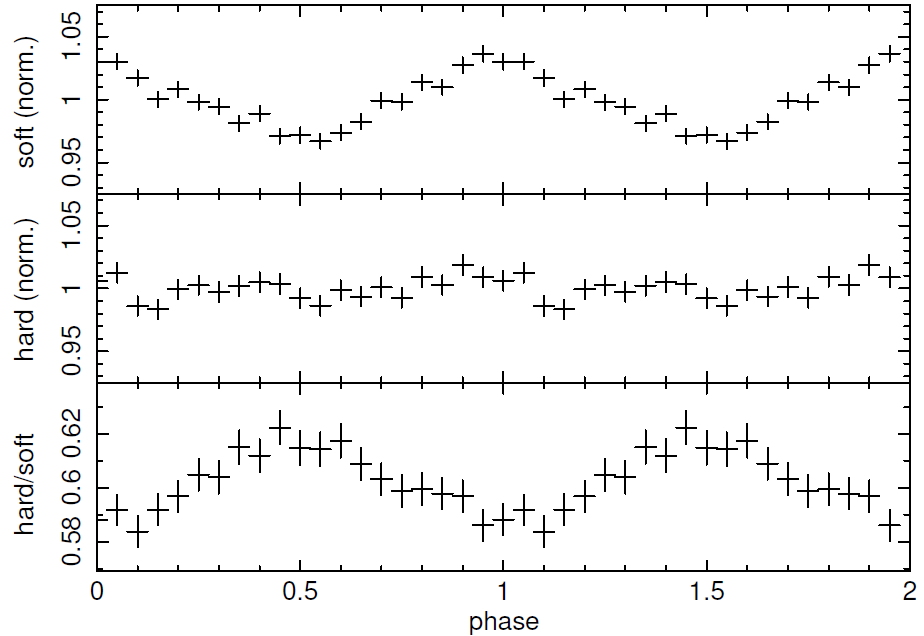}
\caption{\textit{Top:} from top to bottom, soft RGS1 light-curve, hard RGS1 light-curve, and the hardness ratio. The energy bands for the soft and hard light-curves are 23.5\,--\,37.0\,$\mathrm{\AA}$ (0.3325\,--\,0.528\,keV) and 15.0\,--\,23.5\,$\mathrm{\AA}$ (0.528\,--\,0.827\,keV), respectively. \textit{Bottom:} same as top but all folded at the 565\,s spin period and normalized to their respective mean rates. Therefore, the $y$-axis represents the relative amplitude of the modulation in both the soft and hard bands.}
\label{Fig:hard_ratio}
\end{center}
\end{figure}

The long-timescale modulation and variability seen in the \textit{XMM-Newton} light-curves, the time between the dips, and the weak modulation observed in the OM light-curve (see Figs~\ref{Fig:XMM_LC} and~\ref{Fig:LSP_X-ray}) are all consistent with the 3.57\,h signal seen in the \textit{Swift} UV light-curve. In addition, the absence of any variability in the \textit{XMM-Newton} RGS and OM light-curves related to the 1.1\,h signal (seen in the \textit{Swift} UV light-curve) has the potential to rule out the modulation at this short period. Therefore we assume that the 3.57\,h is most likely the $P_{\mathrm{orb}}$ of the binary.

\begin{figure}
\begin{center}
  \includegraphics[width=\columnwidth]{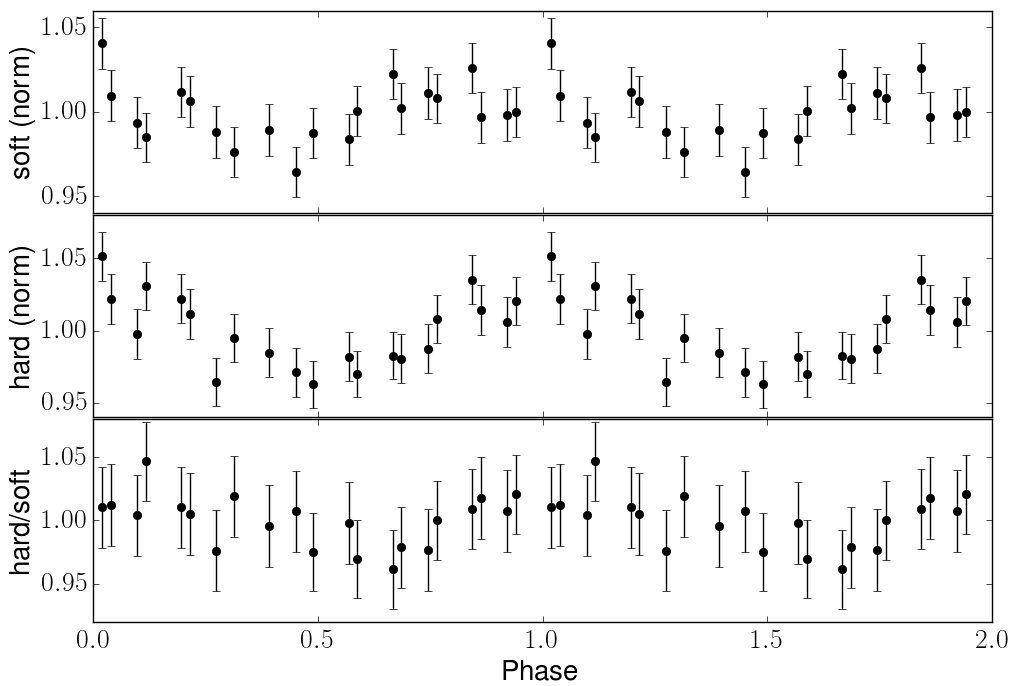}
\caption{From top to bottom, soft \textit{Chandra} HRC light-curve, hard \textit{Chandra} HRC light-curve, and the hardness ratio, all folded over the spin period and normalized to their respective mean rates. Therefore, the $y$-axis represents the relative amplitude of the modulation in both the soft and hard bands. The energy bands for the soft and hard light-curves are 30\,--\,50.0\,$\mathrm{\AA}$ (0.248\,--\,0.413\,keV) and 15.0\,--\,30.0\,$\mathrm{\AA}$ (0.413\,--\,0.827\,keV), respectively.}
\label{Fig:HRC_hard_ratio}
\end{center}
\end{figure}

Ignoring this long-term variability seen in the low frequency part of \emph{XMM-Newton} periodogram, we find significant signals at $\omega$ $\simeq$ 1.77$\times10^{-3}$\,s$^{-1}$ in both (\textit{XMM-Newton} and \textit{Chandra}) light-curves. More precisely, the periodicities we found are at 565.04\,$\pm$\,0.68\,s for the \textit{Chandra} HRC light-curve and 564.96\,$\pm$\,0.50\,s for the \textit{XMM-Newton} RGS light-curve. The \textit{Swift}/XRT observations are almost all too short to usefully constrain the modulation at this short period.

\begin{figure}
\begin{center}
  \includegraphics[width=\columnwidth]{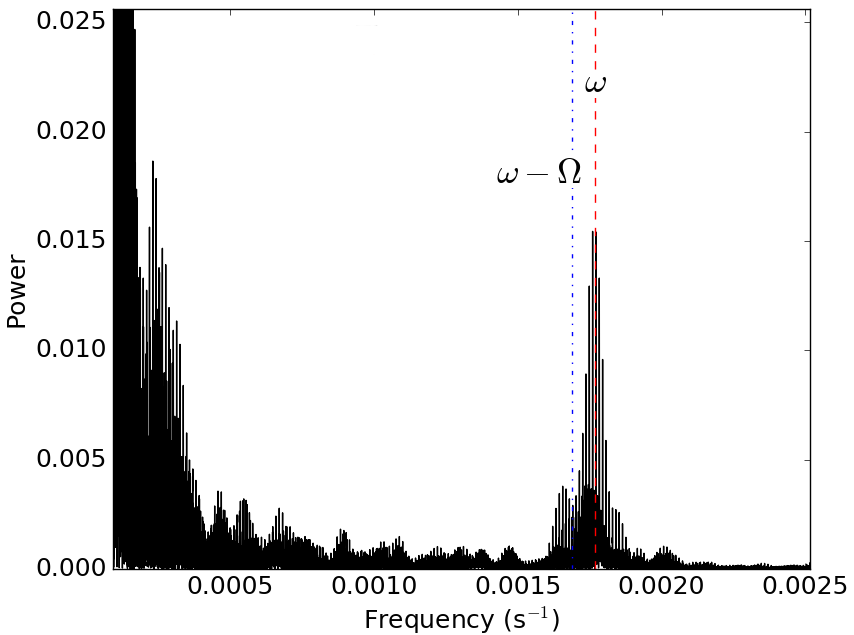}
\caption{Lomb Scargle periodograms of the AAVSO data taken between days 327 and 350 post-eruption. The red dashed line represents the frequency $\omega$ = 1.77$\times10^{-3}$\,s$^{-1}$ ($1/ P_{\mathrm{spin}}$) and the blue dotted line represents the sideband frequency ($\omega - \Omega$).}
\label{Fig:LSP_AAVSO}
\end{center}
\end{figure}

The 565\,s X-ray period and the longer 3.57\,h UV/X-ray period are clearly reminiscent of the typical spin period ($P_{\mathrm{spin}}$) and $P_{\mathrm{orb}}$ seen in IP CVs \citep{Warner_1995}. This implies that the system may host a magnetized WD. The ratio of $P_{\mathrm{spin}}$/$P_{\mathrm{orb}}$ is $\sim$ 0.44, typical for IPs \citep{Warner_1995,Norton_etal_2004}. Therefore, we suggest that the $\sim$ 565\,s period is the $P_{\mathrm{spin}}$ of the WD.
  
While the \emph{Chandra} HRC power spectrum (Fig.~\ref{Fig:LSP_X-ray}) is relatively easy to read with a single dominant peak (at 1.77$\times10^{-3}$\,s$^{-1}$; $P_{\mathrm{spin}}$ = 565\,s), the \emph{XMM-Newton} data shows a more complicated periodogram with some asymmetry in peak amplitudes. This suggests rather complicated signal behaviour. The different peaks around the central one ($\omega$ $\simeq$1.77$\times10^{-3}$\,s$^{-1}$) coincide with $\omega\,\pm\,\Omega$ and $\omega\,\pm\,2\Omega$ sidebands. Such signals are commonly seen in IPs (see e.g. \citealt{Norton_etal_1996,Ferrario_etal_1999}). 

Fig.~\ref{Fig:hard_ratio} represents the \textit{XMM-Newton} RGS1 soft and hard light-curves and the hardness ratio. When folded at the $P_{\mathrm{spin}}$ = 565\,s, the light-curves show near sinusoidal variation. Such variation is expected from an IP (see, e.g., \citealt{Osborne_1988,Norton_Watson_1989,Norton_etal_1992_a,Norton_etal_1992_b,Norton_1993,Warner_1995}), however the RGS data were taken during the post-nova SSS phase and therefore the mechanism responsible for such modulation might be different from that seen usually in IPs (see Section~\ref{acc_res_sec} for further discussion).

Fig.~\ref{Fig:HRC_hard_ratio} represents the folded \textit{Chandra} HRC soft and hard light-curves over the $P_{\mathrm{spin}}$, as well as the hardness ratio. The HRC light-curve (taken 340 days after the eruption) show stronger modulation in the harder bands while the hardness ratio is slightly modulated. This might also suggest that the variation over the $P_{\mathrm{spin}}$ is not simply due to absorption by an accretion curtain as it is usually the case for IPs \citep{Warner_1995}. 

\begin{figure}
\begin{center}
 \includegraphics[width=\columnwidth]{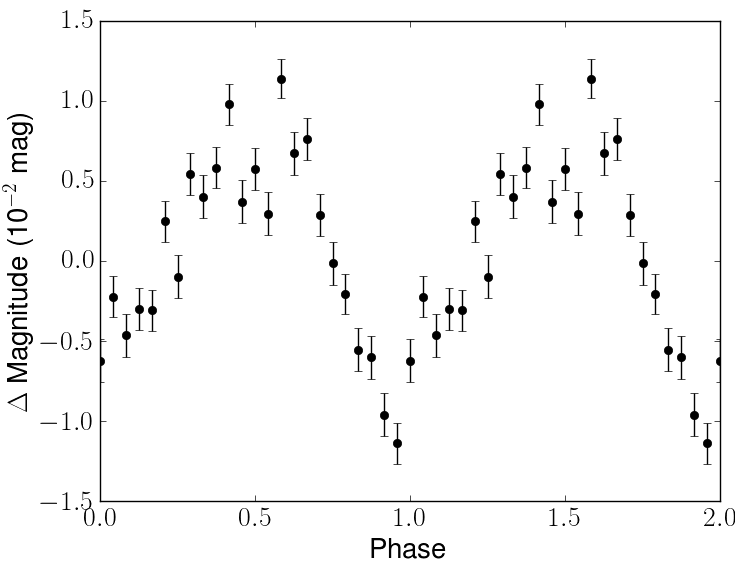}
\caption{The AAVSO optical light-curve folded over the WD spin period of 565\,s, using the same ephemeris used to fold the \textit{XMM-Newton} RGS1 light-curves (Fig.~\ref{Fig:hard_ratio}). The $y$ axis represents the change in magnitude relative to the mean brightness level (positive values mean brighter and negative values mean fainter).}
\label{Fig:LC_AAVSO}
\end{center}
\end{figure}

\begin{figure*}
\centering
\begin{subfigure}{0.48\textwidth}
\centering
\includegraphics[width = \textwidth]{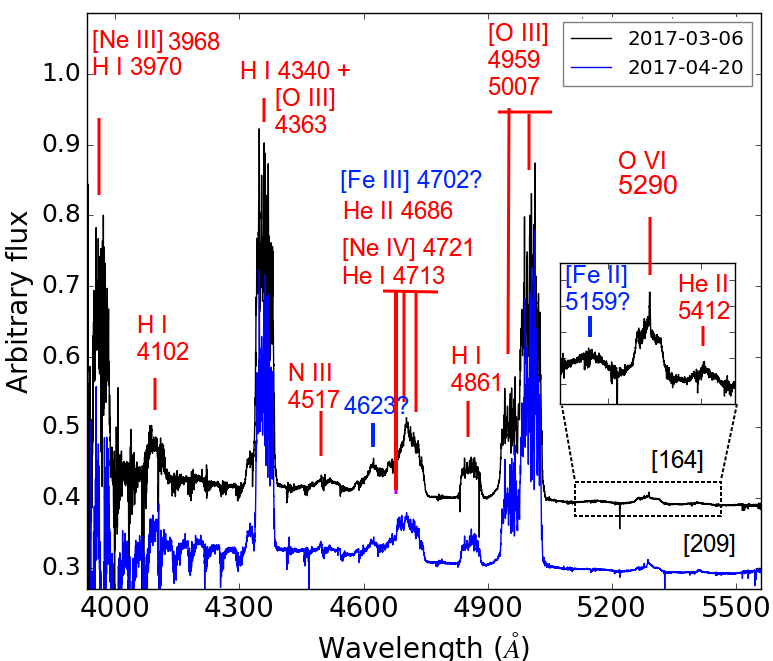}
\label{fig:left}
\end{subfigure}
\begin{subfigure}{0.48\textwidth}
\centering
\includegraphics[width = \textwidth]{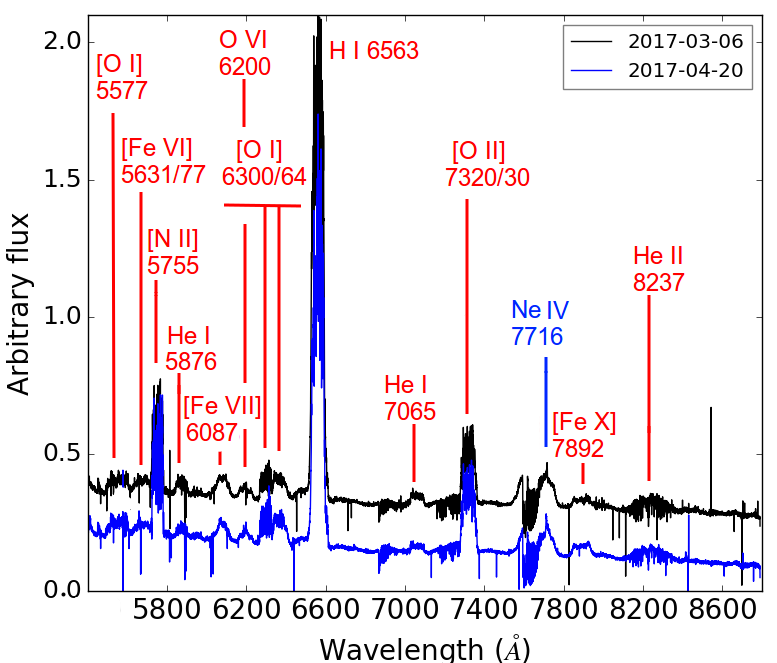}
\label{fig:right}
\end{subfigure}
\caption{The SALT HRS high-resolution spectra of days 164 and 209, plotted between 3900\,--\,5500\,$\mathrm{\AA}$ (left) and 5450\,--\,8800\,$\mathrm{\AA}$ (right) with the flux in arbitrary units. For clarity, the spectra are shifted vertically. The numbers in square brackets are days since $t_0$. We present a zoomed in plot on the range between 5100\,$\mathrm{\AA}$ and 5450\,$\mathrm{\AA}$ to show the possible presence of weak emission lines in this range. Below 4200\,$\mathrm{\AA}$ instrumental noise dominates the spectra.}
\label{Fig:HRS_1}
\end{figure*}

Typically, the physical reason behind an optical/X-ray modulation over the $P_{\mathrm{spin}}$ of the WD seen in IPs is attributed to the variation of the viewing aspect of the accretion curtain as it converges towards the WD surface near the magnetic poles or due to the absorption caused by the accretion curtains. Therefore, such a modulation is usually a sign of accretion impacting the WD surface near the magnetic poles and possibly in this case a sign of accretion restoration. However, the \emph{XMM-Newton} observations were taken during the SSS phase and therefore the soft X-rays are dominated by emission from the H burning on the surface of the WD. Thus, there must be another explanation for the $P_{\mathrm{spin}}$ modulation seen in the X-ray light-curves (see Section~\ref{acc_res_sec} for further discussion about the accretion resumption and the origin of the X-ray modulation).

We also performed Lomb-Scargle analysis of the optical AAVSO data (unfiltered reduced to $V$ band) obtained between 327 and 350 days post-eruption. The observations were performed almost every other night and they consist of multiple exposures separated by $\sim$40\,s and spanning for $\sim$ 4\,h. The power spectrum shows a peak at $\sim$ 1.77$\times10^{-3}$\,s$^{-1}$ the same as that seen in the \textit{Chandra} power spectrum and which is an indication of the $P_{\mathrm{spin}}$ of the WD (Fig.~\ref{Fig:LSP_AAVSO}). The signals seen in the periodogram (Fig.~\ref{Fig:LSP_AAVSO}) at small frequencies are not consistent with either the 
3.57\,h or the 1.1\,h signals seen in the UVOT light-curve and are most likely due to red noise.  The spin phase-folded AAVSO optical light-curve is shown in Fig.~\ref{Fig:LC_AAVSO}. We folded the light-curve using the same ephemeris used to fold the RGS1 light-curves. While the AAVSO optical and \textit{XMM-Newton} X-ray light-curves are out of phase, caution is required when comparing these two light-curves as the \textit{XMM-Newton} observations were done more than 150 days prior to the AAVSO ones (see Section~\ref{acc_res_sec} for further discussion).

\section{Spectroscopic results and analysis}
\label{spec_sec}
\subsection{Optical spectroscopy}
\label{Optical_spec_sec}

\begin{figure*}
\centering
\begin{subfigure}{0.48\textwidth}
\centering
\includegraphics[width = \textwidth]{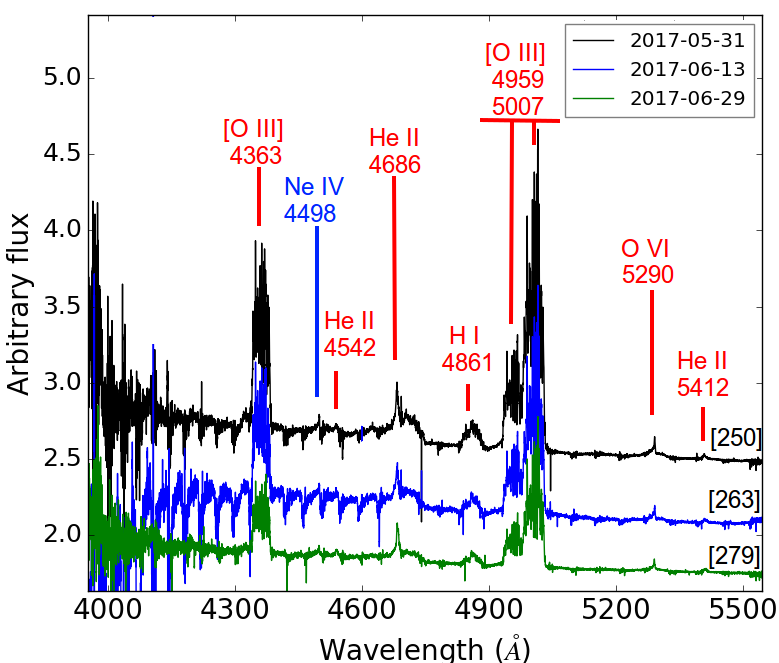}
\label{fig:left}
\end{subfigure}
\begin{subfigure}{0.48\textwidth}
\centering
\includegraphics[width = \textwidth]{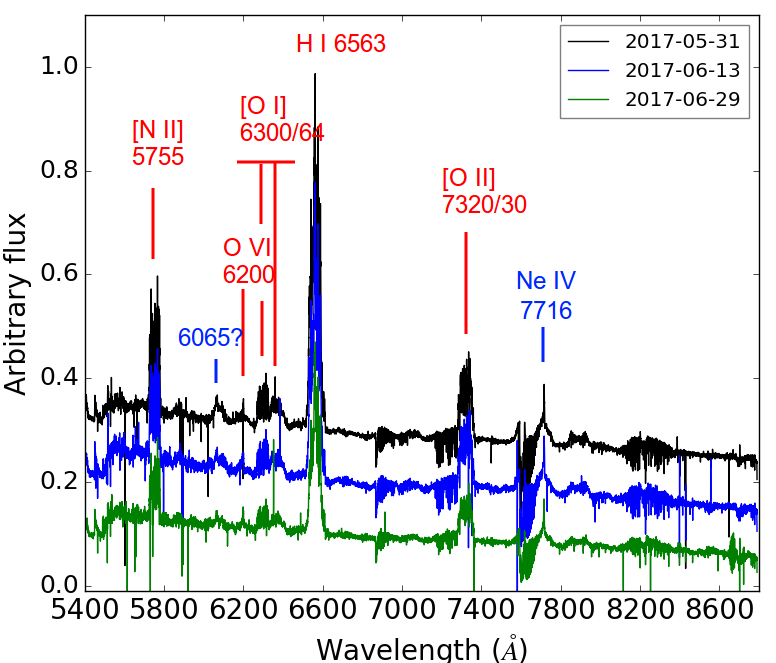}
\label{fig:right}
\end{subfigure}
\caption{The SALT HRS high-resolution spectra of days 250, 263, and 279, plotted between 3900\,--\,5500\,$\mathrm{\AA}$ (left) and 5450\,--\,8800\,$\mathrm{\AA}$ (right) with the flux in arbitrary units. For clarity, the spectra are shifted vertically. The numbers in square brackets are days since $t_0$. Below 4200\,$\mathrm{\AA}$, instrumental noise dominates the spectra.}
\label{Fig:HRS_2}
\end{figure*}

\begin{figure*}
\centering
\begin{subfigure}{0.48\textwidth}
\centering
\includegraphics[width = \textwidth]{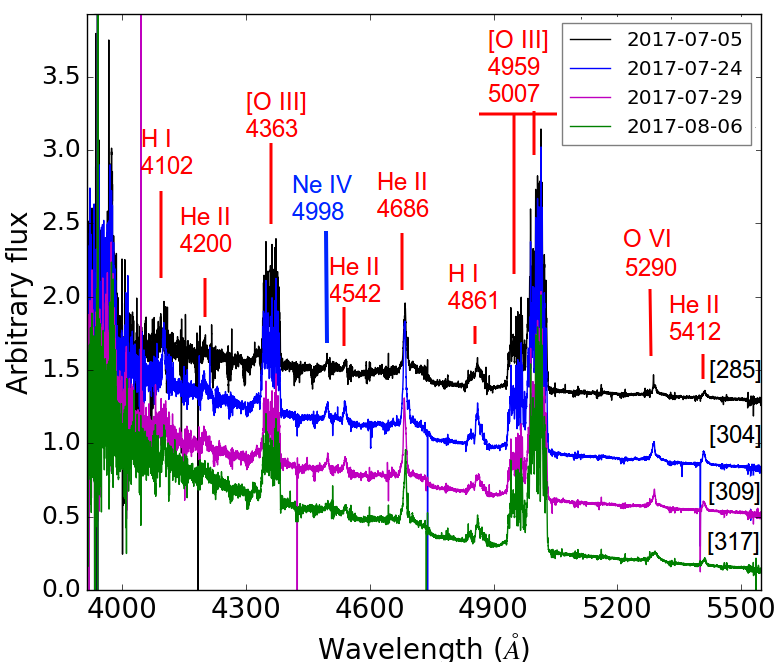}
\label{fig:left}
\end{subfigure}
\begin{subfigure}{0.48\textwidth}
\centering
\includegraphics[width = \textwidth]{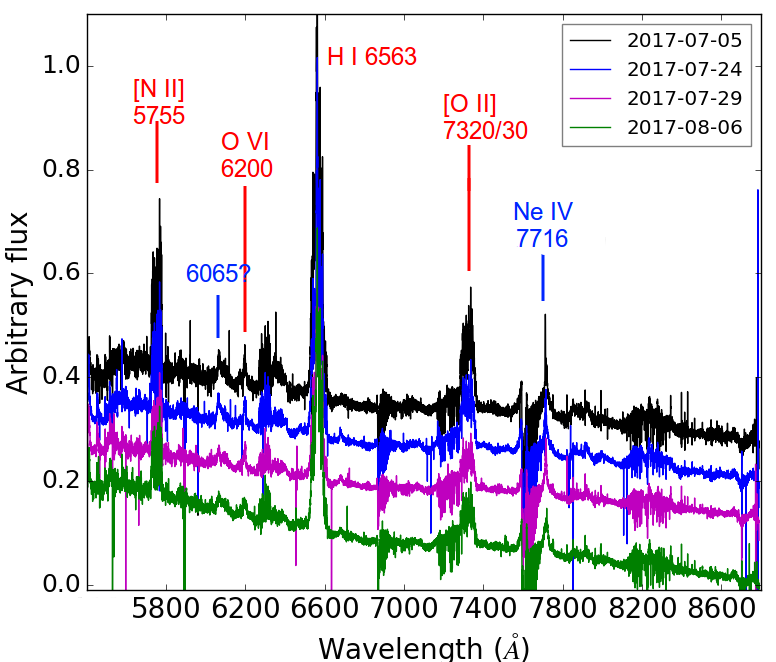}
\label{fig:right}
\end{subfigure}
\caption{The SALT HRS high-resolution spectra of days 285, 304, 309, and 317, plotted between 3900\,--\,5500\,$\mathrm{\AA}$ (left) and 5450\,--\,8800\,$\mathrm{\AA}$ (right) with the flux in arbitrary units. For clarity, the spectra are shifted vertically. The numbers in square brackets are days since $t_0$. Below 4200\,$\mathrm{\AA}$, instrumental noise dominates the spectra.}
\label{Fig:HRS_3}
\end{figure*}

\begin{figure*}
\centering
\begin{subfigure}{0.48\textwidth}
\centering
\includegraphics[width = \textwidth]{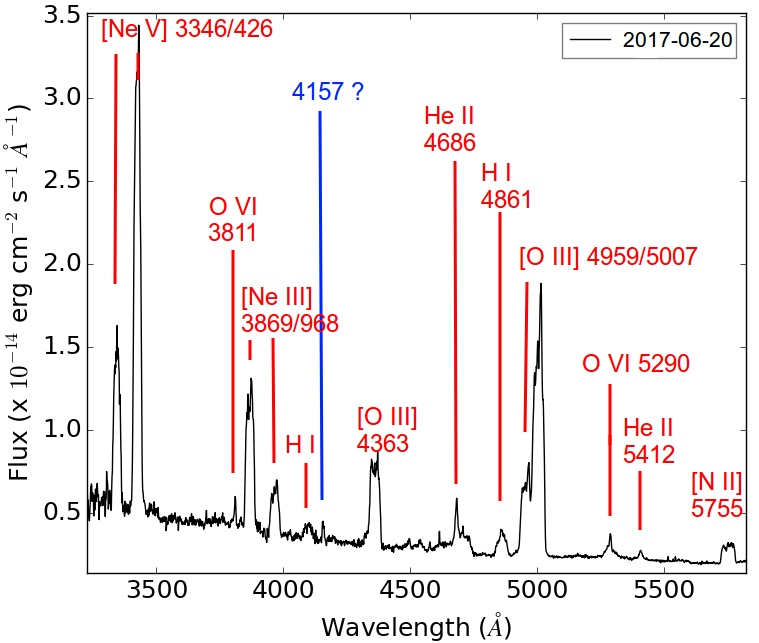}
\label{fig:left}
\end{subfigure}
\begin{subfigure}{0.48\textwidth}
\centering
\includegraphics[width = \textwidth]{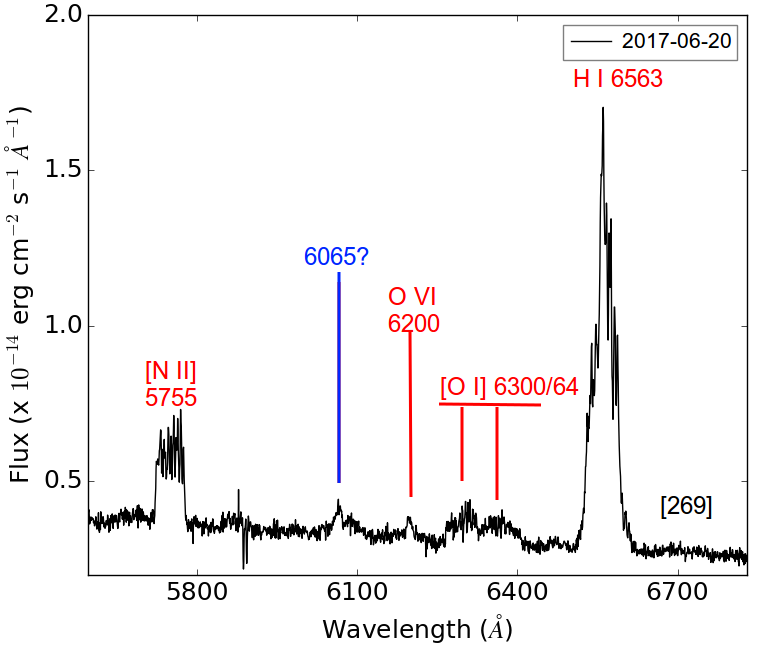}
\label{fig:right}
\end{subfigure}
\caption{The SOAR medium-resolution spectrum of day 270, plotted between 3200\,--\,5800\,$\mathrm{\AA}$ (left) and 5600\,--\,6800\,$\mathrm{\AA}$ (right) with the flux in erg\,cm$^{-2}$\,s$^{-1}$\,$\mathrm{\AA}^{1}$.} 
\label{Fig:SOAR_spec}
\end{figure*}

\begin{figure}
\begin{center}
 \includegraphics[width=\columnwidth]{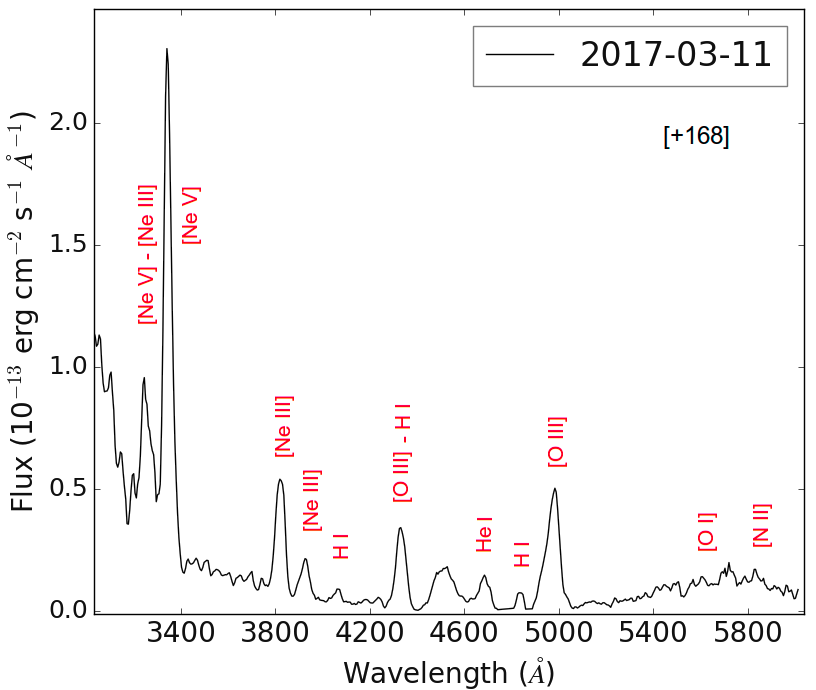}
\caption{The OM (Optical Monitor on board \textit{XMM-Newton}) grism spectrum plotted between 3000 and 6000\,$\mathrm{\AA}$. The numbers in square brackets are days since $t_0$.}
\label{Fig:OM_spec}
\end{center}
\end{figure}

\begin{figure}
\begin{center}
  \includegraphics[width=\columnwidth]{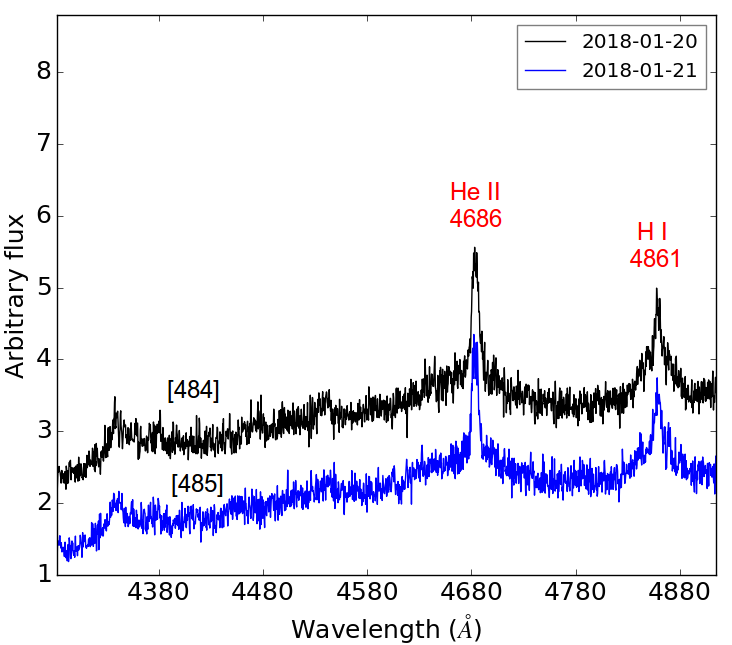}
\caption{The SOAR medium-resolution spectra taken on day 484 and day 485, plotted between 4300 and 5000\,$\mathrm{\AA}$ with the flux in arbitrary units.}
\label{Fig:late_SOAR_spec}
\end{center}
\end{figure}

\subsubsection{Line identification}
\label{spec_lines}

The strongest emission features in the first two SALT HRS spectra and the \textit{XMM-Newton} OM grism spectrum, on days 164, 168, and 209 (Fig.~\ref{Fig:HRS_1} and~\ref{Fig:OM_spec}), are the forbidden oxygen lines, \feal{O}{III} 4363\,$\mathrm{\AA}$, 4959\,$\mathrm{\AA}$, 5007\,$\mathrm{\AA}$, and \feal{O}{II} 7320/30\,$\mathrm{\AA}$, along with the Balmer lines. The lines are very broad with flat-topped, jagged profiles. Forbidden neon lines are also present (\feal{Ne}{III} 3698\,$\mathrm{\AA}$ blended with H$\epsilon$ and \feal{Ne}{IV} 4721\,$\mathrm{\AA}$, which might be blended with other neighbouring lines). Other  forbidden neon lines, such as \feal{Ne}{III} 3343\,$\mathrm{\AA}$, 3869\,$\mathrm{\AA}$, \feal{Ne}{V} 3346\,$\mathrm{\AA}$ and 3426\,$\mathrm{\AA}$, are also present in the SOAR spectrum of day 269 (Fig.~\ref{Fig:SOAR_spec}) and in the \textit{Swift} UVOT spectra (Section~\ref{UV_spec}). Also present are weak permitted lines of helium (\eal{He}{I} 4713\,$\mathrm{\AA}$, 5876\,$\mathrm{\AA}$, \eal{He}{II} 4686\,$\mathrm{\AA}$, 5412\,$\mathrm{\AA}$, and 8237\,$\mathrm{\AA}$ - the lines of the Pickering series at 4860\,$\mathrm{\AA}$ and 6560\,$\mathrm{\AA}$ might be blended with H$\beta$ and H$\alpha$, respectively), high excitation oxygen lines (\eal{O}{VI} 5290\,$\mathrm{\AA}$), and weak \feal{O}{I} lines at 6300\,$\mathrm{\AA}$ and 6364\,$\mathrm{\AA}$. The forbidden \feal{N}{II} line at 5755\,$\mathrm{\AA}$ is strong and broad, while the permitted \eal{N}{III} line at 4517\,$\mathrm{\AA}$ is relatively weak and the one at 4638\,$\mathrm{\AA}$ is possibly blended with other lines. The \feal{N}{II} doublet at 6548 and 6584\,$\mathrm{\AA}$ might be blended with broad H$\alpha$. Relatively weak high ionization, coronal lines of iron might also be present (\feal{Fe}{VII} 6087\,$\mathrm{\AA}$, \feal{Fe}{XI} 7892\,$\mathrm{\AA}$, and possibly \feal{Fe}{II} 5159\,$\mathrm{\AA}$ and \feal{Fe}{III} 4702$\mathrm{\AA}$). The spectra also show lines with  a FWHM of $\sim$ 100\,km\,s$^{-1}$ at $\sim$ 4498\,$\mathrm{\AA}$, 5290\,$\mathrm{\AA}$, and 7716\,$\mathrm{\AA}$ (see Section~\ref{very_narrow}). 

The optical spectra show three different type of lines characterized by significantly different widths, therefore, in the remainder of the paper we will use the following terms:  
\begin{itemize}
\item ``Very narrow lines" to denote those lines with a FWHM 

$\sim$ 100\,km\,s$^{-1}$ (see Section~\ref{very_narrow}).

\item ``Moderately narrow lines'' to denote those lines with a 

FWHM $\sim$ 450\,km\,s$^{-1}$, such as \eal{He}{II} 4686\,$\mathrm{\AA}$ (see

Sections~\ref{narrow_lines} and~\ref{OVI_MNL}).

\item Broad lines to denote those lines with a FWHM $\sim$ 

3000\,km\,s$^{-1}$, that originate from the ejecta, such as 

the Balmer and \feal{O}{III}lines (see Sections~\ref{Balmer_lines} 

and~\ref{forbidden_oxygen_lines}).
\end{itemize}

From day 250 to day 317, the SALT HRS spectra are still dominated by the broad forbidden oxygen lines (see Figs~\ref{Fig:HRS_2} and~\ref{Fig:HRS_3}). The Balmer lines are fading gradually, while other lines such as forbidden Ne and Fe start to disappear. At day 250, a ``moderately narrow'' and sharp \eal{He}{II} line at 4686\,$\mathrm{\AA}$ emerges and is accompanied by less prominent \eal{He}{II} lines at 4200\,$\mathrm{\AA}$, 4542\,$\mathrm{\AA}$, and 5412\,$\mathrm{\AA}$. Similar features of \eal{O}{VI} 5290\,$\mathrm{\AA}$ and 6200\,$\mathrm{\AA}$ emerge simultaneously and become more prominent $\sim$ 30 days later. On top of these two \eal{O}{VI} lines, the aforementioned ``very narrow lines'' are still present. We also detect similar emission features at $\sim$ 4498 and 6050\,$\mathrm{\AA}$ that we could not identify.

All the ``very narrow'' and ``moderately narrow'' lines show changes in radial velocity and structure, unlike the broad lines. At day 285 \eal{He}{II} 4686\,$\mathrm{\AA}$ becomes as strong as \feal{O}{III} 4363\,$\mathrm{\AA}$ and surpasses it after $\sim$ 300 days. At this stage, ``moderately narrow'' Balmer features emerge, superimposed on the pre-existing broad features. 

Further to the blue (below 4200\,$\mathrm{\AA}$) the SOAR medium-resolution spectrum of day 269  (Fig.~\ref{Fig:SOAR_spec}) shows ``very narrow'' emissions of \eal{O}{VI} at 3811\,$\mathrm{\AA}$ and $\sim$ 4157\,$\mathrm{\AA}$. Relatively strong and broad lines of \feal{Ne}{V} at 3426\,$\mathrm{\AA}$ and 3346\,$\mathrm{\AA}$ and \feal{Ne}{III} at 3869\,$\mathrm{\AA}$ and 3968\,$\mathrm{\AA}$ are also present.

The late SOAR medium-resolution spectra of days 484 and 485 were of limited-range, centred near to \eal{He}{II} at 4686\,$\mathrm{\AA}$, which appears stronger than H$\beta$ (Fig.~\ref{Fig:late_SOAR_spec}). The latter has a significant broad base. In Table~\ref{line_det} we list the line identifications along with the FWHM, Equivalent Widths (EWs),
and fluxes of those lines for which an estimate was possible.

\subsubsection{Spectral classification and evolution}

\citet{Izzo_etal_2018} obtained spectra as early as day 5 post-eruption. These spectra show characteristics of the optically thin He/N spectral class with ejecta velocity $\sim$ 2000\,km\,s$^{-1}$. They also point out the presence of \eal{Fe}{II} lines which they attribute to a very rapid iron-curtain phase. 

The HRS spectra taken on days 164 and 209 were entirely emission lines and dominated by broad nebular forbidden oxygen lines, along with Balmer and forbidden neon and iron lines. The spectra show that the nova is well into the nebular phase by then. We expect that this phase started around a month from the eruption, based on the fast light-curve evolution, consistent with the spectra of \citet{Izzo_etal_2018}. The presence of neon lines in the spectrum also suggest that V407 Lup might be a ``neon nova'', showing that the eruption occurred on a ONe WD, which was confirmed by \citet{Izzo_etal_2018} after deriving a Ne abundance $\sim$ 14 times Solar. The highlight of that study was the detection of the $^7$\eal{Be}{II} 3130\,$\mathrm{\AA}$ doublet and that V407 Lup, an ONe nova, has produced a considerable amount of $^7$Be, which decays later into Li. They concluded that not only CO but also ONe novae produce Li, confirming that CNe are the main producers of Li of stellar origin in the Galaxy.   

The optical spectra show the presence of high ionization coronal lines (e.g. \feal{Fe}{VII} 6087\,$\mathrm{\AA}$ and \feal{Fe}{XI} 7892\,$\mathrm{\AA}$). Such lines can be attributed to photoionization from the central hot source during the post-eruption phase, to a hot coronal-line-region, physically separated from the ejecta responsible for the low ionization nebular lines, or possibly to shocks within the ejecta (see e.g. \citealt{Shields_Ferland_1978,Williams_etal_1991,Wagner_Depoy_1996} and references therein). 

In the spectra of day 250 onwards, ``moderately narrow'' lines of \eal{He}{II} and \eal{O}{VI} emerge, the strongest being \eal{He}{II} 4686\,$\mathrm{\AA}$. These lines show changes in their radial velocity between $\sim -210$\,km\,s$^{-1}$ and $0$\,km\,s$^{-1}$. Similar narrow and moving lines have been observed in a few other novae (e.g. nova KT Eri, U Sco, LMC 2004a and 2009) and their origins have been debated (see e.g. \citealt{Mason_etal_2012,Munari_etal_2014,Mason_Munari_2014} and references therein;  see Sections~\ref{narrow_lines} and~\ref{OVI_MNL} for further discussion).

\subsubsection{Line profiles}

\begin{figure}
  \includegraphics[width=\columnwidth]{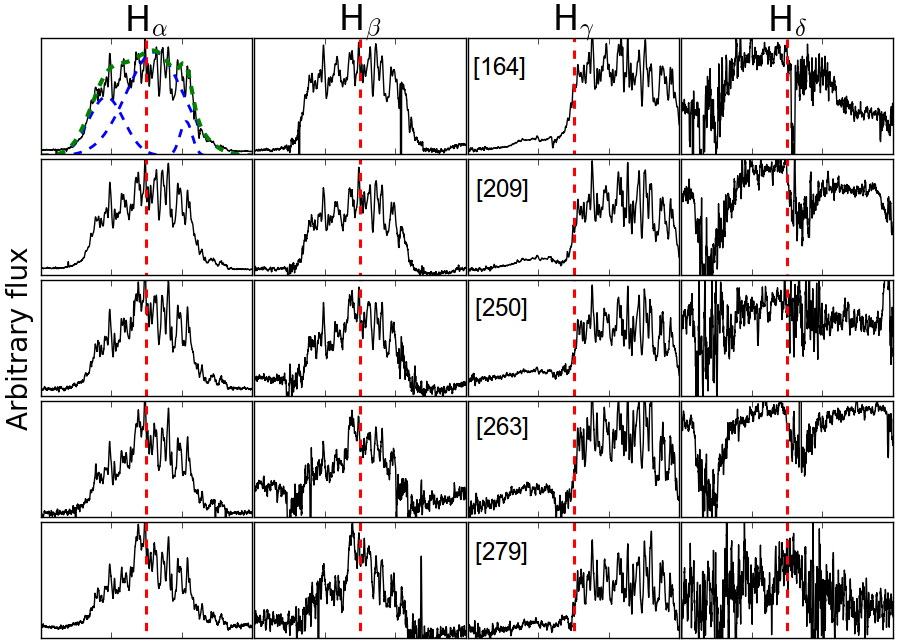}
  \includegraphics[width=\columnwidth]{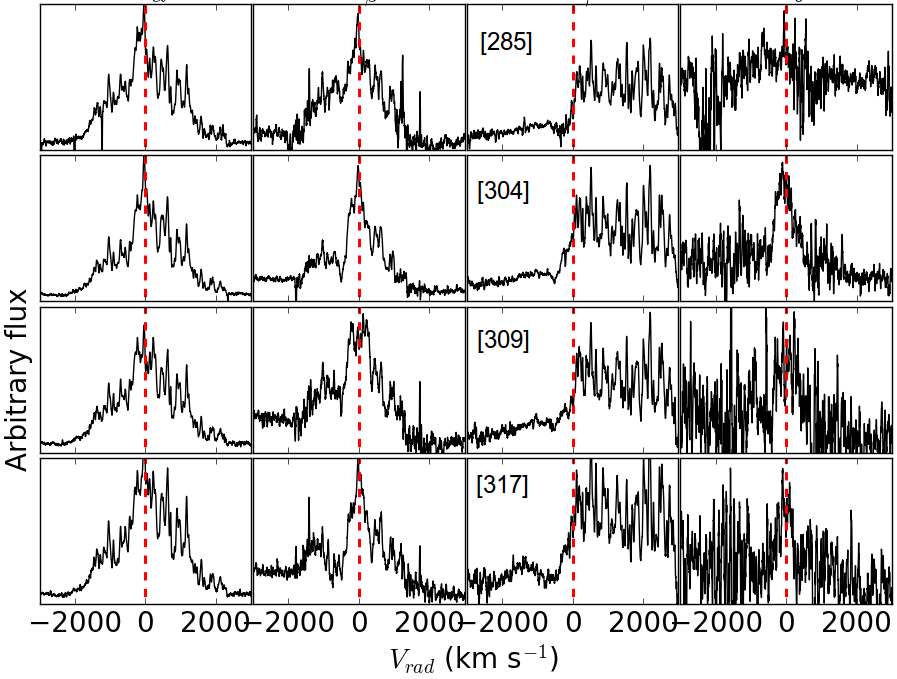}
\caption{The evolution of the line profiles of the Balmer emission. From left to right: H$\alpha$, H$\beta$, H$\gamma$, and H$\delta$. From top to bottom: days 164, 209, 250, 263, 279, 285, 304, 309, and 317. A heliocentric correction has been applied to the radial velocities. The flux is in arbitrary units. The red dashed lines represent the rest-wavelength of each line ($V_{\mathrm{rad}}$ = 0\,km s$^{-1}$). The top-left panel shows an example of multiple
Gaussian fitting that was used to derive the FWHM.}
\label{Fig:Balmer_line_profiles}
\end{figure}

\paragraph{Balmer lines:}
\label{Balmer_lines}
in the early spectra of nova V407 Lup (days 5, 8, and 11), the Balmer lines showed a FWHM of $\sim$ 3700\,km\,s$^{-1}$ \citep{ATel_9587,Izzo_etal_2018}. This FWHM had decreased  to $\sim$ 3000\,km\,s$^{-1}$ by the first HRS spectra at day 164. The lines then show a systematic narrowing with time (Fig.~\ref{Fig:Balmer_line_profiles}). We measure the FWHM of these lines by applying multiple component Gaussian fitting in the Image Reduction and Analysis Facility (IRAF; \citealt{Tody_1986}) and Python ({\sc scipy} packages\footnote{\url{https://www.scipy.org/}}) environments separately. The values are given in Table~\ref{table:BalmerFWHM}. 

This line narrowing has been observed in many other novae (see, e.g., \citealt{Della_Valle_etal_2002,Hatzidimitriou_etal_2007,Shore_etal_2013,Darnley_etal_2016}) and can be attributed to the distribution of the velocity of the matter at the moment of ejection. The density of the fastest moving gas decreases faster than that of the slower moving gas, leading to a decrease in its emissivity and in turn to the line narrowing \citep{Shore_etal_1996}. 

While most novae occur in systems hosting a main-sequence secondary, some nova systems have a giant secondary. For such systems, an alternative explanation for the line narrowing has been presented by \citet{Bode_etal_1985} after efforts to model the 1985 eruption of RS Oph. These authors suggested that shocks and interaction between the high velocity ejecta and low-velocity stellar wind from the companion (red giant in case of RS Oph) are responsible for decelerating the ejecta, which manifests as line narrowing.

The Balmer lines have flat-topped, jagged profiles (probably due to clumpiness in the ejecta) with no changes in radial velocity, indicating an origin associated with the expanding ejecta. At day 250 ``moderately narrow'' Balmer emission features (FWHM$\sim$ 500\,km\,s$^{-1}$), superimposed on the broad  emission profiles, start to emerge. They become prominent in later spectra (day 304 onwards; see Fig.~\ref{Fig:Balmer_line_profiles}). These emission features show variability in structures and radial velocity (see Table~\ref{table:BalmerMNL}), which is difficult to measure accurately due to the contamination by the broad emission component. However, due to their width and the change in radial velocity, it is very likely that these features originate from the inner binary system, possibly associated with an accretion region (as it is unreasonable to associate such ``moderately narrow'' features, which show such a change in radial velocity, with emission from the expanding ejecta). 

\begin{figure*}
\begin{center}
  \includegraphics[width=\textwidth]{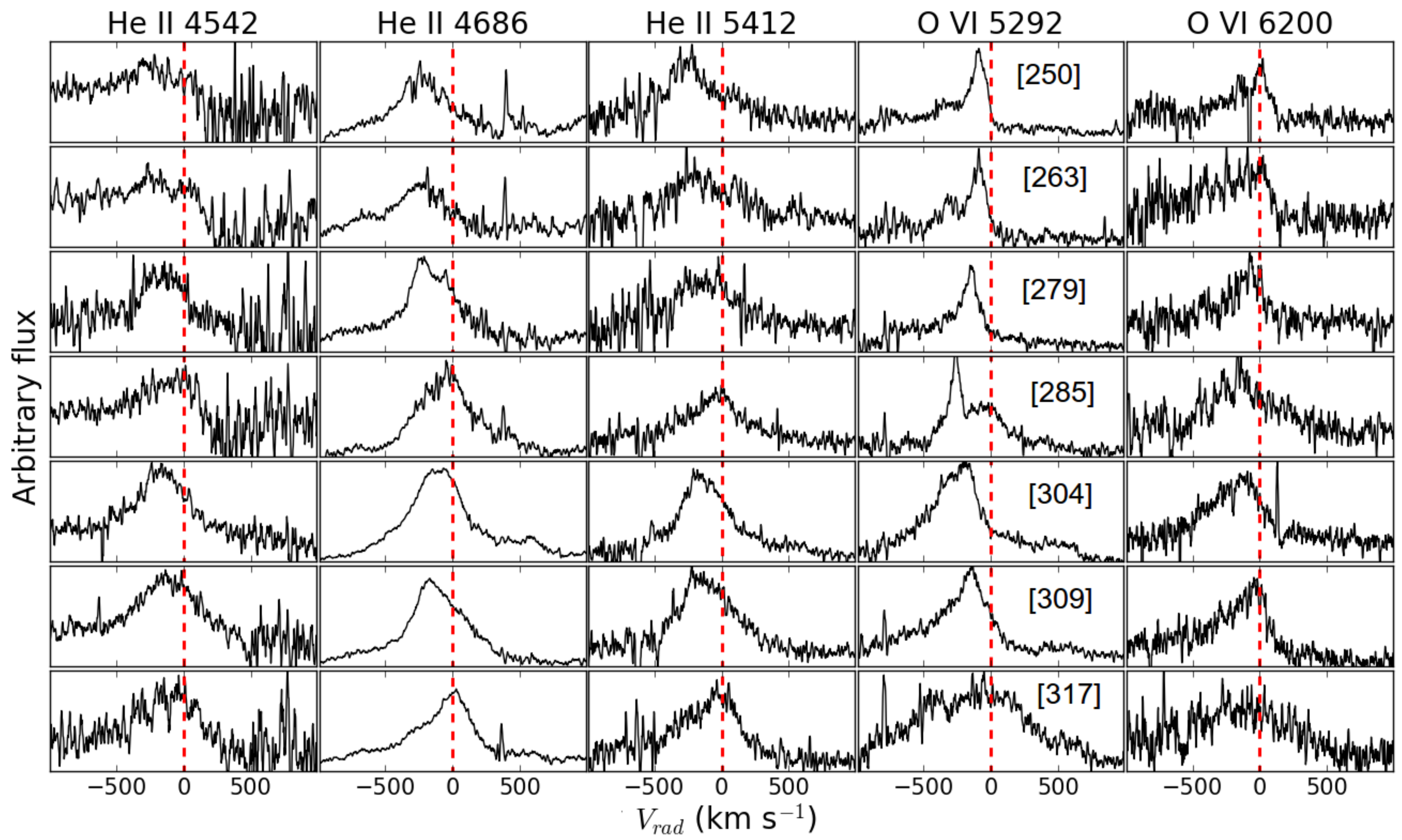}
\caption{The evolution of the profiles of the ``moderately narrow'' emission lines. From left to right: \eal{He}{II} 4542\,$\mathrm{\AA}$, \eal{He}{II} 4686\,$\mathrm{\AA}$, \eal{He}{II} 5412\,$\mathrm{\AA}$, \eal{O}{VI} 5290\,$\mathrm{\AA}$, and \eal{O}{VI} 6200\,$\mathrm{\AA}$. From top to bottom: days 250, 263, 279, 285, 304, 309, and 317. A heliocentric correction is applied to the radial velocities. The flux is in arbitrary units. The red dashed lines represent the rest-wavelength of each line ($V_{\mathrm{rad}}$ = 0\,km s$^{-1}$).}
\label{Fig:MNL}
\end{center}
\end{figure*}

At this stage, H$\gamma$ is still completely dominated by the \feal{O}{III} line at 4363\,$\mathrm{\AA}$. However,  ``moderately narrow'' H$\delta$ emission became prominent while its broad base has completely faded. We also note a dip or absorption feature to the blue of H$\beta$, H$\gamma$, and H$\delta$. This dip becomes very prominent in the last spectrum at $\sim$ $-$650\,km\,s$^{-1}$ (Fig.~\ref{Fig:Balmer_line_profiles}).

These ``moderately narrow'' Balmer features became prominent $\sim$ 30 days after the emergence of the narrow \eal{He}{II} lines (see Section~\ref{narrow_lines}). This is expected because such features, which may originate from the inner binary system, can only be seen clearly once the broad Balmer emission has weakened significantly. They also show single-peak emission in most of the spectra, indicating that if they are originating from the accretion disk, the system is possibly to have a low inclination (e.g. close to face-on; between 0$^{\circ}$ and 15$^{\circ}$; see fig. 2.39 in \citealt{Warner_1995}).

\paragraph{Forbidden oxygen lines:}
\label{forbidden_oxygen_lines}
broad, nebular emission features of forbidden oxygen dominated the late spectra of nova V407 Lup (day 164 onwards). The \feal{O}{III} 4363\,$\mathrm{\AA}$ is superimposed over H$\gamma$, hence the blue-shifted pedestal feature. The forbidden oxygen emission features are broad (FWHM $\sim$ 3000\,km\,s$^{-1}$) and show jagged profiles, possibly an indication of clumpiness in the ejecta.

We measured the FWHM of the \feal{O}{III} 4363\,$\mathrm{\AA}$ by applying multiple component Gaussian fitting in the IRAF and Python environments separately. The values are listed in Table~\ref{table:OIIIFWHM}. The other \feal{O}{III} lines are blended with other lines or with each other. Similar to the Balmer lines, the \feal{O}{III} lines show a systematic narrowing with time (Section~\ref{Balmer_lines}), but no changes in radial velocity.

\paragraph{The He~II moderately narrow lines:}
\label{narrow_lines}
several novae have shown relatively narrow He components superimposed on a broad pedestal  during the early nebular stages. While the spectra evolve, the narrow components become narrower and stronger (e.g. nova KT Eri, U Sco, LMC 2004a and 2009). For these novae it has been suggested that during the early nebular stage, when the ejecta are still optically thick, the relatively narrow components originate from an equatorial ring while the broad pedestal components originate from polar caps (see, e.g., \citealt{Munari_etal_2011,Munari_etal_2014} and references therein). Then, when the ejecta become optically thin, these components become narrower by a factor of around 2 (FWHM changing from $\sim$ 1000\,km\,s$^{-1}$ to $\sim$ 400\,km\,s$^{-1}$) and show cyclic changes in radial velocity. Such cyclic changes in radial velocity can only be associated with the binary system, most likely the accretion disk \citep{Munari_etal_2014}.

\begin{figure}
\centering
\includegraphics[width = \columnwidth]{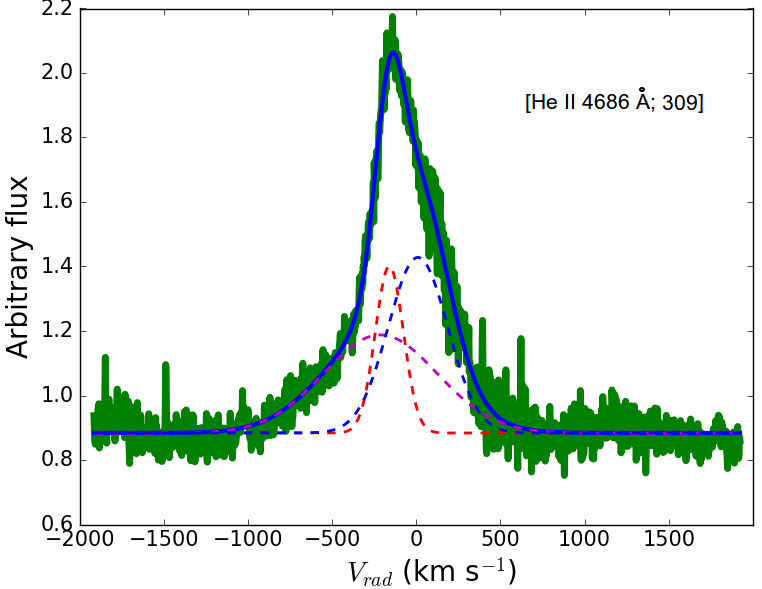}
\caption{A sample of the complex \eal{He}{II} 4686\,$\mathrm{\AA}$ ``moderately narrow'' line from day 309. The line is possibly a complex of 3 components; the green line represents the observation, while the blue solid line is the total fit of the 3 Gaussian components (dashed lines).} 
\label{Fig:comp}
\end{figure}

This is not the case for V407 Lup as only at day 250 the ``moderately narrow lines'' of \eal{He}{II} start to emerge (it is possible that early in the nebular stage, these lines have been blended and dominated by the broad and prominent lines coming from the ejecta). 
We detect \eal{He}{II} 4686\,$\mathrm{\AA}$, 4542\,$\mathrm{\AA}$, 5412\,$\mathrm{\AA}$, and 4200\,$\mathrm{\AA}$. The latter is heavily affected by the increased noise at the edge of the blue arm of HRS. The profiles of the three higher S/N \eal{He}{II} lines are presented in Fig.~\ref{Fig:MNL}. We measure the radial velocity and FWHM of the \eal{He}{II} lines by fitting a single Gaussian component. The measurements are illustrated in Table~\ref{table:HeIIsingle}. The radial velocities of the \eal{He}{II} lines range between $\sim$ $-$210\,km\,s$^{-1}$ and $ 0$\,km\,s$^{-1}$ and they have an average FWHM of $\sim$ 450\,km\,s$^{-1}$. This range of velocities indicates that the system might have a negative systemic velocity of around $-$100\,km\,s$^{-1}$. The three lines show consistent velocity and structure changes across the different spectra, suggesting that they are originating from the same regions. 

The \eal{He}{II} ``moderately narrow lines'' are complex and can be subdivided into at least three components: (1) a medium width component (MWC) with FWHM $\sim$ 300\,km\,s$^{-1}$; (2) a small width component (SWC) with FWHM $\sim$ 100\,km\,s$^{-1}$, blended with the MWC in most of the spectra; (3) and a broad base component (BBC; Fig.~\ref{Fig:comp}). Such complex structures of \eal{He}{II} lines are characteristic of mCVs, and they are variously associated with emission from the accretion stream, the flow through the magnetosphere, and the secondary star \citep{Rosen_etal_1987,Warner_1995,Schwope_etal_1997}. We also note the presence of a very weak red-shifted emission at $\sim$ +600\,km\,s$^{-1}$ from \eal{He}{II} 4686\,$\mathrm{\AA}$ that strengthens and weakens from one spectrum to another (Fig.~\ref{Fig:MNL}).

We measure the velocity of the different components of the \eal{He}{II} 4686\,$\mathrm{\AA}$ line, by applying multiple Gaussian component fitting (Fig.~\ref{Fig:comp}). The radial velocity and FWHM of the MWC and SWC of the \eal{He}{II} 4686\,$\mathrm{\AA}$ line are listed in Tables~\ref{table:HeIIMNL} and~\ref{table:HeIIVNL}, respectively. Although the two components are clearly out of phase, their velocity measurements should be regarded as uncertain and caution is required when interpreting them, since it is not straightforward to deconvolve the different components. We also measure the full width of the BBC (see Table~\ref{table:HeIIBBC}).

Although the complexity of the \eal{He}{II} line profiles makes it difficult to draw conclusions about the origins of the different components, we suggest that the SWC, characterized by a FWHM $\sim$ 100\,km\,s$^{-1}$, must originate from an area of low velocity (possibly the heated surface of the secondary). However, the other two components are possibly associated with emission from an accretion region.
It is possible that a fourth component is also present in the \eal{He}{II} lines, which adds to the complexity. 

\paragraph{The O~VI moderately narrow lines:}
\label{OVI_MNL}

from day 250 we detect relatively weak, ``moderately narrow lines'' of \eal{O}{VI} 5290\,$\mathrm{\AA}$ and 6200\,$\mathrm{\AA}$. These two lines are initially dominated by ``very narrow lines'' (see Fig.~\ref{Fig:MNL} and Section~\ref{very_narrow}) possibly associated with a different element and certainly originating from a different region. At day 304 the intensity of the \eal{O}{VI} lines becomes comparable to the neighbouring ``very narrow lines'' and they form a blend. In addition we detect a similar line at 6065\,$\mathrm{\AA}$ (possibly \eal{N}{II}). We derive the radial velocity of the ``moderately narrow'' \eal{O}{VI} lines by fitting a single Gaussian. The values are listed in Table~\ref{table:NOVI}.

There is no clear correlation between the velocity and the structure of the ``moderately narrow'' \eal{O}{VI} lines with those of their \eal{He}{II} counterparts. It is worth noting that the ionization potential of \eal{He}{II} is $\sim$ 54\,eV while that of \eal{O}{VI} is $\sim$ 138 eV.

\begin{figure}
  \includegraphics[width=\columnwidth]{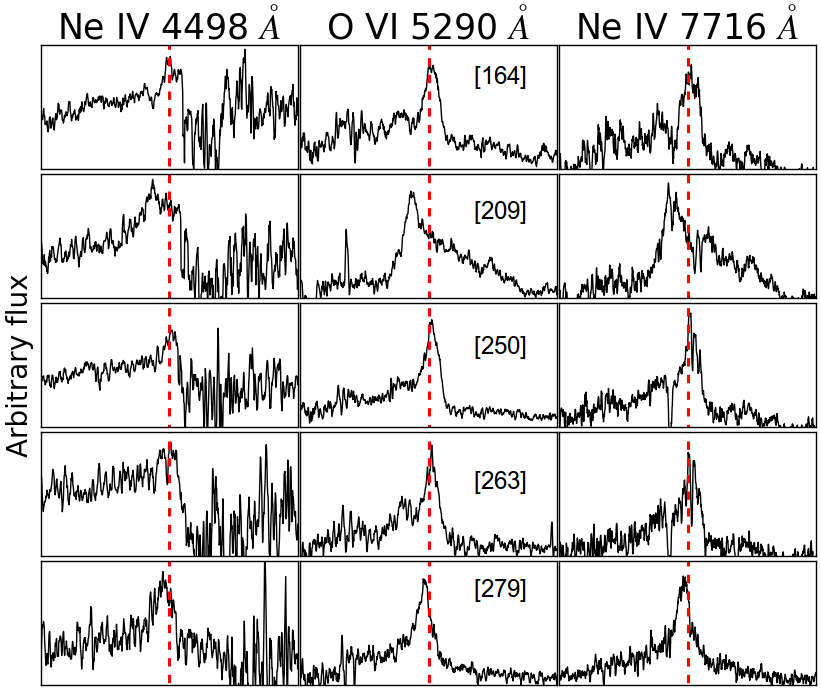}
  \includegraphics[width=\columnwidth]{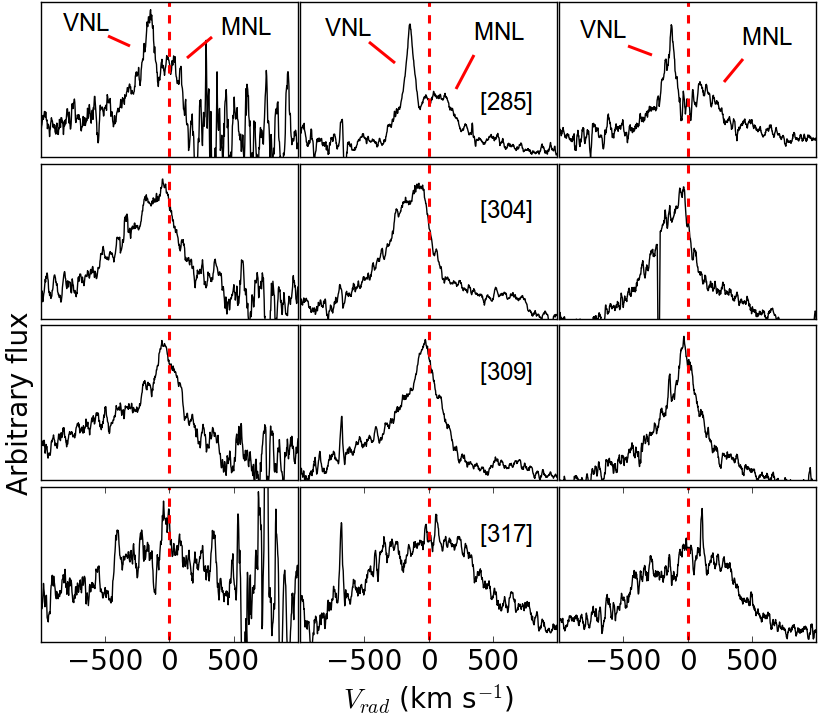}
\caption{The evolution of the profiles of the three ``very narrow lines'' (VNL) plotted against wavelength. From left to right: \eal{Ne}{IV} 4498\,$\mathrm{\AA}$, \eal{O}{VI} 5290\,$\mathrm{\AA}$, and \eal{Ne}{IV} 7716\,$\mathrm{\AA}$. From top to bottom: days 164, 209, 250, 263, 279, 285, 304, 309, and 317. At day 285, the ``very narrow lines'' stand out very clearly from the neighbouring ``moderately narrow lines'' (MNL) as indicated on the plot. A heliocentric correction is applied to the radial velocities.}
\label{Fig:VNL_profiles}
\end{figure}

\paragraph{Very narrow lines:}
\label{very_narrow}
a remarkable aspect of the optical HRS spectra is the presence of very narrow, moving lines. These ``very narrow lines'' have an average FWHM of $\sim$ 100\,km\,s$^{-1}$  and show variation in their velocity (ranging between $-150$ and +20\,km\,s$^{-1}$), width, and intensity. The most prominent lines are at $\sim$ 5290\,$\mathrm{\AA}$ (possibly \eal{O}{VI} 5290\,$\mathrm{\AA}$) and $\sim$ 7716\,$\mathrm{\AA}$ (possibly \eal{Ne}{IV} 7715.9\,$\mathrm{\AA}$). A less prominent line is at $\sim$ 4998\,$\mathrm{\AA}$ (possibly \eal{Ne}{IV} 4498.4\,$\mathrm{\AA}$). The line identification is done using the CMFGEN atomic data\footnote{\url{http://kookaburra.phyast.pitt.edu/hillier/web/CMFGEN.htm}}. These lines stand out in the first few spectra, while a broader neighbouring component emerges later so they form a blend. In Fig.~\ref{Fig:VNL_profiles} we present the evolution of the lines. 

We measured the radial velocity and width of these lines by applying single Gaussian fitting. The values are listed in Table~\ref{table:4VNL}. Their width associates them with a region of low expansion velocity. If these are indeed high ionization \eal{O}{VI} and \eal{Ne}{IV} lines, they must originate from a very hot region. \citet{Belle_etal_2003} found similar lines of \eal{N}{V} in the spectra of the IP EX Hydrae and they have attributed these to an emission region close to the surface of the WD. Not to rule out that disk wind might also be responsible for the formation of such lines (see, e.g., \citealt{Matthews_etal_2015,Darnley_etal_2017}).

\subsubsection{Radial velocities}
\label{velo_sec}
Despite detecting changes in radial velocity of the \eal{He}{II} and \eal{O}{VI} lines, we failed to derive any periodicity from their radial velocities. This is expected as the SALT spectra are taken on different nights, separated by a few days, up to a month. With such a cadence we would not expect to find modulation in the radial velocity curves. Phase-resolved spectroscopy is needed to do this and to investigate the structure of the system via Doppler tomography (see, e.g., \citealt{Kotze_etal_2016}). In addition, the emission from different components (such as the accretion, stream, disk, and curtain and the secondary) adds to the complexity of the line profiles.

The absence of any spin-period-related modulation in the velocity curves is also expected, due to the long exposure time and cadence of the spectra. The exposure time of the SALT HRS spectra is more than three times that of the $P_{\mathrm{spin}}$, thus, any possible WD spin-dependant effects will be smeared out in the spectra.

\begin{figure}
\begin{center}
  \includegraphics[width=\columnwidth]{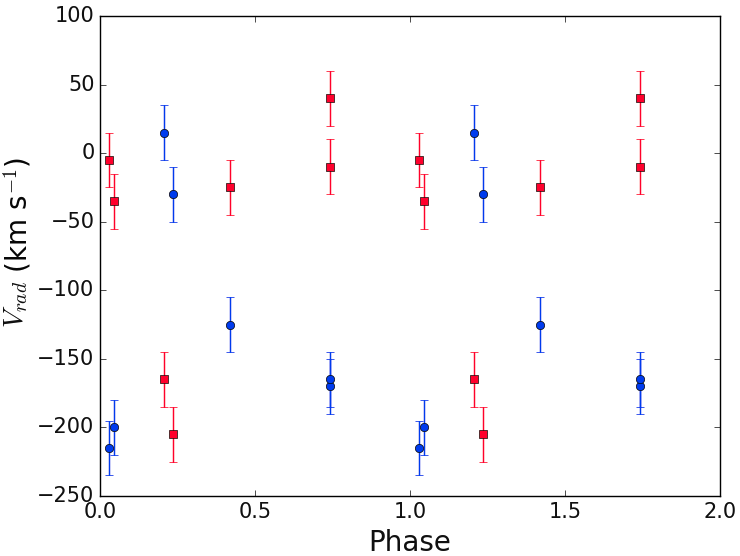}
\caption{Radial velocities of the \eal{He}{II} 4686\,$\mathrm{\AA}$ medium width component (MWC; blue circles) and small width component (SWC; red squares) plotted against the orbital phase. The evolution of the velocities is anti-correlated for the two components.}
\label{Fig:phased_velocities}
\end{center}
\end{figure}

Fig.~\ref{Fig:phased_velocities} shows the phase-folded ( over the $P_{\mathrm{orb}}$) radial velocities of the MWC and SWC of the \eal{He}{II} 4686\,$\mathrm{\AA}$. The velocities of these two components are anti-correlated. The SWC and MWC are out of phase and they show opposite radial velocity curves, neither of which is consistent with the 3.57\,h period. 
The opposite phase is probably an indication for the origin of these components where the SWC might be associated with emission from the surface of the secondary and the MWC is originating from an accretion region around the WD (accretion disk; see, e.g., \citealt{Rosen_etal_1987,Schwope_etal_1997}). However, the lack of periodicity in the velocity curves, related to the orbital period, makes it difficult to confirm this claim. 

The radial velocity amplitude of the emission lines from CVs is strongly dependent on the inclination of the system, as well as the $P_{\mathrm{orb}}$ and the mass ratio. \citet{Ferrario_etal_1993} have modelled IPs with a truncated accretion disk and a dipolar magnetic field, resulting in two accretion curtains above and below the orbital plane, which are responsible for most of the line emission. For such systems, the amplitude of the radial velocities can range from $\sim$ 200\,km\,s$^{-1}$ up to $\lesssim$ 1000\,km\,s$^{-1}$ depending mainly on the inclination of the system. They also showed that velocity cancellation can occur if both curtains are visible, leading to velocity amplitudes in the range of $\sim$ 200\,--\,300\,km\,s$^{-1}$. This happen in the case of either a low-inclination system where both curtains are visible through the centre of the truncated disk or if the system is eclipsing (edge-on). The amplitude of the radial velocities derived from the spectral emission lines of V407 Lup is around 200\,km\,s$^{-1}$. This suggests that the system is possibly at a low inclination.

\begin{figure}
\begin{center}
  \includegraphics[width=\columnwidth]{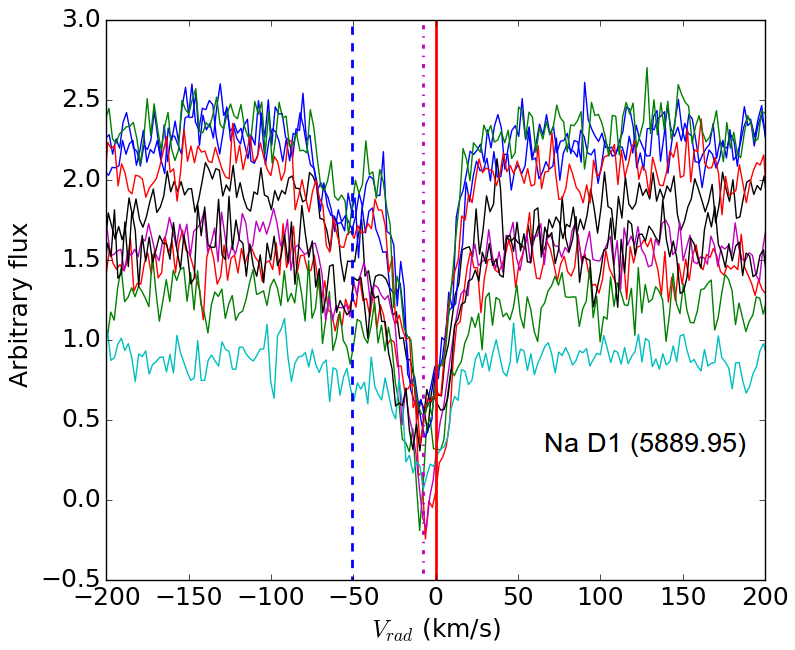}
  \includegraphics[width=\columnwidth]{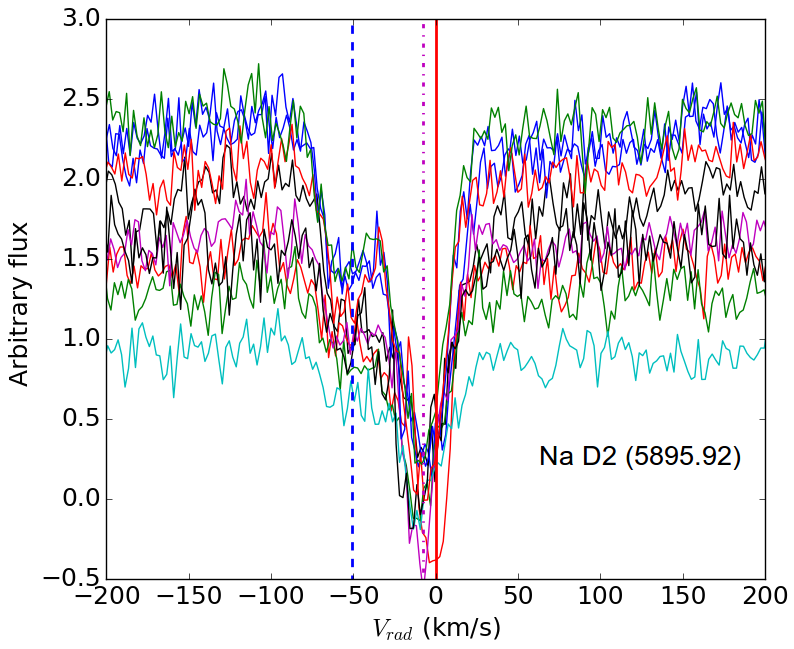}
\caption{The profiles of the \eal{Na}{I} D1 (top) and  D2 (bottom) absorption lines. Each colour corresponds to a different observation. At least two main components can be identified in the lines. The blue dashed line represents the average radial velocity of the weaker component. The magenta dotted line represents the average radial velocity of the stronger component. The red solid line represents the null radial velocity.  A heliocentric correction has been applied to the radial velocities.}
\label{Fig:NaD}
\end{center}
\end{figure}

\subsubsection{Sodium D absorption lines}
\label{NaD}
The sodium D absorption lines, D1 (5896\,$\mathrm{\AA}$) and D2 (5890\,$\mathrm{\AA}$), are well-known tracers of interstellar material (see e.g. \citealt{Munari_Zwitter_1997,Poznanski_etal_2012}). In some cases, circumstellar material around the system can also contribute to \eal{Na}{I}D absorption. For recurrent novae this material has been proposed to be due to ejecta from previous eruptions.

In the high-resolution spectra the \eal{Na}{I}D lines are a complex of at least two components, both blue-shifted (Fig.~\ref{Fig:NaD}). There is a relatively weak component at $\sim -50\pm$ 3\,km\,s$^{-1}$ and a more prominent one at $\sim -8\pm$ 3\,km\,s$^{-1}$. Taking into account a presumable systemic velocity of the system ($\sim-100$\,km\,s$^{-1}$), both components would be red-shifted. We measure a FWHM of $\sim$ 35\,km\,s$^{-1}$ for each component.
We measure an EW of $\sim$ 0.94$\pm$0.02\,$\mathrm{\AA}$ for the D1 line and $\sim$ 0.63$\pm$0.02\,$\mathrm{\AA}$ for the D2 line. These values are used to derive the reddening in Section~\ref{reddening}.


\subsection{\textit{Swift} UVOT spectroscopy}
\label{UV_spec}

\begin{figure*}
\begin{center}
  \includegraphics[width=\textwidth]{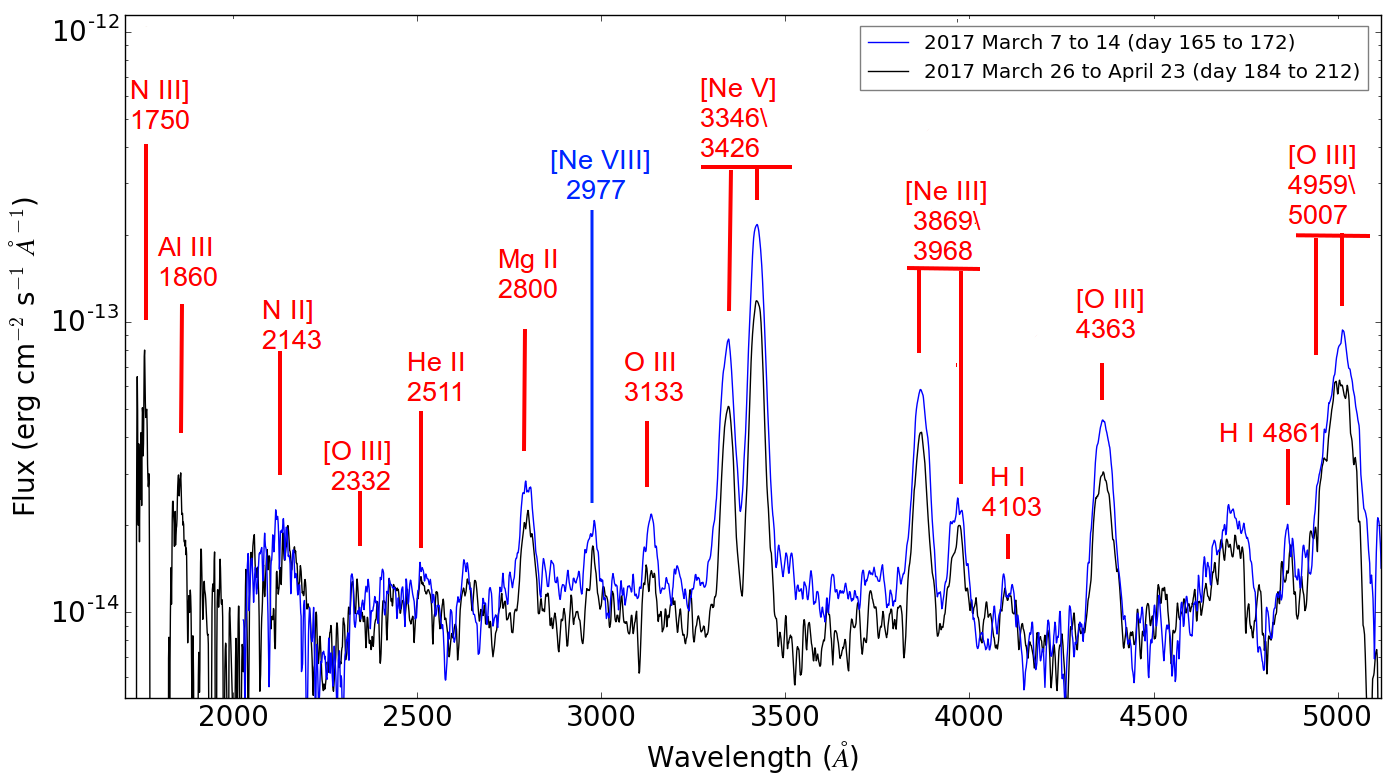}
\caption{The \textit{Swift} UV-optical grism spectra plotted between 1700 and 5100\,$\mathrm{\AA}$. The blue spectrum represents the first four spectra combined (from 2017 March 7 to 14) and the black spectrum represents the second four spectra combined (from 2017 March 26 to April 23).}
\label{Fig:UVOT_spec}
\end{center}
\end{figure*}

The UVOT grism spectra (Fig.~\ref{Fig:UVOT_spec}) are dominated by broad emission lines of forbidden O and Ne, along with H Balmer and \eal{Mg}{II} resonance lines. A list of the observed lines and identifications is given in Table~\ref{uv_line_ids}.
Since the \textit{Swift} UVOT spectra  from the grism exposures have an uncertainty in the position of the wavelengths on the image, the individual spectra were shifted to match the \feal{Ne}{V} 3346\,$\mathrm{\AA}$ and 3426\,$\mathrm{\AA}$ lines. The shifts were 
at most 15\,$\mathrm{\AA}$. Any intrinsic shift of the line spectrum is thus undetermined. 
The emission seen at 1759\,$\mathrm{\AA}$ and 1855\,$\mathrm{\AA}$ in the first order spectrum is uncertain due to the low S/N. 
However, in the grism image we can see the 1750\,$\mathrm{\AA}$ line in the second order spectrum which lies alongside the first order.
The 1860\,$\mathrm{\AA}$ line in the second order is affected by the wings of a nearby first order line, and cannot be confirmed that way.  
There is no significant line of \hfeal{O}{II} at 2471\,$\mathrm{\AA}$, suggesting that the higher ionization state dominates. The \eal{He}{II} 2511\,$\mathrm{\AA}$ line may be blended with an unidentified feature around 2533\,$\mathrm{\AA}$. The line at $\sim$ 2978\,$\mathrm{\AA}$ is possibly \eal{Ne}{VIII} 2977\,$\mathrm{\AA}$ (see, e.g., \citealt{Werner_etal_2007}). 
A likely identification of the broad blend at 4714\,$\mathrm{\AA}$ has been given with the SALT spectrum (\eal{He}{I} 4713\,$\mathrm{\AA}$; see Section~\ref{spec_lines}). Inspection of the spectrum in Fig.~\ref{Fig:UVOT_spec} shows that the Ne and O lines are much stronger than the H and He lines in this nova, which is consistent with the ONe nature of the WD as concluded by \citet{Izzo_etal_2018}.

\begin{figure}
\begin{center}
  \includegraphics[width=80mm]{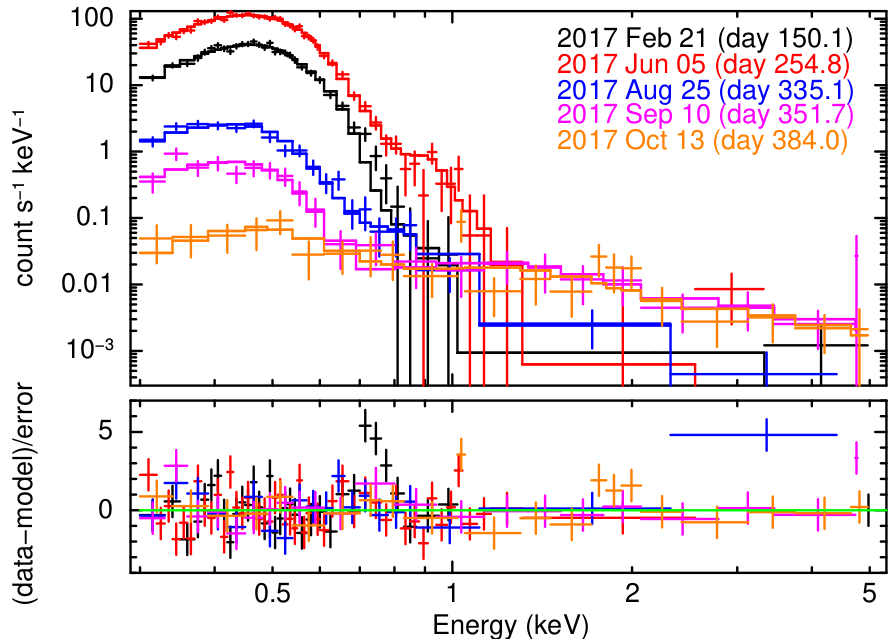}
\caption{The top panel shows a sample of \textit{Swift} X-ray instrumental spectra from days 150.1, 254.8, 335.1, 351.7, and 384.0 since eruption ($t_0$). Note that the FWHM of the energy resolution at 0.5\,keV is $\sim$ 125\,eV. These sample spectra have been modelled with a combination of an atmosphere (as considered throughout the rest of the paper) and a fainter optically-thin component to account for the higher energy photons. The lower panel shows the residuals plotted in terms of sigma.}
\label{Fig:finalcomp}
\end{center}
\end{figure}

\begin{figure*}
\begin{center}
  \includegraphics[width=124mm]{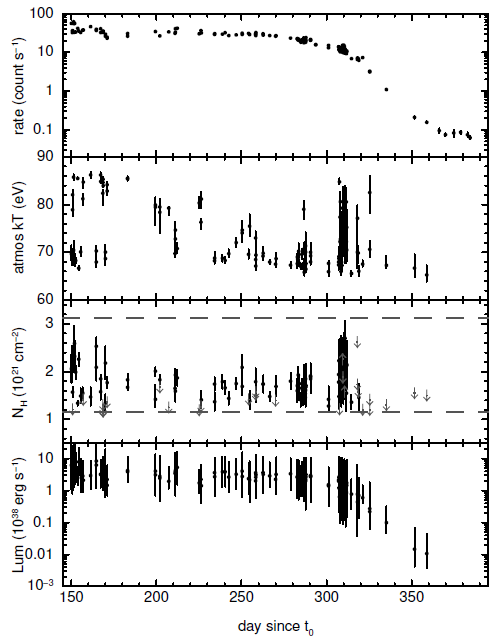}
\caption{Results from fitting the \textit{Swift} XRT spectra (with TMAP atmosphere model; see text for more details), plotted against days since $t_0$. From top to bottom: the X-ray count rate; the temperature of the SSS emission; the best-fit absorbing column density, constrained to lie in the range of (1.15\,--\,3.12)$\times$10$^{21}$\,cm$^{-2}$ from the fits to the \textit{Chandra} and \textit{XMM-Newton} grating spectra, marked by the horizontal dashed lines; bolometric luminosity, for which a distance of 10\,kpc is assumed (Section~\ref{distance_sec} -- this distance is likely an upper limit for what the actual distance might be). For comparison, for a 1.25\,M$_{\odot}$ WD, the Eddington luminosity is 1.6 $\times 10^{38}$\,erg\,s$^{-1}$.}
\label{Fig:highnh}
\end{center}
\end{figure*}

The flux and width of the stronger lines of \eal{Mg}{II}, \eal{O}{III}, \feal{Ne}{V}, and 
\feal{Ne}{III}, have been measured for each of the eight UV spectra and the values are shown in Table~\ref{uv_line_measurements}. The method used was to plot the line, select the points at which the line merges into the background, and then sum the flux over the line minus the background. The flux uncertainty is around 15\% so the forbidden line ratios are as expected in the low density case. The Full Width at Zero Intensity (FWZI) clearly depends on the strength of the line. Weaker lines merge earlier with the noise. We also see larger FWZI for longer wavelengths up to $\sim$ 7000\,km\,s$^{-1}$.

\subsection{\textit{Swift} X-ray spectral evolution}
\label{swift_xray}

The initial detection of an X-ray source, once V407 Lup had emerged from behind the Sun (day 150), showed it to be in the supersoft regime already, at a consistently high flux (Fig.~\ref{Fig:highnh}). Therefore, if there was a high-amplitude flux variability phase as seen in some well-monitored
novae (e.g. V458 Vul -- \citealt{Ness_etal_2009}; RS Oph -- \citealt{Osborne_etal_2011};
Nova LMC 2009a -- \citealt{Bode_etal_2016}; Nova SMC 2016 -- \citealt{Aydi_etal_2018}), it occurred while V407 Lup was behind the Sun and hence unobservable to {\it Swift}.

The early X-ray spectra could be acceptably well modelled below 1 keV with a TMAP\footnote{TMAP: T{\" u}bingen NLTE Model Atmosphere Package:\url{
http://astro.uni-tuebingen.de/~rauch/TMAF/flux_HHeCNONeMgSiS_gen.html}} absorbed plane parallel, static, Non-Local Thermal Equilibrium (NLTE) stellar atmosphere model (\citealt{Rauch_etal_2010}; grid 003 was used). The {\sc tbabs} absorption model \citep{Wilms_etal_2000} within {\sc XSPEC} was applied, using the Wilms abundances and Verner cross-sections; this parameter was allowed to vary in the range (1.15\,--\,3.12) $\times$ 10$^{21}$\,cm$^{-2}$, based on the analysis of the \textit{XMM-Newton} RGS and \textit{Chandra} LETG spectra (Sections~\ref{XMM_sec} and~\ref{Chandra_sec}). A sample of the spectra obtained during the monitoring is shown in Fig.~\ref{Fig:finalcomp}, while Fig.~\ref{Fig:highnh} shows the results of the modelling of the supersoft component. The luminosity was derived using an assumed distance of 10\,kpc, which is currently unknown (see Section~\ref{distance_sec}).

We note, however, that an absorbed blackbody model with a \eal{N}{VII} edge at $\sim$0.65\,keV (see Section~\ref{XMM_sec}) provides a statistically better fit than the more physically-based atmosphere model. For example, considering the spectrum obtained on day 200, while a simple blackbody is a poor fit with C-stat/dof = 611/66, including the oxygen absorption edge decreases this to C-stat/dof = 109/65; the TMAP atmosphere grid leads to C-stat/dof = 533/66, underestimating the observed X-ray flux between 0.8\,--\,1\,keV. Fitting the spectra with the blackbody+edge model, there is no evidence of a temporal trend for the optical depth of the edge over time, with $\tau$ typically lying in the range 1\,--\,4. The absorbing column required for such blackbody fits is higher than for the atmosphere parameterisation, at $\sim$~3~$\times$~10$^{21}$\,cm$^{-2}$. While the blackbody+edge model is statistically preferred, the luminosities from these fits are around a factor of ten higher than those from the atmosphere grids, being $\sim$ a few $\times$ 10$^{39}$\,erg\,s$^{-1}$ between days 150 and 300.

\begin{figure*}
\begin{center}
  \includegraphics[width=180mm]{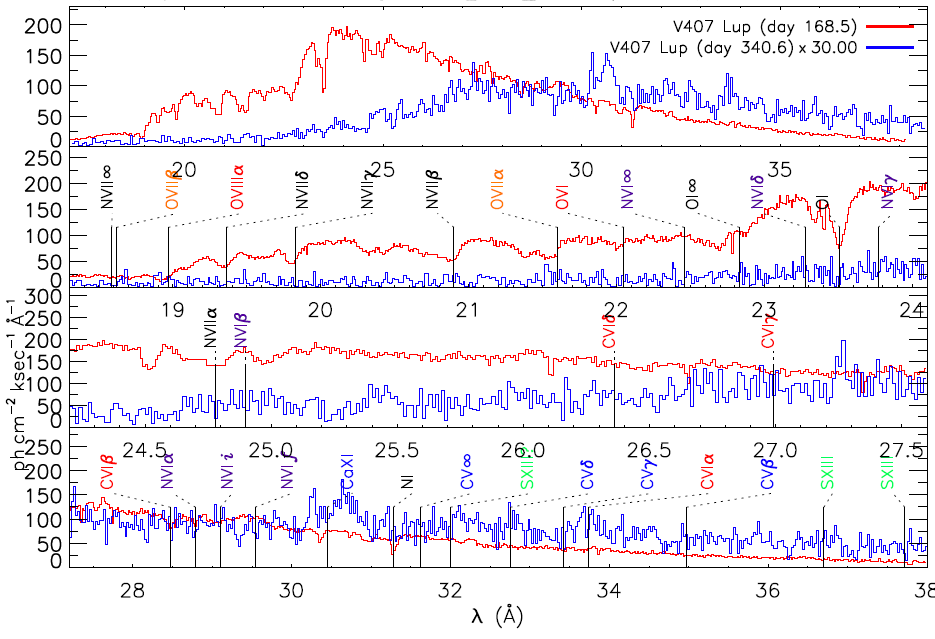}
\caption{The \textit{XMM-Newton} RGS spectrum (red) taken on day 168.5. The spectrum is combined from RGS1 and RGS2 using the SAS tool \textsc{rgsfluxer}. The \textit{Chandra} spectrum (blue) taken on day 340.6 is also plotted for comparison and has been multiplied by a factor of 30. It is worth noting that the \textit{Chandra} LETG spectral range is much larger (see text for more details).}
\label{Fig:RGS_spec}
\end{center}
\end{figure*}

\begin{figure*}
\centering
\includegraphics[width =178mm]{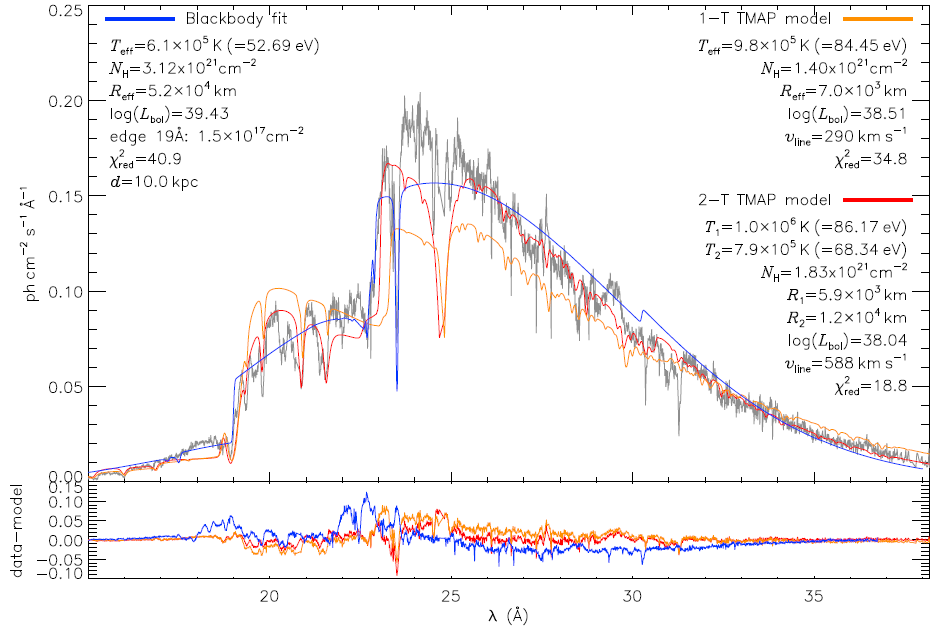}
\caption{The top panel shows a blackbody model (blue) including an edge at $\sim$ 18.6\,$\mathrm{\AA}$ (to reproduce the \eal{N}{VII} edge; see text for more details), a one-component TMAP model (orange), and a two-components TMAP model (red) all fitted to the \textit{XMM-Newton} RGS spectrum (gray). The parameters of each model are given in the plot. The TMAP models are plane parallel, thus without a radius in the model itself. Therefore, the radii shown in the legend of the TMAP models (1-T and 2-T) were derived assuming spherical symmetry. The bolometric luminosities in the legend were derived using the values of the radii and $T_{\mathrm{eff}}$ via Stefan-Boltzmann law. These differ from the luminosities derived in the text based on the X-ray fluxes (see text for more details). The abundances of the interstellar O and N were reduced to 0.6 solar in the blackbody model. The bottom panel shows the residuals of the fits as they vary against wavelength. The luminosities and radii derived here are distance dependent and therefore a range of these values are presented in Appendix~\ref{app_dist}.}
\label{Fig:RGS_BB_TMAP}
\end{figure*}

It is noticeable that the later spectra (day 325 onwards) show evidence for harder X-ray emission above $\sim$~1.5\,keV. This corresponds to the time at which the X-ray source had faded sufficiently that PC mode could be used. The WT mode suffers from a higher background level\footnote{http://www.swift.ac.uk/analysis/xrt/digest\_cal.php\#trail shows the typical count rate per column}; in the case of the data collected for V407 Lup, the WT background begins to dominate the source emission around 1\,keV, meaning that any such harder component cannot be easily measured.
\\

Using the dataset collected on 2018 January 23 (day 486), at which point the SSS emission has completely faded away, this harder spectral component can be parameterized as optically thin emission with kT $>$ 12\,keV, and a 2\,--\,10\,keV luminosity of $\sim$~4$\times$10$^{33}$\,erg\,s$^{-1}$, assuming a very uncertain distance of 10\,kpc (Section~\ref{distance_sec}). This is the order of magnitude expected for the luminosity from an intermediate polar assuming a typical accretion rate for a CV with a period of $\sim$~3\,hr of $\sim$~10$^{-8}$\,--\,10$^{-7}$\,M$_{\odot}$\,yr$^{-1}$ \citep{Patterson_etal_1994,Pretorius_Mukai_2014}. Even at a shorter distance ($d \sim$ 3\,--\,5\,kpc), the hard X-ray luminosity would still be in the right range expected from IPs.

Other novae observed by \textit{Swift} have also shown harder ($>$1\,keV) emission -- see \citet{Osborne_2015}, as well as the sample of novae discussed by \citet{Schwarz_etal_2011}. However, these objects typically showed evidence for this harder component from early on in the outburst, before the start of the SSS phase, rather than "switching on" part way through the nova evolution [although V959 Mon \citep{Page_etal_2013_May} did reveal a slow rise and fall of the 0.8\,--\,10\,keV emission]. In these cases, this harder component was explained as shocks, either within the nova ejecta or with external material, such as the wind from a red giant secondary.

Assuming the same spectral shape as above (optically thin component with kT $>$ 12\,keV), we can place a limit on the luminosity of this component during the WT observations of up to a factor of ten below that measured at late times (after day 325). This suggests that this harder X-ray component has turned on significantly after the nova outburst, and could therefore be a signature of restarting accretion. However, with the available data, it is not possible to rule out that the emission may be caused by shocks.


\subsection{\textit{XMM-Newton} high-resolution X-ray spectroscopy}
\label{XMM_sec}

The absorbed flux at Earth, as measured with the RGS at day 168 over the 14\,--\,38\,$\mathrm{\AA}$ range, was (1.2$\pm$0.5)$\times10^{-9}$\,erg\,cm$^{-2}$\,s$^{-1}$. The RGS spectrum (Fig.~\ref{Fig:RGS_spec}) is dominated by a bright SSS continuum with relatively weak absorption lines covering most of the spectral range. The most obvious features are absorption edges from \eal{N}{VII} at 18.6\,$\mathrm{\AA}$ and \eal{O}{I} 22.8\,$\mathrm{\AA}$ and absorption lines from \eal{N}{VII} 1s-\{2p,3p,4p,5p\} (rest-wavelengths at 24.74\,$\mathrm{\AA}$, 20.9\,$\mathrm{\AA}$, 19.83\,$\mathrm{\AA}$, and 19.36\,$\mathrm{\AA}$), \eal{O}{VII} 1s-2p (rest-wavelength at 21.6\,$\mathrm{\AA}$), and \eal{N}{VI} 1s-{2p,3p} (rest-wavelengths at 28.78\,$\mathrm{\AA}$ and 24.9\,$\mathrm{\AA}$). These lines are shifted by at most $-$400\,km\,s$^{-1}$. In addition, the \eal{O}{VII} and \eal{N}{VII} 1s-2p lines contain a fast component of $-$3200\,km\,s$^{-1}$. A line at 28.5\,$\mathrm{\AA}$ may either be \eal{C}{VI}, shifted by $-$400\,km\,s$^{-1}$, or \eal{N}{VI}, shifted by $-$3200\,km\,s$^{-1}$. Further we find interstellar absorption lines of \eal{O}{I} 1s-2p (23.5\,$\mathrm{\AA}$) and \eal{N}{I} 1s-2p (31.3\,$\mathrm{\AA}$).

The continuum can be parameterized surprisingly well by a blackbody fit plus an absorption edge at $\sim$ 18.6\,$\mathrm{\AA}$ to reproduce the \eal{N}{VII} edge ($\tau$=0.94 equivalent to a column density 1.5$\times10^{17}$\,cm$^{-2}$), yielding $T_{\mathrm{eff}}$=6.1$\times 10^{5}$\,K (kT=53\,eV), $N_{\rm H}$ = 3.12$\times 10^{21}$\,cm$^{-2}$ (Fig.~\ref{Fig:RGS_BB_TMAP}). The abundances of the interstellar O and N were reduced to 0.6 solar. The column density of the \eal{N}{VII} edge suggests that it arises from a hot plasma (the ionization potential of \eal{N}{VII} is $\sim$ 667\,eV). The presence of such a deep absorption edge, while seeing ``shallow" absorption lines, could indicate that the plasma is highly ionized. A high degree of ionization makes the plasma more transparent, and possibly this can explain why the continuum is best fitted by a blackbody model. Note that the \textit{Chandra} spectrum on day 340 shows evidence for even fewer absorption lines (see Section~\ref{Chandra_sec}).
The continuum normalization corresponds to a radius of 5.2$\times$10$^{4}$\,km (assuming spherical symmetry; this is an order of magnitude larger than a typical WD radius) at an assumed distance of 10\,kpc (Section~\ref{distance_sec}). While a bloated WD is a possible interpretation, overestimates of the radius are quite common when using blackbody fits. In Appendix~\ref{app_dist} we present conversion equations for the parameters that are distance dependent (e.g. radii and luminosities). We also present a range of values for these parameters using different distance assumptions.

We also experimented with the NLTE atmosphere models from TMAP (\citealt{Rauch_etal_2010}; Fig.~\ref{Fig:RGS_BB_TMAP}) and with the synthetic models for expanding atmospheres ``wind-type" model (\citealt{Van_Rossum_2012}; Fig.~\ref{Fig:WT}). Neither of these models reproduced the observed RGS spectrum of V407 Lup. The closest approximation with a plane-parallel, static, NLTE TMAP model yields $T_{\mathrm{eff}}$=9.8$\times10^{5}$\,K (kT=85\,eV), $N_{\rm H}$ = 1.4$\times10^{21}$\,cm$^{-2}$, and an absorption line velocity of 290\,km\,s$^{-1}$ (see Fig.~\ref{Fig:RGS_BB_TMAP}). The absorbed and unabsorbed X-ray (15\,--\,38\,$\mathrm{\AA}$; 0.33\,--\,1\,KeV) fluxes derived from this model are 1.19 $\times 10^{-9}$\,erg\,cm$^{-2}$\,s$^{-1}$ and 5.45 $\times 10^{-9}$\,erg\,cm$^{-2}$\,s$^{-1}$, respectively. The X-ray flux derived from the TMAP model is more than 95\% the bolometric flux, therefore this flux equates to an absolute bolometric luminosity of $\sim$ 6.5$\times 10^{37}$\,erg\,s$^{-1}$ at a distance of 10\,kpc (see Section~\ref{distance_sec} and Appendix~\ref{app_dist}). We note that this TMAP model also contains a few weak absorption lines. The comparison with the wind-type atmosphere models suggests $T_{\mathrm{eff}}$ between 5.5$\times10^{5}$\,K and 6.0$\times10^{5}$\,K (kT=47\,--\,52\,eV).

Due to the complexity of the spectrum, we consider the idea that the atmosphere is not homogeneous, possibly due to hotter emitting regions towards the poles. Therefore, we carried out modelling using two TMAP model components from the grid 003. The parameters of the models are given in Table~\ref{table:two_mod_para}. The combination of these two models results in a reasonably good fit to the continuum and it is presented in Fig.~\ref{Fig:RGS_BB_TMAP}. The total flux derived from the two-component TMAP models is 1.22$\times 10^{-9}$\,erg\,cm$^{-2}$\,s$^{-1}$ and the total unabsorbed flux is 9.30$\times 10^{-9}$\,erg\,cm$^{-2}$\,s$^{-1}$. At a distance of 10\,kpc (see Section~\ref{distance_sec} and Appendix~\ref{app_dist}), this equates to an absolute bolometric luminosity of $\sim$ 1.1$\times 10^{38}$\,erg\,s$^{-1}$.

\subsection{\textit{Chandra} high-resolution X-ray spectroscopy}
\label{Chandra_sec}

The \textit{Chandra} LETG spectrum taken on day 340 is presented in Fig.~\ref{Fig:RGS_spec}, together with the \textit{XMM-Newton} RGS spectrum. The observed (absorbed) integrated flux measured with the \textit{Chandra} LETG is 2.89$\times 10^{-10}$\,erg\,cm$^{-2}$\,s$^{-1}$, which is a factor of $\sim$ 45 lower than the observed (absorbed) integrated flux measured with \textit{XMM-Newton} RGS on day 168 (see Section~\ref{XMM_sec}). It is worth noting that the RGS spectrum is sensitive to a shorter range (6\,--\,38\,$\mathrm{\AA}$) compared to the LETG spectrum (1.2\,--\,175\,$\mathrm{\AA}$; i.e. a part of the flux is not included in the RGS spectrum). Even if $N_{\rm H}$ has not decreased, this would imply that the effective temperature, which scales as $L^{-1/4}$, has decreased by a factor of 2.6. This is can be ruled out by examining Fig.~\ref{Fig:RGS_spec} which does not show such a shift in temperature (the peaks of the RGS and LETG spectra show a decrease of temperature by a factor of $\lesssim$1.4). We conclude that we are possibly only observing a small portion of the WD surface. This might be an indication that at this stage ($\sim$ day 340) accretion has resumed and the WD is partially hidden by an accretion disk and/or accretion curtains (see Section~\ref{acc_res_sec} for further discussion) or it may instead mean that the temperature is not homogeneous on the WD surface and supersoft X-rays are emitted only in a restricted region of the surface, possibly on the polar caps (see, e.g., \citealt{Zemko_etal_2015,Zemko_etal_2016}). Another intriguing element in the spectrum is that the
 \eal{N}{VII} absorption feature appears blue-shifted by $\sim -$4600\,km\,s$^{-1}$, which  is a larger velocity compared to the blue-shift measured on day 168 in the RGS spectrum (Fig.~\ref{Fig:RGS_spec}). 

\begin{figure}
\centering
\includegraphics[width = \columnwidth]{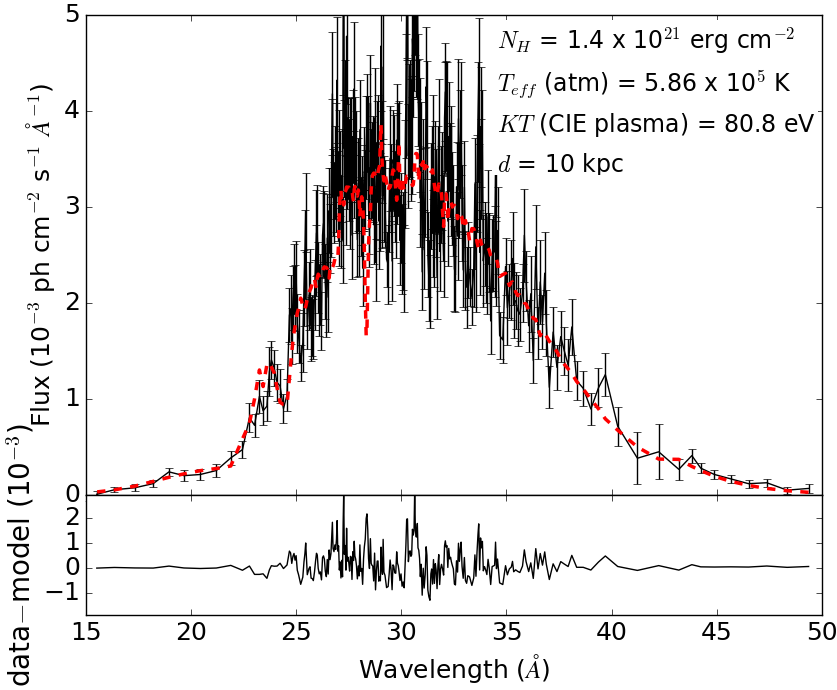}
\caption{The top panel shows Model 1 (red) fit to the \textit{Chandra} LETG spectrum (black). The model consists of two components: a TMAP atmosphere model and a \texttt{bvapec} Collisional Ionization Equilibrium (CIE) plasma model. The carbon abundance in the CIE model are enhanced to 19.6 times the solar abundance. See text for more details. The bottom panel shows the residuals of the fit as they vary against wavelength.} 
\label{Fig:chandra_mod}
\end{figure}

We tried two models to fit the \textit{Chandra} LETG spectrum. In the first model (Model 1; parameters given in Table~\ref{table:chandra_mod_1}) we assumed that the bulk of the flux comes from the WD surface, with an additional component of plasma in collisional ionization equilibrium. For this purpose we used a TMAP atmosphere model with a \texttt{bvapec} model based on the \textit{Astrophysical Plasma Emission Code} (APEC; \citealt{Smith_etal_2001}), with enhanced carbon abundances (19.6 $\times$ solar abundance) and assuming solar abundances for the other neutral absorbing elements. The fit is not perfect and underestimates the observed flux by 10\%, so we conclude that there must be an additional component, possibly a plasma at a different temperature and electron density, since the residual flux seems to be in emission lines.
We experimented by varying the abundances or the wind velocity as red-shift, but this did not result in a better fit. The emission lines appear to be broadened by $\sim$ 5000\,km\,s$^{-1}$, although we had to fix this value to avoid obtaining an unreasonably high velocity. Similarly to the RGS spectrum, we also cannot fit well the shallow absorption line profiles. The total absorbed and unabsorbed X-ray (in the range 15\,--\,50\,$\mathrm{\AA}$; 0.248\,--\,0.827\,KeV) fluxes derived from this model are 2.74$\times 10^{-11}$\,erg\,cm$^{-2}$\,s$^{-1}$ and 4.76$\times 10^{-10}$\,erg\,cm$^{-2}$\,s$^{-1}$. The X-ray flux derived from Model 1 are more than 95\% the bolometric flux, therefore this equates to an absolute bolometric luminosity of $\sim$ 5.7$\times$10$^{36}$\,erg\,s$^{-1}$ at 10\,kpc (see Section~\ref{distance_sec} and Appendix~\ref{app_dist}). Fig.~\ref{Fig:chandra_mod} represents the fit of Model 1.

In the second model (Model 2; parameters given in Table~\ref{table:chandra_mod_2}), we assumed that there is a homogeneous WD atmosphere, and that the polar caps become heated in excess because of accretion, and emit as an additional blackbody component. Because a blackbody energy distribution is broader than just the atmospheric spectral energy distribution, this results in the assumption that there is more absorption of soft flux, partially lifting the discrepancy in $N_{\rm H}$ value (see Section~\ref{reddening}). The best fit in this case yields $N_{\rm H}$ =1.87$\times 10^{21}$\,cm$^{-2}$. The absorbed flux coming from the blackbody component alone would be $2.15 \times 10^{-11}$\,erg\,cm$^{-2}$\,s$^{-1}$ and an unabsorbed flux of $6.53 \times 10^{-9}$\,erg\,cm$^{-2}$\,s$^{-1}$. At a distance of 10\,kpc (see Section~\ref{distance_sec}), this would imply an absolute X-ray luminosity of $ \sim 7.8 \times 10^{37}$\,erg\,s$^{-1}$ (in the range 15\,--\,50\,$\mathrm{\AA}$; 0.248\,--\,0.827\,KeV), which seems absurd because it is four orders of magnitude higher compared to heated polar caps of magnetic CVs (e.g. \citealt{Zemko_etal_2017} and references therein; see also Section~\ref{acc_res_sec}). It is worth noting that even at shorter distances (3\,--\,5\,kpc; see Appendix~\ref{app_dist}) this luminosity is still three orders of magnitude higher compared to heated polar caps of magnetic CVs.

We also experimented with ``wind-type'' models of \citet{Van_Rossum_2012} for an expanding atmosphere to fit the \textit{Chandra} spectrum. These did not result in a good fit. At a distance of 10\,kpc, the best fit parameters are: $T_{\mathrm{eff}} = 4.5 \times 10^5$\,K,  $N_{\rm H}$ =  $1.7 \times 10^{21}$\,cm$^{-2}$, and $\log g$ (logarithm of the surface gravity) = 8.24. Fig.~\ref{Fig:WT} shows a sample of the best fit models for various wind asymptotic velocities $v_{\infty}$ and mass-loss rates $\dot{M}$.

\section{Discussion}
\label{Disc_sec}
\subsection{Reddening, distance, and eruption amplitude}
\subsubsection{Reddening}
\label{reddening}
Estimating the distance to CNe is not straightforward, but it is a crucial step in deriving the properties of the eruption and its energetics. Estimating the reddening is vital for establishing the distance, and we investigate this using various methods below. 

The reddening maps of \citet{Schlafly_Finkbeiner_2011} indicate $A_V = 0.88$ in the direction of nova 407 Lup ($l=330.1,\ b=9.57$), which should be regarded as an approximate lower limit for any object that is sufficiently distant. From the modelling of the X-ray grating spectra, we derived $N_{\rm H}$ ranging between 1.4 and 3.12 $\times$ 10$^{21}$\,cm$^{-2}$. Using the relation between $N_{\rm H}$ and $A_V$ from \citet{Zhu_etal_2017}, we derive $A_V$ between $\sim$ 0.67 and 1.50 ($\pm$ 0.02). Caution is required when drawing conclusions from these values as they are model dependent.

\citet{Poznanski_etal_2012} presented empirical relations between the extinction in the Galaxy and the EW of the \eal{Na}{I}D absorption doublet. Using these and the measured EW of the \eal{Na}{I}D absorption lines at 5990.0\,$\mathrm{\AA}$ (D1) and 5896.0\,$\mathrm{\AA}$ (D2, see Section~\ref{NaD}), we derive $E(B-V) = 0.93 \pm0.25$ and $A_V$ = 2.89 $\pm$ 0.25. 

Using colours of novae around maximum has also been suggested as a way to derive reddening \citep{Van_den_Bergh_Younger_1987}. These authors  derived a mean intrinsic colour $(B-V)_0$ = +0.23 $\pm$ 0.06 for novae at maximum light and $(B-V)_0$ = $-0.02$ $\pm$ 0.04 at $t_2$. We lack broadband observations exactly at maximum and $t_2$, however, we measure $(B-V)\,\sim\,0.8\,\pm\,0.04$ at $t_{\mathrm{max}} + 0.5$\,d using SMARTS. This points towards $E(B-V)$ and $A_V$ values close to that derived from the \eal{Na}{I}D EWs. We consider this interpretation as uncertain due to the fast decline and rapid change of the colours around maximum in the case of V407 Lup.

The reddening values we derive from the optical spectroscopy and photometry are much higher than the values derived from the X-ray modelling (Sections~\ref{XMM_sec} and~\ref{Chandra_sec}) and the ones from the reddening maps (see above). This discrepancy suggests that either \citet{Poznanski_etal_2012} relations are unreliable or the nova is embedded in a cloud of interstellar dust. Note that the IP identity of V407 Lup (see Section~\ref{IP_sce_sec}) also adds to the discrepancy between the reddening derived from the optical and X-ray spectra. If the spin modulation seen in the X-ray data is indeed due to absorption from the system (Section~\ref{acc_res_sec}), this means that the X-ray total absorption is due to two components, the interstellar medium (ISM) and the system. Therefore, the absorption due to the ISM is less than the total absorption and the $A_V$ derived from the X-ray spectra might be even smaller. 

\citet{Izzo_etal_2018} derived an intrinsic extinction of $E(B-V) = 0.24 \pm 0.02$ and therefore $A_V = 0.74 $ for V407 Lup, in good agreement with the reddening we derive from the X-ray spectra. These authors have used the diffuse interstellar bands as well as the \citet{Van_den_Bergh_Younger_1987} relations to derive the reddening. The assumption of different $t_{\mathrm{max}}$, $t_2$, and the use of the broadband photometry reported in  \citet{ATel_9550} led \citet{Izzo_etal_2018} to derive low values of reddening from the relations of \citet{Van_den_Bergh_Younger_1987}. The broadband photometry reported in \citet{ATel_9550} around maximum are different (by $\sim$ 0.8\,mag) from simultaneous SMARTS measurements (see Table~\ref{table:photo_max}).

For the purpose of our analysis we will use the value of $A_V = 0.67 \pm 0.02$ derived from the fitting of the LETG \textit{Chandra} spectra which is taken at a late epoch (day 340) and is most sensitive to the column density. In some cases we make use of both extreme values ($A_V$ = 0.67 and 2.89) derived from the X-ray and optical spectroscopy.

\subsubsection{Distance and eruption amplitude}
\label{distance_sec}
The so-called ``Maximum Magnitude-Rate of Decline" (MMRD) relations (see e.g. \citealt{McLaughlin_1939,Livio_1992,Della_Valle_etal_1998,Downes_etal_2000}) were considered useful for deriving distances to novae. These are based on the assumption that the mass of the WD ($M_{\mathrm{WD}}$) is the only factor that controls the brightness of the eruption, which led to the use of novae as standard candles. The relations have been questioned theoretically  (see e.g. \citealt{Ferrarese_etal_2003,Yaron_etal_2005}) and subsequently proved unreliable following  the discovery of faint-fast novae in the M31 \citep{Kasliwal_etal_2011} and  M87 \citep{Shara_etal_2016,Shara_etal_2017_apr}. It is now accepted that several factors influence the eruption of a nova, including the accretion rate, the mass of the accreted envelope, the chemical composition and the temperature of the WD, in addition to  $M_{\mathrm{WD}}$. \citet{Buscombe_1955} have suggested that all novae decline to the same magnitude around 15 days after maximum light. This method is also is also considered uncertain, especially for fast novae (see e.g. \citealt{Darnley_etal_2006,Shara_etal_2017_feb} and references therein).
Using the $M_{V,15}$ method we obtain a distance estimate $d \simeq 3.4^{+1.3}_{-1.0}$\,kpc, with $A_V$ = 2.89 $\pm$ 0.25 and $d \simeq 9.4^{+2.3}_{-1.8}$\,kpc, with $A_V$ = 0.67 $\pm$ 0.02. We do not adopt any of the distances derived using the methods discussed above, due to their great uncertainty. 

The nova is included in Gaia DR1 (data between 2014 July 25 and 2015 September 16) at $G=18.95$ and in DR2 (data between 2014 July 25 and 2016 May 23) at $G=20.66$. DR2 also provides a parallax of $1.27\pm1.51$ which does not contribute anything to our discussion of the distance.

For the purpose of this discussion we assume a distance of 10 kpc, which should probably be regarded as an upper limit. Distance dependent parameters (e.g. radius and luminosity; see Appendix~\ref{app_dist}) can be easily adjusted for other values. Future Gaia data releases will probably improve the parallax and its associated uncertainty.

Data collection for DR2 stopped just before the nova eruption. The change of $\Delta G \sim 1.7$ between DR1 and DR2 suggests the nova may have faded before the eruption.
At maximum, nova V407 Lup was at $V \lesssim 6.0$. Therefore, the amplitude of the eruption is $\sim$ 15\,mag. Note that the amplitude of the eruption is determined by both the energetics of the eruption itself and by the luminosity of the donor during quiescence. Such a large amplitude is expected for novae of the same speed class as V407 Lup, with a main-sequence companion (see e.g. fig. 5.4 in \citealt{Warner_1995}). Large amplitude eruptions have been observed in a few other novae, including V1500 Cyg, CP Pup, and GQ Mus (see e.g. \citealt {Warner_1985,Diaz_Steiner_1989}). 

\subsection{Colour-magnitude evolution}
\label{CMD_sec}
\begin{figure}
\begin{center}
  \includegraphics[width=\columnwidth]{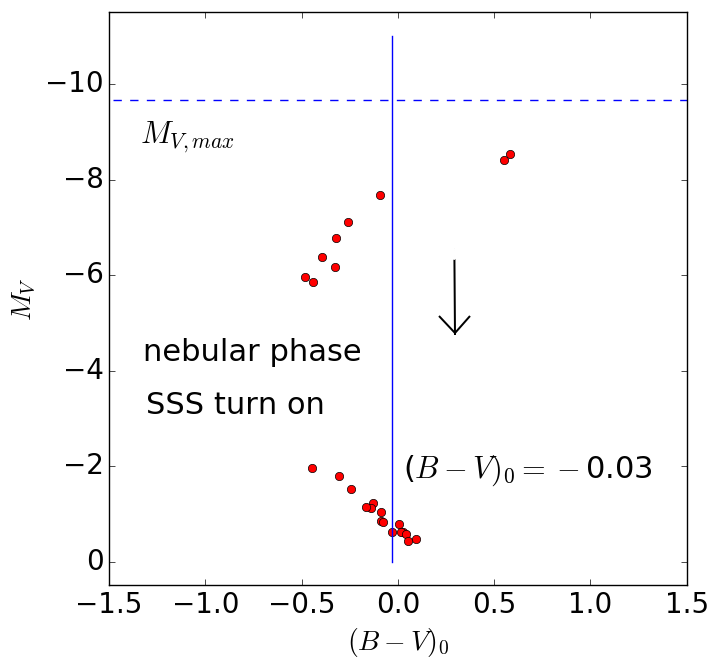}
\caption{Colour-magnitude diagram of V407 Lup showing $M_V$ against $(B-V)_0$ (red dots) at a distance $d$ = 10\,kpc and $E(B-V) = 0.22 (A_V = 0.67)$. The evolution of $M_V$ against time is going vertically from top to bottom (as indicated by the black arrow) and it spans from day $\sim$ 2 to day $\sim$ 13 post-eruption and then from day $\sim$ 117 to day $\sim$ 283. Between these two intervals, a turning-point took place in the evolutionary track of the colours, switching from a blue-ward track to a red-ward track. Around this turning point the nova must have entered the nebular phase and the SSS phase must have started, as indicated in the diagram. The blue horizontal, dashed line indicates the estimated maximum absolute magnitude. The blue vertical line indicates the intrinsic colours for an optically thick wind free-free emission (see e.g. \citealt{Hachisu_Kato_2014,Darnley_etal_2016} and references therein).}
\label{Fig:CMD}
\end{center}
\end{figure}

In Fig.~\ref{Fig:CMD} we present a colour-magnitude diagram (CMD) illustrating the evolution of $M_V$ as a function of $(B-V)_0$. Such diagrams have been used to constrain the nature of the secondary star in CNe. \citet{Darnley_etal_2012} proposed a new classification system for CNe based exclusively on the evolutionary state of the secondary star: a main-sequence star (MS-Nova), a sub-giant star (SG-Nova), or a red giant star (RG-Nova). \citet{Hachisu_Kato_2016_Jan} have shown that there is a difference between the evolutionary track on the CMD of MS- and SG-novae compared to that of RG-novae. For the latter, the track follows a vertical trend between $(B-V)_0$ = $-$0.03 and $(B-V)_0$ = 0.13. The first value represents the intrinsic colour of optically thin free-free emission (free-free emission from an ``optically thin wind" -- winds that are accelerated outside the photosphere), while the second value represents the intrinsic colour of optically thick free-free emission (free-free emission from an ``optically thick wind" -- winds that are accelerated deep inside the photosphere; see fig. 5 in \citealt{Hachisu_Kato_2014}). Novae with main-sequence or sub-giant companions show a track that evolves blue-ward initially going from maximum light to reach a turning point around the start of the nebular phase. After the turning point, the track evolves red-ward in the opposite direction (see fig. 7 in \citealt{Darnley_etal_2016}).

Fig.~\ref{Fig:CMD} demonstrates that the evolution of the colours on the CMD follows a similar track to that shown by MS- and SG-novae. Initially, the colours evolve blue-ward between day 2 and day 13, when the broadband observations stopped for more than a hundred days (Sun constrained) and only resumed at day 117, when the colours had already started evolving red-ward. Hence, the turning-point and eventually the start of the nebular phase and the SSS phase have occurred between day 13 and day 117.

Although we lack broadband observations of the source during quiescence, the evolution of the colour-magnitude track suggests that the companion star is likely to be a main-sequence or a sub-giant star. This is consistent with: (1) the quiescent magnitude measured by Gaia ($G \sim$ 19). At this magnitude and with a giant secondary, the distance to the nova would be unreasonable $(d \sim$ 30\,--\,40\,kpc); (2) the short orbital period (3.57\,h) of the binary system. Such a short orbital period rules out a giant or sub-giant secondary. 

It is worth noting that the first two data points on the CMD, measured at $\sim$ 0.5\,-\,1.5\,d after maximum, represent an intrinsic colour $(B-V)_0 \sim 0.6$. This is inconsistent with the intrinsic colours of novae around maximum and is an indication that the adopted value of $A_V = 0.67$ might be too low a value (see Section~\ref{reddening}).

\subsection{Mass of the WD and ejected envelope}
\label{WDmass_sec}
The behaviour of the nova eruption is governed by several factors such as the WD mass, temperature, and chemical composition, as well as the chemical composition of the accreted matter \citep{Yaron_etal_2005,Hillman_etal_2014,Shara_etal_2017_apr}. However, the mass of the WD and consequently the mass and velocity of the ejected envelope are considered the main factors that control the decline rate of the light-curve. The more massive the WD is, the less accreted material is required to trigger a TNR and hence, the mass of the ejected envelope is small and the light-curve decline time is short. The opposite is equally true. With $t_2 \leq$ 2.9\,days, the mass of the WD of V407 Lup is expected to be $\gtrsim$ 1.25\,M$_{\odot}$ \citep{Hillman_etal_2016}.

\citet{Shara_etal_2017_apr} derived an empirical relation between the mass of the accreted envelope and the decline time $t_2$, using nova samples from M87, M31, the large Magellanic cloud, and the Galaxy:
$\log M_{\mathrm{env}} = 0.825 \log (t_2) - 6.18$. Using this relation, we derive $M_{\mathrm{env}}$ $\sim 1.6 \times 10^{-6}$\,M$_{\odot}$, typical for very fast novae \citep{Hachisu_etal_2016}.

\citet{Sala_Hernanz_2005} derived relations between the SSS temperature and the mass of the WD. For $kT \sim$ 80$\pm$10\,eV (see Fig.~\ref{Fig:highnh}) and for $M_{\mathrm{env}}$ $\sim 1.6 \times 10^{-6}$\,M$_{\odot}$, we estimate a $M_{\mathrm{WD}} \sim$ 1.1\,--\,1.3\,M$_{\odot}$ (see fig.~1 in \citet{Sala_Hernanz_2005}). We also estimate a value of  $M_{\mathrm{WD}} \sim$ 1.2\,--\,1.3\,M$_{\odot}$ based on fig. 1 in \citet{Wolf_etal_2014_erra}. Note that the temperature values given in Fig.~\ref{Fig:highnh} are model dependent, therefore caution is needed when interpreting the aforementioned estimates. 

The SSS turn-on time is also a useful indication for the mass of the ejected envelope. However, it is not possible to constrain when the SSS emission emerged as the \textit{Swift} monitoring was interrupted due to solar constraints and only resumed after the SSS has already turned on. On the other hand, the SSS turn-off time ($t_{\mathrm{off}}$), which is strongly correlated with the WD mass, is better constrained by the X-ray observations ($t_{\mathrm{off}} \sim 300$\,d). This time is inversely proportional to the WD mass as $t_{\mathrm{off}} \propto$ M$_{\mathrm{WD}}^{-6.3}$ \citep{MacDonald_1996}. After an X-ray survey in M31, \citet{Henze_etal_2014_Mar} derived empirical relations between the SSS parameters. Their relations suggest that very fast novae, such as V407 Lup (with $t_2 < 5$\,d), are expected to have $t_{\mathrm{on}} \lesssim$ 25\,d and $t_{\mathrm{off}} \lesssim$ 100\,d. This is much shorter than the observed $t_{\mathrm{off}}$ for V407 Lup (see Section~\ref{acc_res_sec} for further discussion). 

Based on the optical light-curve decline rate and the temperature of the SSS, we estimate a WD mass  between 1.1 and 1.3\,M$_{\odot}$.

\subsection{Resumption of accretion}
\label{acc_res_sec}

Although the \textit{XMM-Newton} X-ray spectrum (taken on day 168) is originating from H burning on the WD surface, the spin-modulated signal observed in the \textit{XMM-Newton} X-ray data is possibly indirect evidence for accretion restoration onto the polar regions of the WD, as early as 168 days post-eruption. This accretion is either igniting H burning at the poles or resulting in an inhomogeneous atmosphere, leading to the observed modulation.

Typically, the spin-modulated signals seen in the X-ray light-curves of IPs are due to the obscuration caused by the accretion curtains and possibly by the changing projected area of the hard X-ray accretion-shock-regions near the pole of the WD. However, the spin-modulated \textit{XMM-Newton} light-curve of V407 Lup is observed during the SSS where the energy output is likely to be much higher than what could result from accretion (i.e. the release of gravitational potential energy by accretion on to the WD gives a much lower luminosity per unit mass accreted than nuclear burning does). Hence, there is a possibility that the spin-modulated X-ray emission seen in the \textit{XMM-Newton} RGS data is not simply due to the supersonic accretion shock (or absorption by an accretion curtain), however it is originating from the burning of the accreting material as it arrives at the WD in magnetically confined regions close to the magnetic poles, possibly making the atmosphere hotter near the poles. This might be responsible for the observed small modulation fraction of $\pm$5\% in the RGS data (Fig.~\ref{Fig:hard_ratio}), while simultaneously the post-nova residual H burning on the whole surface of the WD is likely responsible for the majority of the SSS flux. This idea was examined by \citet{King_etal_2002} for a SSS in M31, as IP-like magnetically controlled accretion deposits fuel near to the magnetic poles of the WD, which burns immediately so giving rise to local hot/luminous regions on the WD surface, possibly contributing to the modulation seen in the optical and X-rays. 

Such a conclusion is well supported by the luminosity derived from the different models in Section~\ref{XMM_sec}. These luminosities exceed the accretion luminosity  by orders of magnitude (the accretion X-ray luminosity of an IP does not exceed a few $10^{33}$\,erg\,s$^{-1}$; e.g. \citealt{Bernardini_etal_2012}). This means that most of the luminosity cannot originate only from accretion onto a restricted area on the WD surface. Therefore, the majority of the SSS luminosity is most likely coming from the whole WD surface while a small fraction is coming from a hot/luminous region near the poles leading to the observed modulation. However, this does not rule out the possibility that absorption from an accretion curtain can also contribute to the observed modulation.

Deriving an accurate value for the fraction of the WD surface which is responsible for the modulation is not achievable with the available data. This value is dependent on constraining the inclination of the system relative to the observer, the temperature ratio between the whole WD surface and the hot/luminous region, and the phenomena responsible for the modulation, which are all still uncertain/unknown. Eclipses in the light-curves are usually used to derive the dimension of the fractional area emitting in the X-ray and causing the modulation. Such eclipses are absent in the light-curves of V407 Lup. An alternative approach is to use the X-ray luminosity derived during the SSS phase, noting the above caveat. Fig.~\ref{Fig:hard_ratio} shows that the amplitude of the modulation is $\sim$ 10\%, therefore we assume that 10\% of the SSS luminosity is emitted by the hot/luminous region (due to burning newly accreted material near the poles). In Section~\ref{XMM_sec} we derive an X-ray luminosity of $\sim$ 1.1$\times$10$^{38}$\,erg\,s$^{-1}$ and a $T_{\mathrm{eff}}$ $\sim$ 80 $\pm$ 10\,eV. Thus, the emitting area is $A \sim L_{(10\%)}/\sigma T_{\mathrm{eff}}^4 \sim 2.6 \pm 1.0 \times 10^{16}$\,cm$^{2}$. Hence, using the effective radius derived from the 1 component T-MAP model (Fig.~\ref{Fig:RGS_BB_TMAP}), the fractional area responsible for the modulation would be $f \sim$ 1.7$\pm 0.6 \times 10^{-2}$. These values are within the normal expectation of 0.001 $\lesssim f \lesssim$ 0.02 for an IP (e.g. \citealt{Rosen_1992,James_etal_2002}). However, using the effective radius derived from the blackbody model (Fig.~\ref{Fig:RGS_BB_TMAP}) implies a fractional area of only $f \sim 3 \pm 1.2 \times 10^{-4}$. 

The $P_{\mathrm{spin}}$ modulation seen in the \textit{Chandra} data at day 340 might also be due to localized H burning near the poles of freshly accreted material, while the residual H burning on the rest of the WD surface is fading. This is supported by the little variation in the hardness ratio and the modulation of the harder X-rays (Fig.~\ref{Fig:HRC_hard_ratio}), which suggest that the $P_{\mathrm{spin}}$ modulation might not be due to typical absorption by an accretion curtain. \citet{Lanz_etal_2005} first noticed that the WD model atmosphere that best fitted the spectrum of the non-nova CAL 83 supersoft source indicated a luminosity almost a factor of 10 higher than observed, and they hypothesized that perhaps the emission originated in hot polar caps. In a study of the decline and return to quiescence of novae that are suspected IPs in which the disk is disrupted and a moderately strong (few 10$^6$ Gauss) magnetic field channels the accretion stream to the WD poles, \citet{Zemko_etal_2015} and \citet{Zemko_etal_2016} found that such novae seem to decline in flux while remaining SSS, as if the emission region shrinks before any final cooling. These authors have attributed the supersoft-emission to H burning at heated regions near the poles where accreting material impinges. Nova V4743 Sgr in eruption also showed supersoft X-ray flux modulated with the WD $P_{\mathrm{spin}}$ (see \citealt{Dobrotka_Ness_2017} and numerous references therein).

Signatures of accretion have often been found in both optical and X-rays observations of post-eruption novae (e.g. \citealt{Leibowitz_etal_1992,Retter_etal_1998,Beardmore_etal_2012,Ness_etal_2012,Orio_etal_2013}) and it has also been hypothesized that the disk may not be fully disrupted in the eruption. It is also possible that nuclear burning is prolonged by renewed accretion, and the supersoft X-ray source lasts for longer because it is refueled by irradiation induced mass transfer \citep{Greiner_etal_2003}.Therefore, we assume that during the SSS phase of V407 Lup, the strongly irradiated secondary is heated by the X-ray emission, leading to an expansion of its outer layers. This will eventually enhance the mass transfer and increase the accretion rate \citep{Osborne_2015}. The material can flow from the L1 Lagrangian point via a ballistic accretion stream to (possibly) a reformed truncated accretion disk. Then it follows the magnetic field lines and finally impacts the WD surface near the magnetic poles, possibly feeding the SSS (see \citealt{Darnley_etal_2017,Henze_etal_2018} and references therein for a discussion about the surface hydrogen burning feeding from a re-established mass accretion in the case of the non-magnetic system M31N 2008-12a). In the case of V407 Lup, this might also explain the long-lasting SSS emission for what is likely a massive WD (see Section~\ref{WDmass_sec}).

An X-ray power spectrum with a dominant signal at the frequency of the WD $P_{\mathrm{spin}}$, such as the \emph{Chandra} power spectrum (Fig.~\ref{Fig:LSP_X-ray}) taken on day 340, might be an indication of a disk-fed accretion \citep{Ferrario_etal_1999}. \citet{Norton_etal_1992_a} point out that in this case the spin modulation dominates and the orbital modulation vanishes because ``passage through a disk washes out all dependence on the orbital phase from the accretion process''. Therefore, it is very likely that a truncated accretion disk has reformed by the time of the \emph{Chandra} observation or even before. Nevertheless, a clear indication of a disk-fed accretion would be very-low amplitude variations of the spectral line radial velocities modulated over the spin frequency (\citealt{Norton_etal_1996,Ferrario_etal_1999}; see Section~\ref{velo_sec} for a discussion of the spectral line velocities). It is worth noting that an accretion disk might reform over a few orbital cycles (private communication with B. Warner).

This is consistent with the flux derived from the X-ray grating data at day 340 (see Section~\ref{Chandra_sec}). As mentioned previously in this section, there is a possibility that a portion of the WD burning surface is hidden. This might be due to a reformed accretion disk and an accretion curtain around day 340 or due to localized H burning near the poles, as described previously (or both).

The emergence of narrow and moving \eal{He}{II} lines around 250 days post-eruption, is another indication that accretion has resumed, as these lines are possibly originating from an irradiated accretion stream/curtain (see, e.g., \citealt{Warner_1995,Potter_etal_2004,Zemko_etal_2016}) or from a reformed accretion disk (see, e.g., \citealt{Mason_etal_2012,Mason_Walter_2014,Munari_etal_2014}). In addition, at day $\sim$ 304, Balmer emission components with a FWHM$\sim$ 500 km\,s$^{-1}$ emerge on top of the broad nebular lines, showing changes in radial velocity (see Section~\ref{Balmer_lines}). These lines might also originate from a reformed accretion disk.
Few classical or recurrent novae have shown such narrow moving \eal{He}{II} and Balmer lines after the eruption when the ejecta become optically thin. \citet{Walter_Battisti_2011} suggested that this can be a sign of either the survival of the accretion disk or accretion restoration soon after the eruption.

We conclude that there is a possibility of accretion restoration around 168 days post-eruption causing localized H burning and/or temperature inhomogeneity on the WD surface and leading to the observed spin-modulated signal. The narrow lines observed in the optical spectra, their low velocity amplitude, and the optical/X-ray power spectra dominated by the $P_{\mathrm{spin}}$ frequency point towards resumption of accretion and the possibility of a reformed accretion disk. Although it is possible that the accretion has resumed earlier than 168 days after eruption in V407 Lup, it is not possible to constrain when exactly that happened due to the lack of observations between days $\sim$ 17 and $\sim$ 150 (solar constraints). It is also worth noting that features of accretion resumption might only be observable when the ejecta become optically thin.  

We showed in Section~\ref{timing_study} that the AAVSO optical and \textit{XMM-Newton} X-ray light-curves are out of phase when folded over the $P_{\mathrm{spin}}$. However, caution is required when interpreting this comparison, particularly because: (1) the error (0.5\,s) on the WD $P_{\mathrm{spin}}$ derived from the RGS data is too large for the ephemeris to be extrapolated to the date of the AAVSO observations ($>$ 150 days later) and for the phase to be known during these observations; (2) the \textit{XMM-Newton} X-ray light-curve was taken during the SSS phase while the AAVSO data were taken when the surface H burning had started fading considerably.

\subsection{The Intermediate Polar scenario}
\label{IP_sce_sec}
Several novae have been suggested to occur in IP systems including V533 Her, GK Per, DD Cir, V1425 Aql, V4743 Sgr, and Nova Scorpii 1437 AD (see e.g. \citealt{Warner_Woudt_2002,Bianchini_etal_2003,Woudt_Warner_2003,Woudt_Warner_2004,Zemko_etal_2016,Potter_Buckley_2018} and references therein). These systems either show spin-modulated optical and X-ray light-curves, strong \eal{He}{II} emission lines, or spin-modulated circular polarization. 

There is a striking similarity between the \eal{He}{II} lines of V407 Lup and the complex \eal{He}{II} lines seen in mCVs \citep{Rosen_etal_1987,Schwope_etal_1997}.
 \citet{Zemko_etal_2016} have found multiple components blue and red-shifted with the same velocity in the spectra of V4743 Sgr, which also have been found in GK Per, an old nova and IP, and were attributed to emission from the accretion curtains in the magnetosphere of the WD \citep{Bianchini_etal_2003}. 

The strong \eal{He}{II} 4686\,$\mathrm{\AA}$ line is usually seen in the spectra of IPs, although this line can also be prominent in other CV systems that host a very hot WD. We measure the EW of the narrow component in H$\beta$ in the last three HRS spectra, when the contribution from the broad emission (from the ejecta) is minimal. The ratio of EW(\eal{He}{II} 4686\,$\mathrm{\AA}$) and EW(H$\beta$) is $\sim$ 2. Such a value is in agreement with the ratio seen in magnetic CVs and IPs which is usually $>0.4$ \citep{Silber_1992}.  

\citet{Belle_etal_2003} detected very narrow lines of \eal{O}{V} and \eal{N}{V} in the UV spectra of the IP EX Hydra. These high ionization, very narrow (FWHM $\sim$ 60\,km\,s$^{-1}$) lines are very similar to the ``very narrow lines'' that we detect in the optical spectra of V407 Lup. The authors of this study suggest that such high ionization lines should originate from a very high temperature region which increases the probability that the lines are not contaminated by emission from the accretion disk. Therefore, due to their small widths, they suggested that the lines are formed near the surface of the WD. 

The main feature that supports the IP scenario in V407 Lup is the detection of the two periodicities in the X-ray, UV, and optical light-curves. These periodicities are very likely to represent the $P_{\mathrm{orb}}$ of the binary and the $P_{\mathrm{spin}}$ of a highly magnetized WD and therefore are compelling evidence that V407 Lup is an IP.

The detection of a hard X-ray component in the late ($\gtrsim$ day 300) \textit{Swift} X-ray data also favours the IP scenario. In addition, the luminosity of the hard X-ray component in the \textit{Swift} X-ray data taken around the end of January 2018 is in the range observed in IP systems (see Section~\ref{swift_xray}). However, this has still to be confirmed when a better distance estimate is available.  

All of these point towards the scenario of a nova occurring in an IP system and an evidence for accretion restoration.

\section{Summary and conclusions}
\label{Concl_sec}
We have presented multiwavelength observations of nova V407 Lup, which was discovered on HJD 2457655.5 (2016 September 24.0 UT; \citealt{ATel_9538}). The optical, UV, and X-ray data of this nova have led us to the following conclusions:
\begin{enumerate}[1.]
\item V407 Lup is a very fast nova. With $t_2 \leq$ 2.9\,d it  is one of the fastest known examples. This is also an indication that the eruption took place on a high mass WD ($\gtrsim$ 1.25 M$_{\odot}$).
\item Based on the evolution of the colours after the eruption, the system is very likely to contain a main-sequence companion.
\item Timing analysis of the optical, UV, and X-ray light-curves shows two periodicities of 3.57\,h and 565\,s. These are interpreted as the $P_{\mathrm{orb}}$ of the binary and the $P_{\mathrm{spin}}$ of the WD, respectively, suggesting that the system is an IP. 
\item The presence of modulation at the $P_{\mathrm{spin}}$ of the WD in the X-ray and optical data from 168 days post-eruption suggests that an accretion towards the magnetic poles has resumed leading to localized burning near the polar regions of the WD surface. 
\item The high resolution optical spectra exhibit very narrow and moving lines as early as 164 days post-eruption. Such moving, high-excitation lines are most likely associated with very hot regions within the inner binary system.  
\item At day 250, relatively strong, narrow, and moving lines of \eal{He}{II} emerge in the optical spectra. Their profiles are complex, similar to those observed in mCVs. 285 days post eruption, moderately narrow and moving Balmer emission components also emerge. All of these lines are possibly originating from accretion regions (accretion stream, disk, and curtain) within the inner binary system.
\item Based on the peak temperature of the SSS emission ($\sim$80$\pm$10\,eV) and the time taken for the optical brightness to fade by 2 magnitudes ($t_2 \leq$ 2.9\,d), we conclude that the mass of the WD is in the range of 1.1\,--\,1.3\,M$_{\odot}$
\item We constrain the turn-off time of the SSS to be around 300 days post-eruption, later than expected for a massive WD. The optical and X-ray data show evidence of some form of accretion restoration while the SSS is still on. The accretion of fresh material onto the WD surface may have been responsible for extending the duration of the SSS.\\

As the system is evolving back to quiescence, further observations are required to constrain the following:
(i) IPs might show optical circular polarization originating from cyclotron radiation (see, e.g., \citealt{Buckley_etal_1995,Buckley_etal_1997,Potter_Buckley_2018}), therefore polarimetric observations are encouraged to further investigate the IP identity of V407 Lup; (ii) orbital-period-resolved and spin-period-resolved spectroscopy are both needed to derive any periodicity from the optical emission lines and possibly to reveal the morphology of the system via Doppler tomography; (iii) Future Gaia data releases might provide an accurate distance estimate allowing us to better constrain the properties of the system, such as the absolute maximum magnitude, luminosities, and evolutionary state of the companion. 

\end{enumerate} 

\section*{Acknowledgments}
A part of this work is based on observations made with the Southern African Large
Telescope (SALT), under the Large science Programme on transient 2016-2-LSP-001. EA, DB, PAW, SM, and BM gratefully acknowledge the receipt of research grants from the National Research Foundation (NRF) of South Africa.\\
AK acknowledges the National Research Foundation of South Africa and the Russian Science Foundation (project no.14-50-00043).\\
MO had support from a NASA-Smithsonian grant for Director Discretionary Time \textit{Chandra} observations\\
APB, KLP, NPMK, and JPO acknowledge support from the UK Space Agency.\\
JS acknowledges support from the Packard Foundation. This paper is partially based on observations obtained at the Southern Astrophysical Research (SOAR) telescope, which is a joint project of the Minist\'{e}rio da Ci\^{e}ncia, Tecnologia, Inova\c{c}\~{a}os e Comunica\c{c}\~{a}oes (MCTIC) do Brasil, the U.S. National Optical Astronomy Observatory (NOAO), the University of North Carolina at Chapel Hill (UNC), and Michigan State University (MSU).\\
MJD acknowledges the partial support from the UK Science \& Technology Facilities Council (STFC).\\
SS gratefully acknowledges partial support from NASA grants to ASU.\\
We thank B. Warner and R. E. Williams, and L. Izzo for private discussion.\\
We acknowledge the use of observations from the AAVSO International Database and we thank the observers who obtained these observations.\\
We thank an anonymous referee for useful comments.

\bibliography{biblio}

\appendix

\section{Distance dependent parameters}
\label{app_dist}

The luminosity and effective radius at a certain distance $d$ (in kpc) can be converted using:
$$L(d) = L(10_{\mathrm{kpc}}) \times 0.01\,d^2,$$  $$R(d) = R(10_{\mathrm{kpc}}) \times 0.1\,d,$$ where $L(10_{kpc})$ and $R(10_{kpc})$ are the luminosity and effective radius derived assuming a distance $d$ = 10\,kpc.

Table~\ref{table:diff_dist_RGS} presents the luminosities and effective radii derived from the modelling of the \textit{XMM-Newton} RGS spectrum for different distance assumptions. Table~\ref{table:diff_dist_LETG} presents the luminosities derived from the modelling of the \textit{Chandra} LETG spectrum for different distance assumptions.

\begin{table}
\caption{The luminosities and effective radii derived from the fitting of the blackbody model (BB), 1 component atmosphere model (1-T), and 2 components atmosphere model (2-T) to the \textit{XMM-Newton} RGS spectrum for different distance assumptions.}
\begin{center}
\begin{tabular}{rrrr}
\hline
\centering
$d$ (kpc) & 3 & 5 & 7\\ 
\hline
$L$(BB) (erg\,s$^{-1}$)  & 2.4 $\times$ 10$^{38}$ &  6.7 $\times$ 10$^{38}$ &  1.3$\times$ 10$^{39}$\\ 
$R_{\mathrm{eff}}$(BB) (km) & 1.5 $\times$ 10$^{4}$ & 2.6 $\times$ 10$^{4}$ & 3.6 $\times$ 10$^{4}$ \\ 
\hline
$L$(1-T) (erg\,s$^{-1}$) & 2.9 $\times$ 10$^{37}$ & 8.0 $\times$ 10$^{37}$ & 1.5 $\times$ 10$^{38}$ \\ 
$R_{\mathrm{eff}}$(1-T) (km) & 2.1 $\times$ 10$^{3}$  & 3.5 $\times$ 10$^{3}$ & 4.9 $\times$ 10$^{3}$ \\
\hline
$L$(2-T) (erg\,s$^{-1}$)  & 9.9 $\times$ 10$^{36}$ &  2.7 $\times$ 10$^{37}$& 5.4 $\times$ 10$^{37}$\\ 
$R_1$(2-T) (km) & 1.8 $\times$ 10$^{3}$ & 2.9 $\times$ 10$^{3}$ & 4.1 $\times$ 10$^{3}$\\ 
$R_2$(2-T) (km) & 3.6 $\times$ 10$^{3}$ & 6.0 $\times$ 10$^{3}$ & 8.4 $\times$ 10$^{3}$ \\  
\hline
\end{tabular}
\end{center}
\label{table:diff_dist_RGS}
\end{table}

\begin{table}
\caption{The luminosities derived from the fitting of  models (1) and (2) to the \textit{Chandra} LETG spectrum for different distance assumptions.}
\begin{center}
\begin{tabular}{rrrr}
\hline
\centering
$d$ (kpc) & 3 & 5 & 7\\ 
\hline
$L$(1) (erg\,s$^{-1}$)  &  5.1 $\times$ 10$^{35}$ & 1.4 $\times$ 10$^{36}$ & 2.8 $\times$ 10$^{36}$\\ 
\hline
$L$(2) (erg\,s$^{-1}$)  & 7.0 $\times$ 10$^{36}$ &  1.9 $\times$ 10$^{37}$ & 3.8 $\times$ 10$^{37}$ \\ 
\hline
\end{tabular}
\end{center}
\label{table:diff_dist_LETG}
\end{table}

\section{Complementary plots}
\label{appB}
In this Appendix we present complementary plots.

\begin{figure*}
\includegraphics[width=120mm]{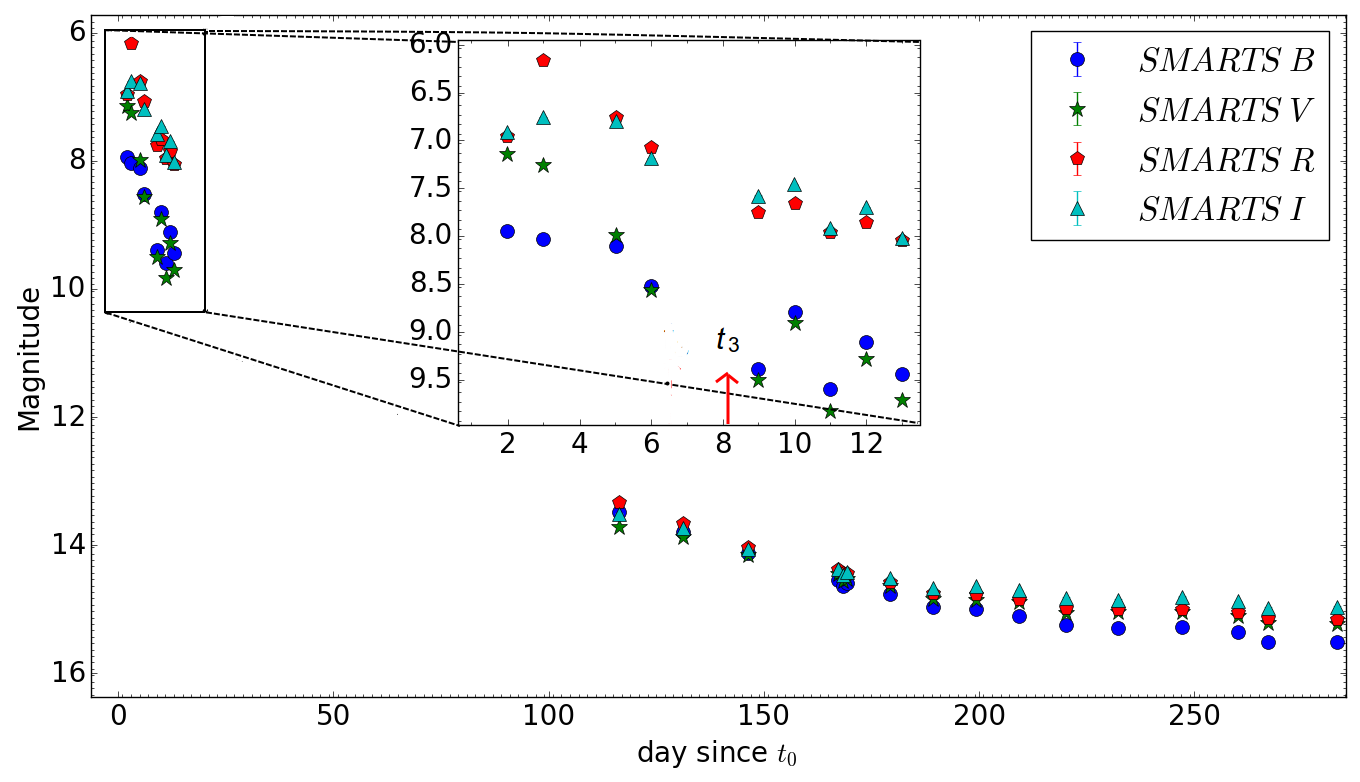}
\includegraphics[width=120mm]{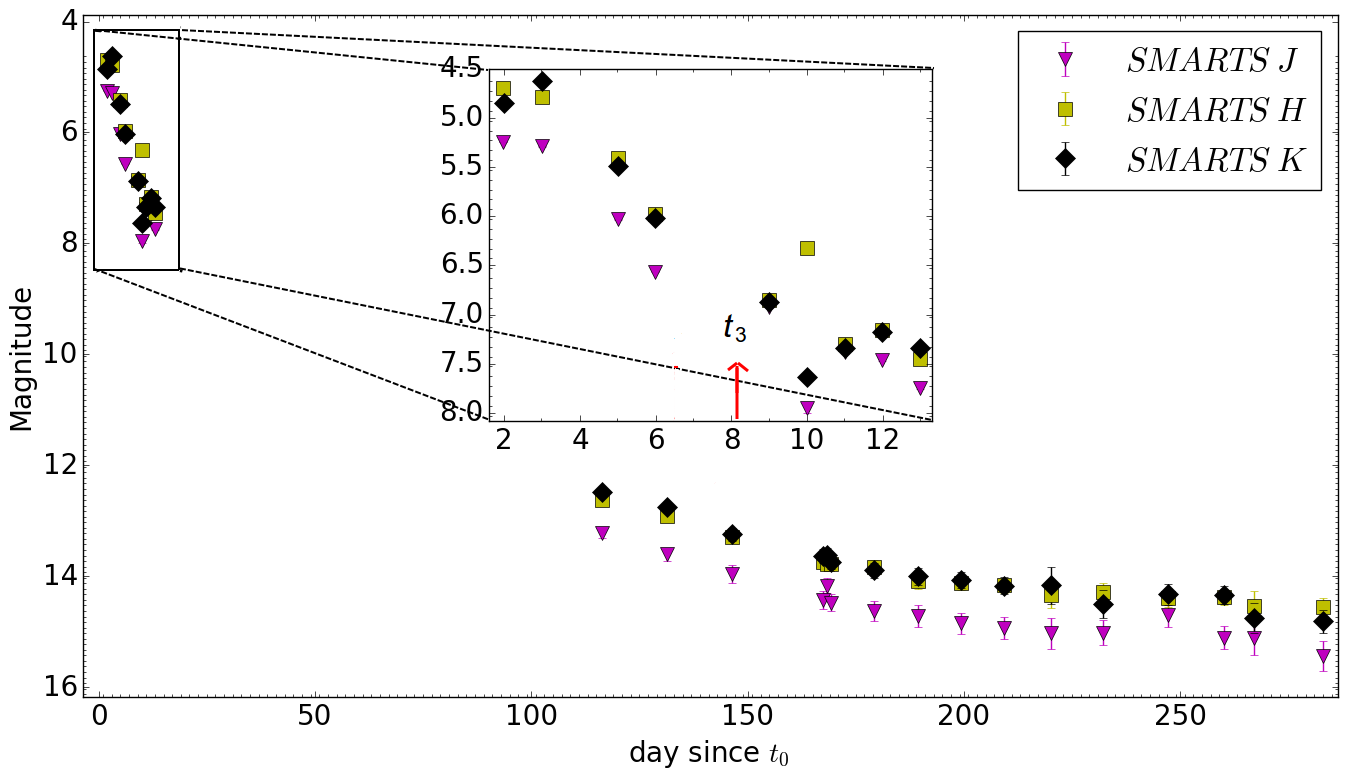}
\caption{The \textit{BVRI} (top) and \textit{JHK} (bottom) photometry from SMARTS as a function of days since eruption, colour and symbol coded as indicated in the legend. We present a zoomed sub-plot to show in more detail the evolution of the light-curve during the early decline. Day to day variability appear in the light-curve after $t_3$ (marked by a red arrow).}
\label{Fig:BVRI_JHK_LC}
\end{figure*}

\begin{figure*}
\begin{center}
  \includegraphics[width=130mm]{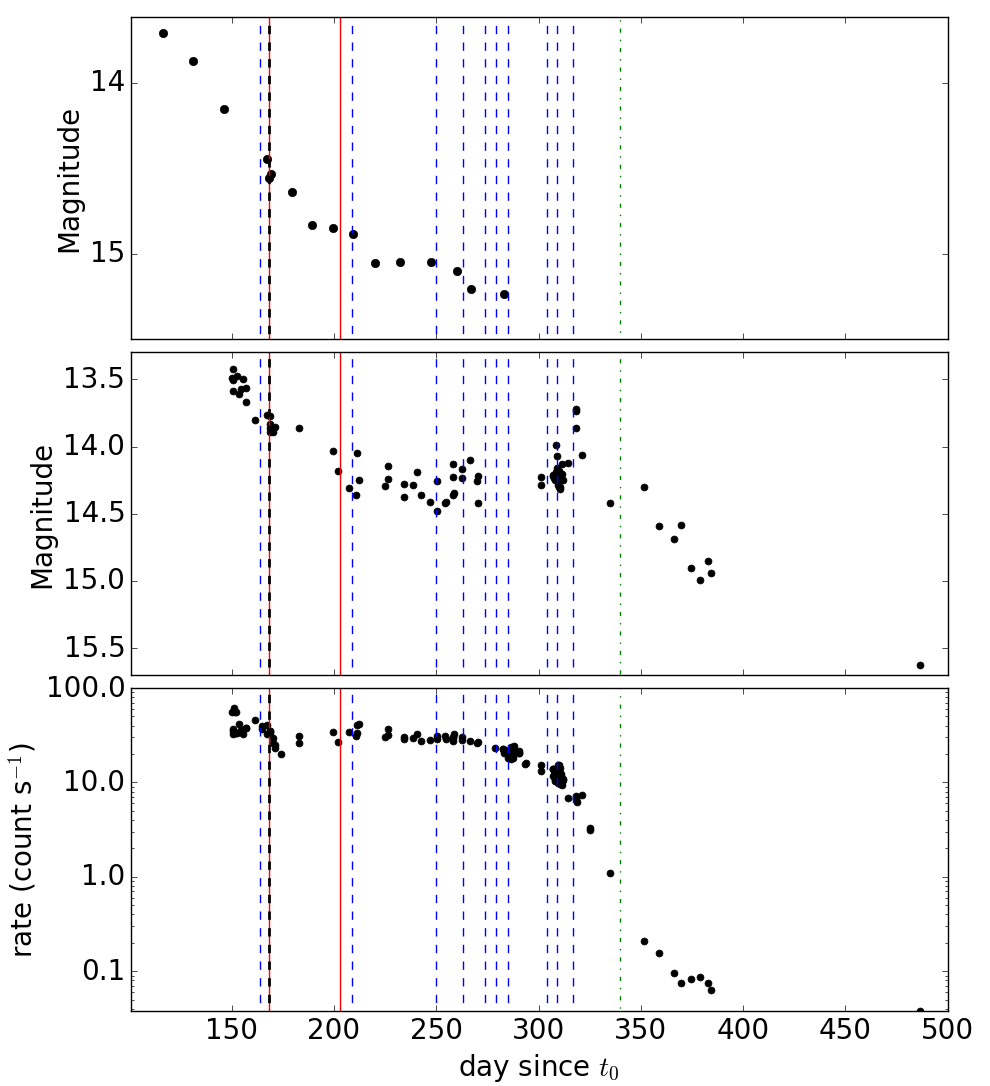}
\caption{A direct comparison between the SMARTS $V$-band (top), \textit{Swift} UVOT \textit{uvw2} (middle), and \textit{Swift} X-ray light-curves (bottom), from day 100. The blue dashed lines represent the dates of the SALT HRS observations respectively. The black dashed line represents the date of the \textit{XMM-Newton} observations. The red solid lines represent the median dates of the first four and second four UV grism spectra. The green dotted line represents the date of the \textit{Chandra} observation.}
\label{Fig:multi_LC}
\end{center}
\end{figure*}

\begin{figure*}
\centering
\includegraphics[width = 120mm]{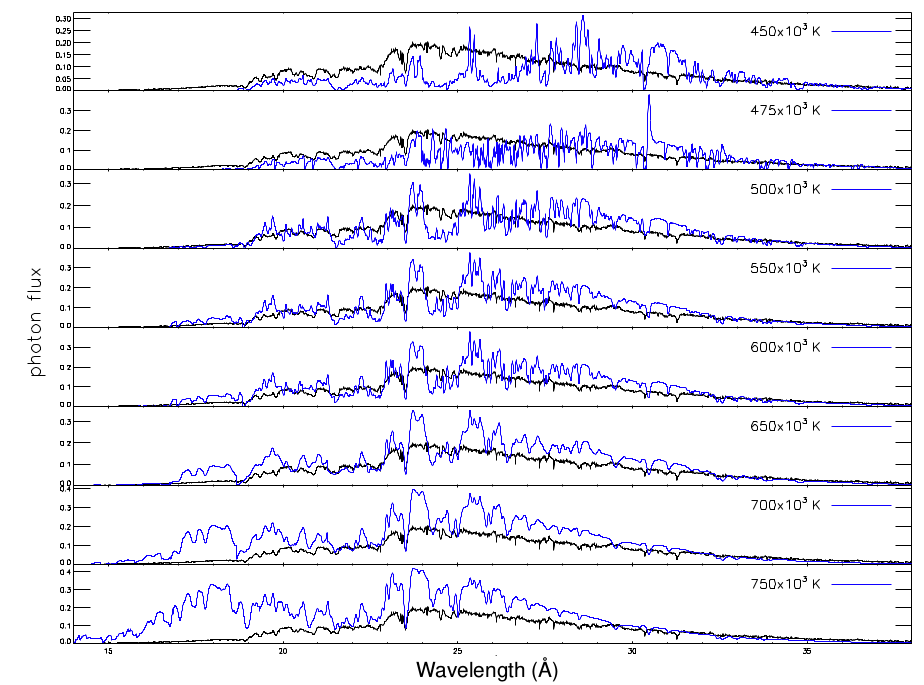}
\includegraphics[width = 120mm]{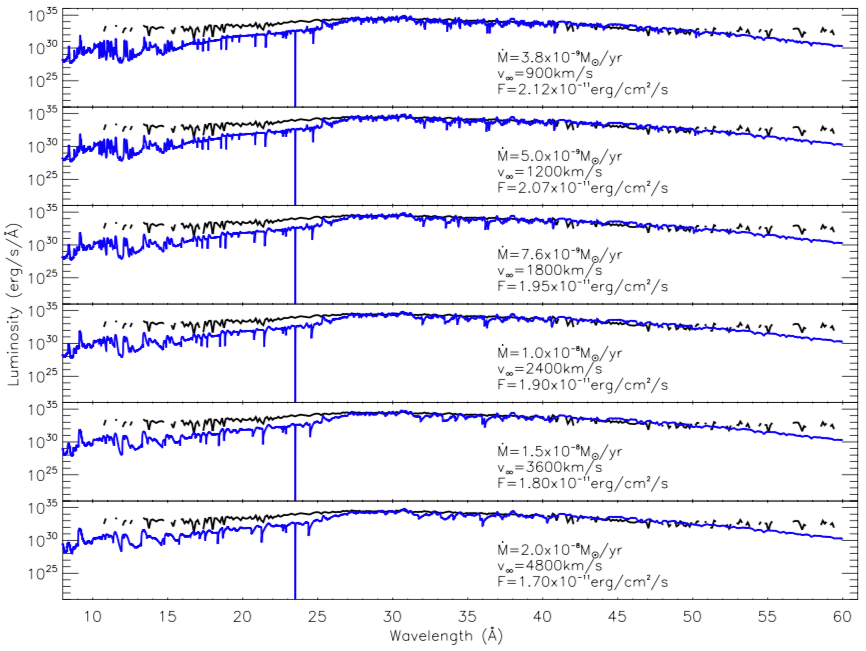}
\caption{\textit{Top:} A sample of the wind-type atmosphere models \citep{Van_Rossum_2012} fitting to the \textit{XMM-Newton} RGS spectrum. The $T_{\mathrm{eff}}$ characterizing each model is indicated on the plot. \textit{Bottom:} best fit wind-type atmosphere model to the \textit{Chandra} LETG spectrum for various wind asymptotic velocities $v_{\infty}$ and mass-loss rates $\dot{M}$ as indicated on the plot. In all panels, the models in blue are compared to the observations in black.} 
\label{Fig:WT}
\end{figure*}

\section{Tables}
\label{appC}
In this Appendix we present the tables. 

\begin{table}
\caption{$V$ and \textit{Vis} measurements around maximum from different detectors.}
\begin{center}
\begin{tabular}{rrrrrr}
\hline
\centering
HJD & $(t-t_0)$ & Magnitude & Uncertainty & Band & Source \\ 
& (days) & & & & \\
 \hline
2457655.50 & 0.0 & 9.10 & 0.010 & $V$ & ATel 9538\\ 
2457655.93 & 0.43 & 6.82 & 0.075 & $V$ & AAVSO\\
2457656.24 & 0.74 & 6.43 & 0.050 & $V$ & ATel 9550\\
2457656.57 & 1.02 & 6.32 &    ---     & \textit{Vis} & AAVSO\\
2457656.90 & 1.10 & 5.60 &    ---     & \textit{Vis} & AAVSO\\ 
2457657.22 & 1.72 & 6.33 & 0.060 & $V$ & ATel 9550\\
2457657.45 & 1.95 & 6.50 &    ---     & \textit{Vis} & AAVSO\\
2457657.46 & 1.96 & 6.84 &    ---     & \textit{Vis} & AAVSO\\
2457657.48 & 1.97 & 6.80 &    ---     & \textit{Vis} & AAVSO\\
2457657.48 & 1.97 & 7.14 & 0.001 & $V$ & SMARTS\\
2457657.51 & 2.01 & 6.70 &    ---     & \textit{Vis} & AAVSO\\
2457657.89 & 2.39 & 6.62 & 0.050 & \textit{Vis} & AAVSO\\
2457657.92 & 2.42 & 6.70 &    ---     & \textit{Vis} & AAVSO\\
2457658.41 & 2.91 & 7.10 &    ---     & \textit{Vis} & AAVSO\\
2457658.46 & 2.96 & 6.92 &    ---     & \textit{Vis} & AAVSO\\
2457658.48 & 2.98 & 6.90 &    ---     & \textit{Vis} & AAVSO\\
2457658.48 & 2.98 & 7.26 & 0.001 & $V$ & SMARTS\\
2457658.99 & 3.49 & 7.20 &    ---     & \textit{Vis} & AAVSO\\
2457659.41 & 3.91 & 7.50 &    ---     & \textit{Vis} & AAVSO\\
2457659.42 & 3.92 & 7.50 &    ---     & \textit{Vis} & AAVSO\\
2457659.44 & 3.94 & 7.60 &    ---     & \textit{Vis} & AAVSO\\
2457659.46 & 3.96 & 7.54 &    ---     & \textit{Vis} & AAVSO\\
2457659.46 & 3.96 & 7.40 &    ---     & \textit{Vis} & AAVSO\\
2457659.98 & 4.48 & 7.85 &    ---     & \textit{Vis} & AAVSO\\
2457660.23 & 4.73 & 7.70 &    ---     & \textit{Vis} & AAVSO\\
2457660.41 & 4.91 & 8.20 &    ---     & \textit{Vis} & AAVSO\\
2457660.42 & 4.92 & 8.00 &    ---     & \textit{Vis} & AAVSO\\
2457660.43 & 4.93 & 8.10 &    ---     & \textit{Vis} & AAVSO\\
2457660.46 & 4.96 & 8.00 &    ---     & \textit{Vis} & AAVSO\\
2457660.46 & 4.96 & 7.70 &    ---     & \textit{Vis} & AAVSO\\
2457660.48 & 4.98 & 7.88 & 0.080 & $V$ & ATel 9564\\
2457660.50 & 5.00 & 7.98 & 0.001 & $V$ & SMARTS\\
2457660.92 & 5.42 & 8.70 &    ---     & \textit{Vis} & AAVSO\\
2457661.02 & 5.52 & 8.10 & 0.012 & $V$ & AAVSO\\
2457661.20 & 5.70 & 8.30 &    ---     & \textit{Vis} & AAVSO\\
2457661.40 & 2.90 & 8.60 &    ---     & \textit{Vis} & AAVSO\\
2457661.41 & 5.91 & 8.50 &    ---     & \textit{Vis} & AAVSO\\
2457661.48 & 5.98 & 8.56 & 0.001 & $V$ & SMARTS\\
 \hline
\end{tabular}
\end{center}
\label{table:photo_max}
\end{table}

\begin{table*}
\centering
\caption{Spectral line identification of the optical SALT and SOAR spectra. We list the EW, FWHM and integrated flux of the lines for which an estimate was possible.}
\begin{tabular}{rrrrrr}
\hline
Line & $\lambda_0$  & EW ($\lambda$) & FWHM  & Flux & ($t-t_0$)\\ 
 & \multicolumn{2}{c}{($\mathrm{\AA}$)} & (km\,s$^{-1}$) & $10^{-13} $erg cm$^{-2}$s$^{-1}$\\
\hline
\feal{Ne}{V} & 3346 & $-55 \pm 5$ & 2400$\pm$100 & 3.0 $\pm 0.2$ & 269\\
\feal{Ne}{V} & 3426 & $-160 \pm 10$ & 2400$\pm$100 & 8.0 $\pm 0.5$ & 269\\
\eal{O}{VI} & 3811& $-1.8 \pm 0.2$ & 400$\pm$ 50 & 0.08 $\pm 0.02$ & 269\\
\feal{Ne}{III} & 3869 & $-75 \pm 10$ & 2400$\pm$100 & 2.7 $\pm 0.2$ & 269\\
\feal{Ne}{III} & 3968 & $-35 \pm 5$ & 2400$\pm$100 & 1.2 $\pm 0.2$ & 269\\
H$\delta$ & 4102 & $-10 \pm 2$ & 2600$\pm$100 & 0.34 $\pm 0.05$ & 269\\
\feal{O}{III} & 4363 & $-450 \pm 50$ & 2400$\pm$100 & -- & 164\\
H$\gamma$ & 4341 & -- & -- & -- & 164\\
\eal{He}{II} & 4542 & $-0.8 \pm 0.2$ & $270 \pm 50$ & -- & 263\\
\eal{He}{II} & 4686 & $-8.7 \pm 1.0$ & 435 $\pm$ 50 &  -- & 250\\
\eal{He}{I} & 4713 & -- & -- & -- & 164\\
\feal{Ne}{VI} & 4721 & -- & -- & -- & 164\\
H$\beta$ & 4861 & $-$100$\pm$20 &2850$\pm$100 & -- & 164\\
\feal{O}{III} & 4959 & -- & -- & -- & 164\\
\feal{O}{III} & 5007 & -- & -- & -- & 164\\
\eal{O}{VI} & 5290 & $0.7 \pm 0.2$ & 120$\pm 30$ & -- & 164\\
\eal{He}{II} & 5412 & $-3.8 \pm 0.5$  & $460 \pm 50$ & -- & 250 \\
\feal{O}{I} & 5577 & -- & -- & -- & 164\\
\feal{Fe}{VI} & 5631 & -- & -- & -- & 164\\
\feal{Fe}{VI} & 5677 & -- & -- & -- & 164\\
\feal{N}{II} & 5755 & $-80 \pm 10$ & 2800$\pm$100 & -- & 164\\
\eal{He}{I} & 5876 & $-20 \pm 5$ & 2400$\pm$100 & -- & 164\\
\feal{Fe}{VII} & 6087 &  $-35 \pm 5$ & 3100$\pm$100 & -- & 164\\
\eal{O}{VI} & 6200 & -- & -- & -- & 164\\
\feal{O}{I} & 6300 & $-30 \pm 5$ & -- & -- & 164\\
\feal{O}{I} & 6364 & $-25 \pm 5$ & -- & -- & 164\\
H$\alpha$ & 6563 & $-$575 $\pm$2 0 & 2900$\pm$100 & -- & 164\\
\eal{He}{I} & 7065 & $-25 \pm 5$ & 2800$\pm$100 & -- & 164\\
\feal{O}{II} & 7320/30 & -- & -- & -- & 164\\
\eal{Ne}{IV} & 7716 & $-1.1 \pm 0.2$ & 150 $\pm$ 30 & -- & 164\\
\feal{Fe}{X} & 7892 & $-30 \pm 5$ & 2800$\pm$100 & -- & 164\\
\eal{He}{II} & 8237 & -- & -- & -- & 164\\
\hline
\end{tabular}
\label{line_det}
\end{table*}

\begin{table}
\caption{The FWHM of the broad H$\alpha$ and H$\beta$. In the last three spectra, the narrow features dominate the broad nebular emission, hence we exclude them from the fitting.}
\begin{center}
\begin{tabular}{rrrr}
\hline
\centering
HJD & $t-t_0$ & FWHM (H${\alpha}$)  & FWHM (H${\beta}$) \\ 
 & (days) &\multicolumn{2}{c}{$\pm$ 100 (km\,s$^{-1}$)} \\
\hline
2457819.49 & 164 & 2900 & 2850\\
2457864.38 & 209 & 2800 & 2800\\
2457905.52 & 250 & 2700 & 2600\\
2457918.47 & 263 & 2650 & 2400\\
2457934.43 & 279 & 2650 & 2300\\
2457940.41 & 285 & 2200 & 2400\\
2457959.36 & 304 &     --    &     --   \\
2457964.35 & 309 &     --    &     --   \\
2457972.32 & 317 &     --    &     --   \\
 \hline
\end{tabular}
\end{center}
\label{table:BalmerFWHM}
\end{table}

\begin{table}
\caption{The radial velocity of the ``moderately narrow'' H$\beta$ and H$\alpha$ features. These measurements were done by cursor position, when possible.}
\begin{center}
\begin{tabular}{rrrr}
\hline
\centering
HJD & $ t-t_0$ & $V_{\mathrm{rad}}$(H$\beta$) & $V_{\mathrm{rad}}$ (H$\alpha$)\\ 
 & (days) & \multicolumn{2}{c}{$\pm$ 50 (km\,s$^{-1}$)} \\
\hline
2457905.52 & 250 & $-$210 & $-$230\\
2457918.47 & 263 & $-$215 & $-$235\\
2457934.43 & 279 & -- & $-$220\\
2457940.41 & 285 & $-$20 & $-$100\\
2457959.36 & 304 & $-$20 & $-$20\\
2457964.35 & 309 & -- & --\\
2457972.32 & 317 & 10 & $-$20 \\
 \hline
\end{tabular}
\end{center}
\label{table:BalmerMNL}
\end{table}

\begin{table}
\caption{The FWHM of \feal{O}{III} 4363\,$\mathrm{\AA}$.}
\begin{center}
\begin{tabular}{rrr}
\hline
\centering
HJD & $t-t_0$ & FWHM \\ 
 & (days) & $\pm$ 100 (km\,s$^{-1}$) \\
\hline
2457819.49 & 164 & 2950 \\
2457864.38 & 209 & 2700 \\
2457905.52 & 250 & 2400 \\
2457918.47 & 263 & 2200 \\
2457934.43 & 279 & 2300 \\
2457940.41 & 285 & 2200 \\
2457959.36 & 304 & 2100 \\
2457964.35 & 309 & 1800 \\
2457972.32 & 317 & 1600 \\
 \hline
\end{tabular}
\end{center}
\label{table:OIIIFWHM}
\end{table}

\begin{table*}
\caption{The radial velocity ($V_{\mathrm{rad}}$), FWHM, and EW derived from a single component fitting of the \eal{He}{II} lines.}
\begin{center}
\begin{tabular}{rrrrrr}
\hline
\centering
HJD & $t = t-t_0$ & Phase & $V_{\mathrm{rad}}$(\eal{He}{II} 4686) & $V_{\mathrm{rad}}$(\eal{He}{II} 5412) & $V_{\mathrm{rad}}$(\eal{He}{II} 4542)\\ 
 & (days) & & \multicolumn{3}{c}{$\pm$ 30 (km\,s$^{-1}$)}\\
\hline
2457905.52 & 250 & 0.088 & $-$210 & $-$205 & $-$240\\
2457918.47 & 263 & 0.072 & $-$210 & $-$177 & $-$210\\
2457924.51 & 270 & 0.583 & $-$175 & -- & -- \\ 
2457934.43 & 279 & 0.269 & $-$155 & $-$140 & $-$135\\
2457940.41 & 285 & 0.464 & $-$25  & $-$20 &  $-$40\\
2457959.36 & 304 & 0.766 & $-$70  & $-$100 & $-$100 \\
2457964.35 & 309 & 0.229 & $-$90  & $-$100  & $-$105\\
2457972.32 & 317 & 0.762 &  0 & $-$20 & $-$60\\
2458139.84 & 484 & 0.033 &  $-$90 & -- & --\\
2458140.85 & 485 & 0.806 &  $-$95 & -- & --\\
\hline
HJD & $t-t_0$ & Phase & FWHM (\eal{He}{II} 4686) & FWHM (\eal{He}{II} 5412) & FWHM (\eal{He}{II} 4542)\\ 
 & (days) & & \multicolumn{3}{c}{$\pm$ 50 (km\,s$^{-1}$)}\\
\hline
2457905.52 & 250 & 0.088 & 435 & 460 & --\\
2457918.47 & 263 & 0.072 & 470 & 550 & 270\\
2457924.51 & 270 & 0.583 & 550 & -- & -- \\ 
2457934.43 & 279 & 0.269 & 470 & 530 & 260 \\
2457940.41 & 285 & 0.464 & 400 & 420 & 360 \\
2457959.36 & 304 & 0.766 & 510 & 420 & 380 \\
2457964.35 & 309 & 0.229 & 430 & 500  & 380\\
2457972.32 & 317 & 0.762 & 410 & 420 & 380\\
2458139.84 & 484 & 0.033 & 570 & -- & --\\
2458140.85 & 485 & 0.806 & 465 & -- & --\\
\hline
HJD & $t-t_0$ & Phase & EW (\eal{He}{II} 4686) & EW (\eal{He}{II} 5412) & EW (\eal{He}{II} 4542)\\ 
 & (days) & & $\pm$ 1 ($\mathrm{\AA}$) & $\pm$ 0.5 ($\mathrm{\AA}$) & $\pm$ 0.2 ($\mathrm{\AA}$)\\
\hline
2457905.52 & 250 & 0.088 & $-$8.7 & $-$3.8 & $-$0.8\\
2457918.47 & 263 & 0.072 & $-$8.8 & $-$4.2 & $-$0.8\\
2457924.51 & 270 & 0.583 & $-$9.0 & -- & -- \\ 
2457934.43 & 279 & 0.269 & $-$11.5 & $-$3.0 & $-$1.1\\
2457940.41 & 285 & 0.464 & $-$19.0 & $-$4.6 & $-$1.5\\
2457959.36 & 304 & 0.766 & $-$18.0 & $-$5.3 & $-$2.0 \\
2457964.35 & 309 & 0.229 & $-$14.1 & $-$5.0  & $-$1.9\\
2457972.32 & 317 & 0.762 &  $-$9.6 & $-$3.0 & $-$1.6\\
2458139.84 & 484 & 0.033 &  $-$90 & -- & --\\
2458140.85 & 485 & 0.806 &  $-$95 & -- & --\\
\hline
\end{tabular}
\end{center}
\label{table:HeIIsingle}
\end{table*}

\begin{table}
\caption{The radial velocity ($V_{\mathrm{rad}}$) and FWHM of the medium width component (MWC) of the \eal{He}{II} 4686\,$\mathrm{\AA}$ line.}
\begin{center}
\begin{tabular}{rrrrr}
\hline
\centering
HJD & $ t-t_0$ & Phase & $V_{\mathrm{rad}}$ & FWHM \\ 
 & (days) & &  \multicolumn{2}{c}{$\pm$ 30 (km\,s$^{-1}$)}\\
\hline
2457819.49 & 164 & 0.142 & -- & --  \\
2457864.38 & 209 & 0.708 &  --  & -- \\
2457905.52 & 250 & 0.043 & -200 & 240\\
2457918.47 & 263 & 0.029 & -215 & 260\\
2457934.43 & 279 & 0.234 & -30 &340\\
2457940.41 & 285 & 0.418 & -125  & 230\\
2457959.36 & 304 & 0.740 & -170  & 265\\
2457964.35 & 309 & 0.205 & +15  & 415 \\
2457972.32 & 317 & 0.741 &  -165 & 300 \\
 \hline
\end{tabular}
\end{center}
\label{table:HeIIMNL}
\end{table}

\begin{table}
\caption{The radial velocity ($V_{\mathrm{rad}}$) and FWHM of the small width component (SWC) of the \eal{He}{II} 4686\,$\mathrm{\AA}$ line.}
\begin{center}
\begin{tabular}{rrrrr}
\hline
\centering
HJD & $ t-t_0$ & Orbital phase & $V_{\mathrm{rad}}$ & FWHM\\ 
 & (days) & & \multicolumn{2}{c}{$\pm$ 30 (km\,s$^{-1}$)}\\
\hline
2457819.49 & 164 & 0.151 & -- & -- \\
2457864.38 & 209 & 0.753 &  --  & --\\
2457905.52 & 250 & 0.088 & -35 & 70 \\
2457918.47 & 263 & 0.072 & -5 & 65 \\
2457934.43 & 279 & 0.269 & -205 & 120 \\
2457940.41 & 285 & 0.464 & -25  & -- \\
2457959.36 & 304 & 0.766 & -10  & 106 \\
2457964.35 & 309 & 0.229 &  -165 & 196 \\
2457972.32 & 317 & 0.762 &  +40 & 200 \\
 \hline
\end{tabular}
\end{center}
\label{table:HeIIVNL}
\end{table}

\begin{table}
\caption{The full base width of the broad base component (BBC) of the \eal{He}{II} 4686\,$\mathrm{\AA}$ line.}
\begin{center}
\begin{tabular}{rrrrr}
\hline
\centering
HJD & $ t-t_0$ & Orbital phase & [$\lambda_i$-- $\lambda_f$] & $\Delta\lambda$\\ 
 & (days) & &  \multicolumn{2}{c}{$\pm$ 1.0 ($\mathrm{\AA}$)}\\
\hline
2457819.49 & 164 & 0.151 & -- & -- \\
2457864.38 & 209 & 0.753 & -- & --\\
2457905.52 & 250 & 0.088 & [4671-- 4691] & 20 \\
2457918.47 & 263 & 0.072 & [4671-- 4691] & 20 \\
2457934.43 & 279 & 0.269 & [4671-- 4691] & 20 \\
2457940.41 & 285 & 0.464 & [4677-- 4697] & 20 \\
2457959.36 & 304 & 0.766 & [4671-- 4693] & 22 \\
2457964.35 & 309 & 0.229 & [4670-- 4692] & 22 \\
2457972.32 & 317 & 0.762 & [4670-- 4693] & 23 \\
 \hline
\end{tabular}
\end{center}
\label{table:HeIIBBC}
\end{table}

\begin{table}
\caption{The radial velocity ($V_{\mathrm{rad}}$) of the ``moderately narrow'' \eal{O}{VI} 5290\,$\mathrm{\AA}$ and 6200\,$\mathrm{\AA}$ lines.}
\begin{center}
\begin{tabular}{rrrrr}
\hline
\centering
HJD & $ t-t_0$ & Phase & $V_{\mathrm{rad}}$(5290) & $V_{\mathrm{rad}}$(6200)\\ 
 & (days) & & \multicolumn{2}{c}{$\pm$ 30 (km\,s$^{-1}$)}\\
\hline
2457819.49 & 164 & 0.151 & -- & -- \\
2457864.38 & 209 & 0.753 &  5 & --\\
2457905.52 & 250 & 0.088 & -260 & -165 \\
2457918.47 & 263 & 0.072 & -265 & -- \\
2457934.43 & 279 & 0.269 & -115 & -100 \\
2457940.41 & 285 & 0.464 & -75 & -65 \\
2457959.36 & 304 & 0.766 & -255  & -185 \\
2457964.35 & 309 & 0.229 &  -80 & -80 \\
2457972.32 & 317 & 0.762 &  -- & -- \\
\hline
\end{tabular}
\end{center}
\label{table:NOVI}
\end{table}

\begin{table}
\caption{The radial velocity, FWHM and EW of the ``very narrow lines'': \eal{Ne}{IV} 4498.4\,$\mathrm{\AA}$ (top), \eal{O}{VI} 5290\,$\mathrm{\AA}$ (middle), and \eal{Ne}{IV} 7715.9\,$\mathrm{\AA}$ (bottom). The last two measurements (days 304 and 309) the lines merge with neighbouring ``moderately narrow lines'' and therefore the FWHM increases considerably.}
\begin{center}
\begin{tabular}{rrrrrr}
\hline
\centering
HJD & $ t-t_0$ & Phase &  $V_{\mathrm{rad}}$ & FWHM & EW\\
 & (days) & & \multicolumn{2}{c}{$\pm$ 30 (km\,s$^{-1}$)} & $\pm$ 0.2 ($\mathrm{\AA}$)\\
\hline
2457819.49 & 164 & 0.151 & $-5$ & 120 & $-0.4$\\
2457864.38 & 209 & 0.753 & $-130$ & -- & --\\
2457905.52 & 250 & 0.088 & 15 & 90 & $-0.4$\\
2457918.47 & 263 & 0.072 & 10 & -- & -- \\
2457934.43 & 279 & 0.269 & $-50$ & 140 & $-0.8$\\
2457940.41 & 285 & 0.464 & $-150$ & 85 & $-0.3$\\
2457959.36 & 304 & 0.766 & $-90$ & 230 & $-0.3$\\
2457964.35 & 309 & 0.229 & $-50$ & 265 & $-0.7$\\
2457972.32 & 317 & 0.762 &  -- & -- & --\\
\hline
\hline
2457819.49 & 164 & 0.151 & 20 & 120 & $-0.7$\\
2457864.38 & 209 & 0.753 & $-125$ & -- & --\\
2457905.52 & 250 & 0.088 & 25 & 110 & $-0.9$\\
2457918.47 & 263 & 0.072 & 25 & 110 & $-0.7$\\
2457934.43 & 279 & 0.269 & $-35$ & 110 & $-0.9$\\
2457940.41 & 285 & 0.464 & $-145$ & 60 & $-0.4$\\
2457959.36 & 304 & 0.766 & $-75$ & 340 & $-2.5$\\
2457964.35 & 309 & 0.229 & $-30$ & 280 & $-1.5$\\
2457972.32 & 317 & 0.762 &  -- & -- & --\\
\hline
\hline
2457819.49 & 164 & 0.151 & 20 & 150 & $-1.1$\\
2457864.38 & 209 & 0.753 & $-120$ & 150 & $-1.0$\\
2457905.52 & 250 & 0.088 & 20 & 105 & $-1.0$\\
2457918.47 & 263 & 0.072 & 20 & 110 & $-0.9$\\
2457934.43 & 279 & 0.269 & $-35$ & 110 & $-1.5$\\
2457940.41 & 285 & 0.464 & $-145$ & 75 & $-1.1$\\
2457959.36 & 304 & 0.766 & $-80$ & 250 & $-2.3$\\
2457964.35 & 309 & 0.229 & $-40$ & 310 & $-3.5$\\
2457972.32 & 317 & 0.762 &  -- & -- & --\\
\end{tabular}
\end{center}
\label{table:4VNL}
\end{table}

\begin{table}
\caption{Spectral line identifications of the UV-optical grism spectra.}
\begin{tabular}{@{}lcrl}
\hline
spectrum   & line  & vacuum lab & notes\\
wavelength (a) &  ID & wavelength   &   \\
\hline
 1759 &  \hfeal{N}{III}  &1750  \\
 1855 &  \eal{Al}{III}  &1855,1863  \\
 2141 &  \hfeal{N}{II}   &2143  \\
 2332 & \feal{O}{III}  &2332  \\
 2518 & \eal{He}{II}   &2511 & blend? \\
 2635 &  unid    &     & \eal{Ne}{VI} 2626 ?\\
 2801 &  \eal{Mg}{II}   &2800  \\
 2978 &  \eal{Ne}{VIII}   & 2977 \\
 3135 &  \eal{O}{III}   &3133  \\
 3346 & \feal{Ne}{V}   &3346  \\
 3424 & \feal{Ne}{V}   &3426  \\
 3868 & \feal{Ne}{III} &3869  \\
 3970 & \feal{Ne}{III} &3968  \\
 4104 &  \eal{H}{I}  & 4103& weak  \\
 4363 & \feal{O}{III}  &4363  \\
 4523 &  \eal{N}{III}  &4517& weak          \\
 4714 &  unid    &    & broad feature \\
 4866 &  \eal{H} {I}     &4863  \\
 5010 & \feal{O}{III}  &4959,5007,5017 & blend  \\ 
\hline
\end{tabular}
(a)\,after shifting the spectra to best match the \feal{Ne}{V} doublet. \\
\label{uv_line_ids}
\end{table}

\begin{table*}
\caption{UV line measurements for the UV-optical grism spectra. The numbers between brackets are days after $t_0$.}
\begin{tabular}{@{}lcccccccccrl}
\hline
   &   & \multispan{7}{$10^{-10}$ Flux (erg\,cm$^{-2}$\,s$^{-1}$\,$\mathrm{\AA}^{-1}$)}\\
nominal & line   &   MJD &   MJD &   MJD &   MJD &   MJD &   MJD &   MJD &   MJD \\
wavelength& ID   & 57820 & 57822 & 57825 & 57826 & 57838 & 57854 & 57857 & 57866 \\
 & & [+165] & [+167] & [+170] & [+171] & [+183] & [+199] & [+202] & [+211]\\
\hline
 2800 &  Mg II   & 9.76& 14.9 & 18.4& 10.6&  7.7&  9.6&   7.6& 11.2 \\
 3133 &  O III   & 12.1&  9.9 &  7.0& 8.1 &  5.3&  5.4&   3.2&  8   \\
 3346 & [Ne V]   & 32.5&  28.9& 36.1& 34.0& 23.6& 16.1&  14.0& 15.5 \\
 3426 & [Ne V]   & 90.3&  76.8& 94.4& 83.5& 59.3& 41.1&  36.7& 36.0 \\
 3869 & [Ne III] & 27.9&  28.9& 27.9& 31.3& 19.7& 19.6&  16.3& 17   \\
 3968 & [Ne III] & 13.8&  12.6& 11.0& 14.6& 11.5& 10.6&   8.3& 12.4 \\
 4363 & [O III]  & 32.7&  28.1& 31.8& 28.5& 20.6& 22.0&  18.4& 19.0 \\
\hline
   &   & \multispan{7}{FWZI (\AA)}\\
\hline
 2800 &  Mg II   & 70.9& 90.2& 113 & 55.2& 56.6&  70.3& 66.7&  87  \\
 3133 &  O III   & 90.0& 77.3& 57.0& 57.5& 50.8&  51.4& 40  &  80  \\
 3346 & [Ne V]   & 64.6& 73.9& 67.0& 77.4& 64.9&  65.9& 54.3&  58  \\
 3426 & [Ne V]   & 79.8& 91.1& 86.1& 86.1& 72.1&  78.3& 71.9&  81  \\
 3869 & [Ne III] & 81.3& 95.2& 74.3& 91.7& 72.1&  85.9& 78.6&  81  \\
 3968 & [Ne III] & 89.0& 77.7& 56.1& 80.6& 82.0&  73.4& 61.8&  98  \\
 4363 & [O III]  &113.6& 88.2& 116 & 97.4& 95.7&  99.5& 90.0& 103  \\
\hline
\end{tabular}
\label{uv_line_measurements}
\end{table*}

\begin{table}
\caption{The parameters of the two-components TMAP model used to fit the RGS spectrum. $F_{\mathrm{(abs)}}$ and $F_{\mathrm{(unabs)}}$ represent the absorbed and unabsorbed fluxes, respectively.}
\begin{center}
\begin{tabular}{rr}
\hline
\centering
Parameter & Value\\
\hline
$N_{\rm H}$ (cm$^{-2}$) & 1.83 $\times 10^{21}$\\
$T_{\mathrm{eff, 1}}$ (K)  & 1.016 $\times 10^{6}$\\
$T_{\mathrm{eff, 2}}$ (K)  & 7.930  $\times 10^{5}$\\
Velocity blue-shift (km\,s$^{-1}$ & 588 \\
$F_{\mathrm{(1, abs)}}$ (erg\,cm$^{-2}$\,s$^{-1}$) & 6.18 $\times10^{-10}$\\
$F_{\mathrm{(1, unabs)}}$ (erg\,cm$^{-2}$\,s$^{-1}$) & 4.04 $\times10^{-9}$\\
$F_{\mathrm{(2, abs)}}$ (erg\,cm$^{-2}$\,s$^{-1}$) & 5.98 $\times10^{-10}$\\
$F_{\mathrm{(2, unabs)}}$ (erg\,cm$^{-2}$\,s$^{-1}$) & 5.26 $\times10^{-9}$\\
Total $F_{\mathrm{(abs)}}$ (erg\,cm$^{-2}$\,s$^{-1}$) & 1.22 $\times10^{-9}$\\
Total $F_{\mathrm{(unabs)}}$ (erg\,cm$^{-2}$\,s$^{-1}$) & 9.30 $\times10^{-9}$\\
 \hline
\end{tabular}
\end{center}
\label{table:two_mod_para}
\end{table}

\begin{table}
\caption{The parameters of the first model (Model 1) used to fit the \textit{Chandra} LETG spectrum. $T_{\texttt{bvapec}}$ represents the temperature of the collisionally ionized plasma around the nova. $F_{\mathrm{(abs)}}$ and $F_{\mathrm{(unabs)}}$ represent the absorbed and unabsorbed fluxes, respectively.}
\begin{center}
\begin{tabular}{rr}
\hline
\centering
Parameter & Value\\
\hline
$N_{\rm H}$ (cm$^{-2}$) & 1.4 $\times 10^{21}$\\
$T_{\mathrm{eff}}$ (K)  & 5.86 $\times 10^{5}$\\
$T_{\texttt{bvapec}}$ (eV) & 80.8\\
Atmosphere $F_{\mathrm{(abs)}}$ (erg\,cm$^{-2}$\,s$^{-1}$) & 1.13 $\times10^{-11}$\\
Atmosphere $F_{\mathrm{(unabs)}}$ (erg\,cm$^{-2}$\,s$^{-1}$) &  1.29 $\times10^{-10}$\\
CIE $F_{\mathrm{(abs)}}$ (erg\,cm$^{-2}$\,s$^{-1}$) & 1.61 $\times10^{-11}$\\
CIE $F_{\mathrm{(unabs)}}$ (erg\,cm$^{-2}$\,s$^{-1}$) &  3.46 $\times10^{-10}$\\
Total $F_{\mathrm{(abs)}}$ (erg\,cm$^{-2}$\,s$^{-1}$) & 2.74 $\times10^{-11}$\\
Total $F_{\mathrm{(unabs)}}$ (erg\,cm$^{-2}$\,s$^{-1}$) &  4.75 $\times10^{-10}$\\
 \hline
\end{tabular}
\end{center}
\label{table:chandra_mod_1}
\end{table}

\begin{table}
\caption{The parameters of the second model (Model 2) used to fit the \textit{Chandra} LETG spectrum.  $F_{\mathrm{(abs)}}$ and $F_{\mathrm{(unabs)}}$ represent the absorbed and unabsorbed fluxes, respectively.}
\begin{center}
\begin{tabular}{rr}
\hline
\centering
Parameter & Value\\
\hline
$N_{\rm H}$ (cm$^{-2}$) & 1.87 $\times 10^{21}$\\
$T_{\mathrm{eff}}$ (K)  & 6.36 $\times 10^{5}$\\
$T_{\mathrm{(atm)}}$ (eV) &  39 \\
Total $F_{\mathrm{(abs)}}$ (erg\,cm$^{-2}$\,s$^{-1}$) & 2.75$\times10^{-11}$\\
Total $F_{\mathrm{(unabs)}}$ (erg\,cm$^{-2}$\,s$^{-1}$) &  6.94 $\times10^{-9}$\\
\hline
\end{tabular}
\end{center}
\label{table:chandra_mod_2}
\end{table}

\section{Observation log}
\label{appD}

In this Appendix we list the log of the observations. The SMARTS and the {\em Swift} UVOT photometry can be found on the electronic version. The SMARTS time series photometry is available on the SMARTS online atlas, at \footnote{\url{http://www.astro.sunysb.edu/fwalter/SMARTS/NovaAtlas/nsmc2016/nsmc2016.html}}.

\begin{table}
\centering
\caption{A sample of the SMARTS \textit{BVRIJHK} photometry. The time series photometry is available on the electronic version and on the SMARTS atlas.}
\begin{tabular}{rrrrr}
\hline
HJD & $(t - t_0)$   & Band & Magnitude & Instrument\\ 
& (days) & & \multicolumn{2}{c}{(mag)}\\
\hline
2457657.48 & 1.98 & J & 5.245 & 0.003\\   
2457657.48 & 1.98 & H & 4.695 & 0.002\\  
2457657.48 & 1.98 & K & 4.850 & 0.003\\   
2457657.48 & 1.98 & I & 6.909 & 0.000\\
2457657.48 & 1.98 & B & 7.944 & 0.000\\ 
2457657.48 & 1.98 & V & 7.145 & 0.000\\ 
2457657.48 & 1.98 & R & 6.953 & 0.000\\   
2457658.48 & 2.98 & J & 5.288 & 0.003\\   
2457658.48 & 2.98 & H & 4.789 & 0.002\\  
2457658.48 & 2.98 & K & 4.629 & 0.003\\    
2457658.48 & 2.98 & I & 6.750 & 0.000\\
2457658.48 & 2.98 & B & 8.027 & 0.000\\ 
2457658.48 & 2.98 & V & 7.260 & 0.000\\  
2457658.48 & 2.98 & R & 6.160 & 0.001\\   
2457660.50 & 3.00 & J & 6.032 & 0.005\\   
2457660.50 & 3.00 & H & 5.409 & 0.003\\    
2457660.50 & 3.00 & K & 5.492 & 0.004\\    
2457660.50 & 3.00 & I & 6.793 & 0.000\\
2457660.50 & 3.00 & B & 8.107 & 0.000\\  
2457660.50 & 3.00 & V & 7.985 & 0.001\\ 
2457660.50 & 3.00 & R & 6.752 & 0.001\\  
\hline
\end{tabular}
\label{table:SMARTS}
\end{table}

\begin{table}
\centering
\caption{The SALT HRS spectroscopy observation log.}
\begin{tabular}{rrrrr}
\hline
  HJD & $ t - t_0$ & Exp. time (blue) & Exp. time (red) & Mode  \\ 
 & (days) &  \multicolumn{2}{c}{(s)} & \\
\hline
2457819.49 & 164 & 1800 & 1800 & LR\\
2457864.38 & 209 & 1800 & 1800 & LR\\
2457905.52 & 250 & 1800 & 1800 & LR\\
2457918.47 & 263 & 1800 & 1800 & LR\\
2457934.43 & 279 & 1800 & 1800 & LR\\
2457940.41 & 285 & 1800 & 1800 & LR\\
2457959.36 & 304 & 1800 & 1800 & LR\\
2457964.35 & 309 & 1800 & 1800 & LR\\
2457972.32 & 317 & 1800 & 1800 & LR\\
\hline
\end{tabular}
\label{table:SALT}
\end{table}

\begin{table}
\centering
\caption{A sample of the \textit{Swift} observation log, providing the observation-averaged XRT and UVOT \textit{uvw2} photometric measurements. The rest of the \textit{Swift} observation log is available on the electronic version.}
\begin{tabular}{rrrrrrr}
ObsID	&	$t-t_0$    &      half-bin width & XRT rate & error & $uvw2$  & $\Delta(uvw2)$\\
& (days) & & & & \multicolumn{2}{c}{(mag)}\\
\hline
00034741001  &	2.916	    &	0.024        &$<$  0.003   *  & -    &    -      &	       -\\
00034741002  &	8.210	    &	0.005        &$<$  0.014   *  & - 	  &    -      &	       - \\
00034741003  &	9.141	    &	0.005        &$<$  0.011   *  & - 	  &    -      &	       - \\
00034741004  &	10.300    &	0.097        &$<$  0.018   *  & - 	  &    -      &	       -\\
00034741005  &	11.492    &	0.035        &$<$  0.010   *  & - 	  &    -      &	       -\\
00034741006  &	12.162    &	0.035        &$<$  0.014   *  & - 	  &    -      &	       -\\
00034741007  &	13.595    &	0.005        &$<$  0.015   *  & - 	  &    -      &	       -\\
00034741008  &	14.392    &	0.006        &$<$  0.012   *  & - 	  &    -      &	       -\\
00034741009  &	15.157    &	0.035        & $<$ 0.013   *  & - 	  &    -      &	       -\\
\\						   
00034741010  &	16.383    &	0.006        & 6.6 * \$ & $(+2.6/-1.9)\times10^{-3}$  & -  &  -\\
00034741012  &	19.096    &	0.066        & $<$ 0.039   *  & -   &    -        &        -        \\
\\
00034741013  &	150.423   &	0.371      & 42.03  &  0.14	&  13.48  &   0.02  \\
00034741014  &	151.657   &	0.328      & 49.74  &  0.17	&      -        &        -\\		
00034741016  &	152.575   &	0.002      & 33.51  &  0.27	&  13.47  &    0.02\\
00034741017  &	153.784   &	0.005      & 41.82  &  0.21	&  13.61  &    0.02 \\
00034741018  &	154.381   &	0.005      & 35.69  &  0.20	&  13.57  &    0.02 \\
00034741019  &	155.512   &	0.002      & 32.81  &  0.28	&  13.49  &    0.02\\
00034741020  &	156.768   &	0.005      & 37.62  &  0.20	&  13.67  &    0.02 \\
00034741021  &	157.025   &	0.001      & 37.86  &  0.51	&  13.56  &    0.02 \\
00034741028  &	161.355   &	0.001      & 46.30  &  0.40	&  13.80  &    0.02\\
00034741030  &	164.670   &	0.001      & 36.88  &  0.37	&      -        &    	-\\	
00034741031  &	164.678   &	0.005      & 39.43  &  0.20	&      -        &    	-\\	
00034975002  &	167.261   &	0.001      & 40.72  &  0.46	&      -        &    	-\\	
00034975001  &	167.270   &	0.007      & 32.51  &  0.17	&  13.76  &    0.03\\
00034741034  &	168.628   &	0.105      & 37.67  &  0.10	&  13.84  &    0.02 \\
00034975003  &	170.054    &	0.001       & 29.65  &  0.35	&  13.89  &    0.02\\			
00034975004  &	170.056    &	0.000       & 26.09  &  0.48	&      -       &    	-\\		
00034975005  &	171.048    &	0.001       & 24.73  &  0.31	&      -       &    	-\\
\hline	   
\end{tabular}
\begin{tabular}{@{}cl}
\centering
* &  Grade 0 events only were considered before the start of the solar observing constraint,\\
 & because of concern about optical loading (see text for details).\\
\$ &  Possible detection.\\
\end{tabular}
\label{table:XRT_log}
\end{table}

\begin{table*}
\caption{UV grism observations with lenticular filters.}
\begin{tabular}{@{}llrrrl}
\hline
date/time        &  MJD    & ($t-t_0$)   & exposure&filter&  notes \\
  (UT)       &            & (days) & time (s)&      &        \\
\hline    
   2017-03-07T16:08 & 57819.67 & 164.67 &   892.5 & UGRISM&   (a)  \\

   2017-03-10T06:15 & 57822.26 &167.26 &   201.1 &   UVM2&        \\
   2017-03-10T06:19 & 57822.26 & 167.26 &  1107.0 & UGRISM&   (a)  \\
   2017-03-10T06:38 & 57822.27 & 167.27 &   103.6 &   UVW2&        \\

   2017-03-13T01:16 & 57825.05 & 170.05 &   240.5 &   UVW2&        \\
   2017-03-13T01:21 & 57825.05 & 170.05 &   103.5 & UGRISM&   (a)  \\ 

   2017-03-14T01:08 & 57826.04 & 171.04 &  285.5 &   UVM2&        \\
   2017-03-14T01:13 & 57826.05 & 171.05 &  1170.0 & UGRISM&   (a)  \\
   2017-03-14T01:33 & 57826.06 & 171.06 &    98.8 &   UVW2&        \\

   2017-03-26T01:57 & 57838.08 & 183.08 &   202.2 &   UVM2&        \\
   2017-03-26T02:01 & 57838.08 & 183.08 &   839.3 & UGRISM&   (b)  \\
   2017-03-26T02:15 & 57838.09 & 183.09 &    76.6 &   UVW2&        \\

   2017-04-11T09:59 & 57854.41 & 199.41 &   181.4 &   UVM2&        \\
   2017-04-11T10:03 & 57854.41 & 199.41 &   901.3 & UGRISM&   (b)  \\
   2017-04-11T10:18 & 57854.42 & 199.42 &    73.2 &   UVW2&        \\

   2017-04-14T00:15 & 57857.01 & 202.01 &    96.1 &   UVW2&        \\
   2017-04-14T00:17 & 57857.01 & 202.01 &   999.7& UGRISM&   (b)  \\
   2017-04-14T00:34 & 57857.02 & 202.02 &    92.0 &   UVW2&        \\

   2017-04-23T04:04 & 57866.16 & 211.16 &   222.8 &   UVM2&        \\
   2017-04-23T04:08 & 57866.17 & 211.17 &   889.5 & UGRISM&   (b)  \\
   2017-04-23T04:23 & 57866.18 & 211.18 &    85.4 &   UVW2&        \\
\hline
\end{tabular}
\begin{tabular}{@{}cl}
(a) & no offset on detector, roll=112$\deg$; First order contamination $<2050\AA$ \\
    & Zeroth orders contaminate [1750-1865],[1950-2185],[2410-2540] $\AA$ \\
(b) & offset around 5.6$\arcmin$ on the detector, roll = 130$deg$ \\
\end{tabular}
\label{uvgrism_obs_table}
\end{table*}

\begin{table*}
\centering
\caption{The \textit{XMM-Newton} observation log taken on day 168.}
\begin{tabular}{rrrrrr}
\hline
Exposure Number  &  Instrument & Observing Mode  & Filter & Start of exposure & Stop of exposure \\
\hline
001 & MOS1 & Full Frame    & MEDIUM FILTER   & 2017-03-11T11:02:17 & 2017-03-11T17:06:57 \\
002 & MOS2 & Timing Uncompressed     & MEDIUM FILTER   & 2017-03-11T11:02:57 & 2017-03-11T17:02:43\\
003 & pn   & Timing    & MEDIUM FILTER   & 2017-03-11T11:37:25 & 2017-03-11T17:07:17 \\
004 & RGS1 & Spectro + Q       & NOT APPLICABLE  & 2017-03-11T11:01:47 & 2017-03-11T17:08:12\\
005 & RGS2 & Spectro + Q             & NOT APPLICABLE  & 2017-03-11T11:01:52 & 2017-03-11T17:08:12\\
011 & OM   & OM Science User Defined & VISIBLE GRISM 2 & 2017-03-11T11:07:29 & 2017-03-11T12:12:35 \\
012 & OM   & OM Sci User Def FAST    & UVW1	       & 2017-03-11T12:12:36 & 2017-03-11T13:31:02 \\
013 & OM   & OM Sci User Def FAST    & UVW1	       & 2017-03-11T13:31:03 & 2017-03-11T14:16:09 \\
014 & OM   & OM Sci User Def FAST    & UVW1	       & 2017-03-11T14:16:10 & 2017-03-11T15:34:28\\
015 & OM   & OM Sci User Def FAST    & UVW1	       & 2017-03-11T15:34:29 & 2017-03-11T16:52:55\\
\end{tabular}
\label{table:XMM-Newton}
\end{table*}

\label{lastpage}
\end{document}